\documentclass[a4paper,11pt]{article}
\pdfoutput=1 % if your are submitting a pdflatex (i.e. if you have
\usepackage{jheppub} % for details on the use of the package, please
\usepackage[T1]{fontenc} % if needed

\usepackage{slashed}
\usepackage{bigints}
\usepackage{amssymb}
\usepackage{physics}
\usepackage{graphicx}
\usepackage[section]{placeins}
\usepackage{subcaption}
\usepackage{hepunits}
\usepackage{xcolor}

\newcommand{\eq}[1]{\begin{align}#1\end{align}}

\newcommand{\MeV}{\ {\rm MeV}}
\newcommand{\keV}{\ {\rm keV}}
\newcommand{\GeV}{\ {\rm GeV}}
\newcommand{\TeV}{\ {\rm TeV}}
\newcommand{\cm}{\ {\rm cm}}
\newcommand{\RNum}[1]{\uppercase\expandafter{\romannumeral #1\relax}}
\newcommand*\Diff[1]{\mathop{}\!\mathrm{d^#1}}
\DeclareRobustCommand{\rchi}{{\mathpalette\irchi\relax}} %Inline Chi
\newcommand{\irchi}[2]{\raisebox{\depth}{$#1\chi$}} 
\graphicspath{{./Images/}}

\usepackage[numbers,sort&compress]{natbib}

\title{Constraints from the Neutron EDM on Subleading Effective
  Operators for Direct Dark Matter Searches}

\author[a]{Manuel Drees,}

\author[a,1]{Rahul Mehra\note{Corresponding author.}}

\affiliation[a]{Bethe Center for Theoretical Physics and
  Physikalisches Institut, Universit\"{a}t Bonn, Nussallee 12, D-53115
  Bonn, Germany}

\emailAdd{drees@th.physik.uni-bonn.de}
\emailAdd{rmehra@physik.uni-bonn.de}

\abstract{Interactions between Dark Matter (DM) and nucleons relevant
  for direct search experiments can be organised in a model
  independent manner using a Galiliean invariant, non--relativistic
  effective field theory (NREFT). Here one expands the interactions in
  powers of the momentum transfer $\vec{q}$ and DM velocity
  $\vec{v}$. This approach generates many operators. The potentially
  most important subleading operators are odd under $T$, and can thus
  only be present in a theory with $CP$ violating interactions. We
  consider two such operators, called $\mathcal{O}_{10}$ and
  $\mathcal{O}_{11}$ in the literature, in simplified models with
  neutral spin$-0$ mediators; the couplings are chosen such that the
  coefficient of the leading spin independent (SI) operator, which
  survives for $\vec{v} \rightarrow 0$, vanishes at tree
  level. However, it is generically induced at the next order in
  perturbation theory. We perform a numerical comparison of the number
  of scattering events between interactions involving the $T-$odd
  operators and the corresponding loop induced SI contributions. We
  find that for ``maximal'' $CP$ violation the former can dominate
  over the latter. However, in two of the three models we consider, an
  electric dipole moment of the neutron (nEDM) is induced at two--loop
  order. We find that the experimental bound on the nEDM typically
  leads to undetectably small rates induced by ${\mathcal O}_{10}$.
  On the other hand, the model leading to a nonvanishing coefficient
  of ${\mathcal O}_{11}$ does not induce an nEDM.}

\begin{document} 
\maketitle
\flushbottom

%%%%%%%%%%%%%%%%%%%%%%%%%%%%%%%%%%%%%%%
\section{Introduction} \label{sec:intro}
%%%%%%%%%%%%%%%%%%%%%%%%%%%%%%%%%%%%%%%

The search for non--gravitational interactions of Dark Matter (DM) has
not yielded a convincing signal so far. This is in stark contrast with
the accumulated evidence for its gravitational interactions across a
broad range of astrophysical length scales. Direct searches aim to
detect non--gravitational interactions by observing the recoil of a
target nucleus after an incoming DM particle has scattered off
it. Terrestrial experiments using this search principle have led to
tight constraints on the DM mass and cross-section parameter space
\cite{XENON:2020rca, XENON:2018voc, PandaX-4T:2021bab, LUX:2016ggv,
  LUX:2017ree, LZ:2022lsv, DEAP:2019yzn, SuperCDMS:2017mbc,
  DarkSide:2018kuk, CRESST:2019jnq, NEWS-G:2017pxg, DAMIC:2020cut,
  PICO:2019vsc, LZ:2023lvz}. For the past few decades such
experimental efforts have focused on particle DM with weak scale
interactions termed as Weakly Interacting Massive Particles
(WIMPs). WIMPs can be produced in the early universe through
freeze--out from the thermal plasma, which yields the observed DM
relic abundance \cite{Planck:2018vyg} for very roughly electroweak
strength effective couplings. This has made WIMPs a particularly
compelling category of DM candidates.

In order to interpret the results of direct search experiments one has
to make assumptions regarding the nature of the interaction between DM
and nucleons. Traditionally it was assumed that DM--nucleon
interactions are dominated by just two operators describing spin
independent (SI) and spin dependent (SD) interactions in the limit of
zero WIMP velocity \cite{Bertone:2004pz}. These SI and SD operators
are the leading terms of an EFT description, which is an expansion in
powers of small parameters such as the DM velocity $v$ and three
momentum transfer scaled by the nucleon mass $\vec{q}/m_N$. Since
$v/c \lesssim \mathcal{O}(10^{-3})$ in the solar neighborhood, the
momentum transfer is restricted to
$\abs{\vec{q}} \lesssim \mathcal{O}(100 \MeV)$. Although the
momentum exchange can be substantial on nuclear physics scales, it is
far below the electroweak scale, and also below the range of WIMP
masses most direct search experiments can probe. Hence, it is
reasonable to expect that the canonical SI and SD interactions (zeroth
order terms) dominate the EFT expansion.

A detailed non--relativistic effective field theory (NREFT)
description of elastic DM--nucleus scattering retains NLO and NNLO
terms by modeling the nucleus as a spatially extended composite
particle with spin and charge densities \cite{Fan:2010gt,
  Fitzpatrick:2012ix, Fitzpatrick:2012ib, Anand:2013yka}. For WIMPs of
spins $0$ or $1/2$, truncating the EFT expansion at second order
generates a total of 14 operators.\footnote{The set of NREFT operators
  is substantially enlarged for WIMPs of higher spin
  \cite{Catena:2019hzw, Gondolo:2020wge}.} Since these higher order
terms vanish for $v \rightarrow 0$ (which implies 
$\vec{q} \rightarrow 0$ as well), they lead to a spectrum of recoil
energies quite different from the usual quasi--exponential fall--off,
often preferring energies larger than the typical range of values
implemented for SI and SD searches in experiments
\cite{Bozorgnia:2018jep}. Multiple experiments have now extended their
recoil energy search window in order to optimize the search strategy
in the NREFT framework \cite{XENON:2017fdd, PandaX-II:2018woa,
  SuperCDMS:2015lcz, CRESST:2018vwt, DarkSide-50:2020swd, LUX:2020oan,
  DEAP:2020iwi, CDEX:2020tkb, LUX:2021ksq, SuperCDMS:2022crd}.

The NREFT contains, at least, $28$ free parameters when the operators
for neutrons and protons are counted separately. Probing this large
parameter space is an arduous task. A number of global analyses using
data from current and planned direct detection experiments have placed
upper limits on the coupling strengths in this multi--dimensional
parameter space \cite{Catena:2014uqa, Catena:2014epa, Catena:2015uua,
Rogers:2016jrx, Liu:2017kmx, Kang:2018odb,Brenner:2022qku,AvisKozar:2023iyb}. 
A common conclusion drawn from these global analyses is that experiments
are nearly as sensitive to some momentum-- or velocity--dependent
operators that are odd under $P$ (parity) and $T$ (time reversal)
transformations as they are to the leading SD operator. These
subleading operators have been scrutinized less in the literature. It
is therefore interesting to study the phenomenology of models where
they are generated in the non--relativistic (NR) limit, and to
understand when these operators can dominate over the traditional SI
and SD interactions.

The CPT theorem \cite{luders1954equivalence, pauli1955exclusion}
implies that every $T-$odd NREFT operator must arise from a
$CP-$violating (CPV) quantum field theory, where $C$ refers to charge
conjugation. However, there are extremely stringent experimental
constraints on $CP$ violation \cite{Workman:2022ynf}, which can be
used to place constraints on such NREFT operators. The electric dipole
moment of the neutron (nEDM) is one such observable. The sensitivity
of current experiments \cite{Abel:2020pzs, Pendlebury:2015lrz,
  Burghoff:2011xk} is several orders of magnitude above the prediction
of the standard model of particle physics (SM), but they provide tight
constraints on flavor diagonal $CP$ violation in extensions of the
SM. The experimental upper limit on the nEDM is
\begin{equation} \label{eq:nEDM_limit}
| d_n | < 1.8 \times 10^{-26}  \; \text{e} \, \cdot \, \text{cm} \quad
\text{(90 \% C.L.)} \; .
\end{equation}

There have been a number of articles linking extensions of the SM to
low energy NREFT operators \cite{Dent:2015zpa, Bishara:2017pfq,
  DelNobile:2018dfg}. In particular, Ref.~\cite{Dent:2015zpa} lists a
set of simplified models for scalar, spinorial and vector DM
candidates and derives the full set of NREFT operators in terms of the
parameters for each simplified model. However, it is worth noting that
most extensions of the SM only generate a small subset of the NREFT
operators. Moreover, in most cases the leading order operators
describing the SI and SD interactions are generated as well, and will
then typically dominate. However, it is conceivable that the standard
operators are strongly suppressed, in which case formally subleading
operators actually provide the dominant contribution to scattering.

In Ref.~\cite{Drees:2019qzi}, we considered simplified models with
charged mediators, which are exchanged in the $s-$channel in
DM--nucleon scattering. We found that suppressing the leading SI
operator required finetuning of couplings; moreover, the bound on the
nEDM, which is generated at one--loop in these models, implies that
the subleading operators lead to undetectably small DM scattering
rates. The $P-$odd, $T-$odd operators can thus be neglected in such
scenarios.

In this article, we consider models that augment the SM by a WIMP
candidate and a mediator particle which does not carry electric or
color charge; DM--nucleon scattering then proceeds via $t-$channel
diagrams. We focus on WIMPs with spin $0$ or $1/2$, 
since in a renormalizable theory a spin$-1$ boson needs to be a gauge boson, 
making the construction of (semi-)realistic models somewhat more 
cumbersome. If one allows all couplings that respect the
$SU(3)_C \times U(1)_{\rm em}$ gauge symmetry of the simplified
Lagrangian, the leading SI term will be generated at tree level.
However, the coefficient of this operator can be ``switched off'' by
setting relevant coupling(s) to zero. This is an ad hoc choice, which
cannot be justified by any symmetry. In this case, at the lowest order
in perturbation theory, these models generate the $P-$odd, $T-$odd
operators $\mathcal{O}_{10}$ or $\mathcal{O}_{11}$ without giving rise
to the leading order operators.

However, since the vanishing of the coefficient of the leading SI
operator is not enforced by a symmetry, it will usually be generated
at the next order in perturbation theory. Since now the contribution
from the leading operator to DM--nucleon scattering is loop
suppressed\footnote{For a purely pseudoscalar mediator, loop--induced
  contributions to $\mathcal{O}_1$ typically dominate the DM
  scattering rate \cite{Arcadi:2017wqi, Bell:2018zra, Li:2018qip,
    Abe:2018emu}. However, the mediator has to have both scalar and
  pseudoscalar couplings in order to generate $\mathcal{O}_{10}$ or
  $\mathcal{O}_{11}$. A light mediator with general CP phases was
  considered in ref.\cite{Ertas:2019dew}, but constraints on these
  phases from EDMs were not considered, minimal flavor violation was
  assumed, and a trilinear coupling of the mediator to the SM Higgs
  boson was introduced.} while the contributions from
$\mathcal{O}_{10}$ and $\mathcal{O}_{11}$ are suppressed by powers of
$v$ or $|\vec{q}|/m_N$, it is not a priori obvious which contribution
is more important. We therefore numerically compare these
contributions to the total number of events for a Xenon target. We
find that for large regions in the parameter space of these models,
the subleading $P-$odd, $T-$odd operators actually dominate over the
SI term if the latter is purely loop--induced.

However, in models where tree level interactions generate only
$\mathcal{O}_{10}$, we find that an nEDM is generically induced at
two--loop order. The resulting upper bound on the couplings then again
leads to unobservably small scattering rates, so that contributions
from $\mathcal{O}_{10}$ to DM--nucleon scattering can be neglected. On
the other hand, in the model where $\mathcal{O}_{11}$ is generated at
tree level, the CP violation is restricted to the dark sector, and no
nEDM is generated (apart from the tiny SM contribution). This case
then provides an example of an NLO operator dominating over the
traditional SI operator in DM--nucleon scattering, at least at the
level of a simplified model.

The remainder of this article is organized as follows. In Section
\ref{sec:nreft-simplified-models}, we provide a brief introduction to
the NREFT formalism, and introduce three $CP$ violating simplified
models yielding the $P-$ and $T-$odd NREFT operators. We also compute
the loop diagrams for WIMP--nucleon scattering that give rise to the
leading SI operator in the non--relativistic limit. We then match the
different scattering contributions to the corresponding four--field
effective operators and finally match these to the set of NREFT
operators. We also compute the two--loop Feynman diagrams that induce
a nEDM and discuss the implications for the corresponding $P-,T-$odd
operator in the NREFT. In Section~\ref{sec:results-and-discussions} we
compute the number of events for a Xenon target for the three
simplified models. We discuss our main numerical results, comparing
the contributions of the NREFT operators to elastic WIMP--nucleon
scattering. We conclude in Section \ref{sec:conclusions}. Details of
our loop computations are given in Appendices
\ref{sec:one-loop-calculations} and \ref{app:two-loop-calculations}.

%%%%%%%%%%%%%%%%%%%%%%%%%%%%%%%%%%%%%%% 
\section{NREFT and Simplified Models} 
\label{sec:nreft-simplified-models}
%%%%%%%%%%%%%%%%%%%%%%%%%%%%%%%%%%%%%%%

A non--relativistic effective field theory (NREFT) of elastic
scattering between DM and nuclei exhaustively categorizes the possible
interactions involved in direct searches. An incoming DM particle
striking a target nucleus on Earth is quite slow in the detector rest
frame, $v/c \sim \mathcal{O}(10^{-3})$, and therefore a
non--relativistic EFT can be used to describe the scattering. This
simplifies the nuclear physics required to compute the scattering
rate, which is nevertheless rather nontrivial
\cite{Fitzpatrick:2012ix}. Traditionally only the leading terms were
kept, which remain finite as $v \rightarrow 0$
\cite{Bertone:2004pz}. The first and second order terms in DM velocity
$v$ and momentum transfer $\vec{q}/m_N$ (in units of the nucleon mass)
were considered only relatively recently \cite{Anand:2013yka,
  Fitzpatrick:2012ix, Fitzpatrick:2012ib, DelNobile:2018dfg,
  DelNobile:2021wmp}. In the following we briefly summarize the
salient points.

In this NREFT elastic DM--nucleon scattering is described using a
basis of operators constructed from the following Hermitian quantities
invariant under Galilean transformations:
\eq{ \label{eq:nreft-building-blocks}
  i \vec{q}, \quad
  \vec{v}^{\; \perp} \equiv \vec{v} + \dfrac{\vec{q}}{2 \mu_N}, \quad
  \vec{S}_N, \quad \vec{S}_\rchi \,.
}
Here $\mu_N = m_N m_{\rm DM} / (m_N + m_{\rm DM}) $ is the reduced
mass of the DM--nucleon system. Energy conservation implies that the
transverse velocity $\vec{v}^\perp$ is orthogonal to the momentum
transfer $\vec{q}$. $\vec{S}_N$ and $ \vec{S}_\rchi$ are the spin of
the nucleon and the WIMP $\chi$; of course, the latter may be
zero. Using these four building blocks, and only imposing Galilean
invariance, one obtains a set of linearly independent operators
$\mathcal{O}_i$ when the EFT is truncated at second order in the
expansion parameter $\vec{q}/m_N$. Table \ref{table:operatorlist}
lists this set of $14$ operators; the operators $\mathcal{O}_1$ and
$\mathcal{O}_4$ describe the traditional leading SI and SD
interaction, respectively. Note that the coefficients of these
operators are in general different for neutrons and protons.

No invariance under any discrete symmetry was imposed in the
construction of the NREFT operators. Their behavior under discrete
transformations, in particular parity ($P$) and time reversal ($T$),
can thus be used to classify the operators. A parity transformation
corresponds to $\vec{q} \rightarrow - \vec{q}, \ \vec{v}^\perp
\rightarrow - \vec{v}^\perp$, while $\vec{S}_N$ and $\vec{S}_\rchi$,
being pseudovectors, remain unchanged. On the other hand, under
time reversal, all four vectors listed in (\ref{eq:nreft-building-blocks})
change sign, and in addition $i \rightarrow -i$ (i.e. the Hermitian
operator $i \vec{q}$ remains unchanged). This leads to the following
classification:
\eq{
%  \label{eq:classi}
 \mathcal{O}_1, \mathcal{O}_3, \mathcal{O}_4, \mathcal{O}_5, \mathcal{O}_6:
  \quad &\text{$P-$even and $T-$even}, \nonumber \\
  \mathcal{O}_7, \mathcal{O}_8, \mathcal{O}_9: \quad
  &\text{$P-$odd and $T-$even}, \nonumber \\
  \mathcal{O}_{13}, \mathcal{O}_{14}: \quad
  &\text{$P-$even and $T-$odd}, \nonumber \\
  \mathcal{O}_{10}, \mathcal{O}_{11}, \mathcal{O}_{12}: \quad
  &\text{$P-$odd and $T-$odd} . \nonumber }

{
 \renewcommand{\arraystretch}{1.5}
\begin{table}[t]
\centering
\begin{tabular}{l@{\hskip 0.15in}l@{\hskip 0.15in}l}
  $\mathcal{O}_1 = 1_\rchi 1_N$;
  & $\mathcal{O}_6 = \left( \dfrac{\vec{q}}{m_N} \cdot \vec{S}_N \right)
    \left( \dfrac{\vec{q}}{m_N}	\cdot \vec{S}_\rchi \right)$;
  &  $\mathcal{O}_{10} = i \dfrac{\vec{q}}{m_N} \cdot \vec{S}_N$; \\
  $\mathcal{O}_3 = i \vec{S}_N \cdot \left( \dfrac{\vec{q}}{m_N} \times
  \vec{v}^{\perp} \right)$;
& $\mathcal{O}_7 = \vec{S}_N \cdot \vec{v}^{\perp} $;
& $\mathcal{O}_{11} = i \dfrac{\vec{q}}{m_N} \cdot \vec{S}_\rchi $; \\
$\mathcal{O}_4 = \vec{S}_{\rchi} \cdot \vec{S}_{N}$;
& $\mathcal{O}_8 = \vec{S}_\rchi \cdot \vec{v}^{\perp} $;
& $\mathcal{O}_{12} = \vec{S}_\rchi \cdot (\vec{S}_N \times \vec{v}^{\perp})$; \\
  $\mathcal{O}_5  = i \vec{S}_\rchi \cdot  \left( \dfrac{\vec{q}}{m_N} \times
  \vec{v}^{\perp} \right)$;
& $\mathcal{O}_9 = i \vec{S}_\rchi \cdot \left(\vec{S}_N \times
	\dfrac{\vec{q}}{m_N} \right)$;
& $\mathcal{O}_{13} = i ( \vec{S}_\rchi \cdot \vec{v}^{\perp} )
\left(\dfrac{\vec{q}}{m_N} \cdot \vec{S}_N \right) $;
\end{tabular}
$\mathcal{O}_{14} = i (\vec{S}_N \cdot \vec{v}^{\perp}) \left(
  \dfrac{\vec{q}}{m_N} \cdot \vec{S}_\rchi \right)$
\setlength{\tabcolsep}{15pt}
\caption{List of operators in the NREFT for elastic WIMP--nucleon
  scattering. We adopt the conventions of \cite{Anand:2013yka} by
  defining the operators normalized by the nucleon mass $m_N$ in order
  to have a dimensionless basis. We omit the invariant
  $\mathcal{O}_2 = v_{\perp}^2$ because it is a second order
  correction to the SI operator $\mathcal{O}_1$, as well as
  $\mathcal{O}_{15} = - \left( \vec{S}_{\rchi} \cdot
    \frac{\vec{q}}{m_N} \right) \left( (\vec{S}_N \times
    \vec{v}^{\perp})\cdot \frac{\vec{q}}{m_N} \right)$ since it
  generates a cross section of order $v_T^6$, which is
  $ \text{N}^3 $LO.}
\label{table:operatorlist}
\end{table} }

The $P$ and $T$ quantum numbers of the NREFT operators must match
those of the relativistic operators generating them. In particular,
the $CPT$ theorem stipulates that only a $CP$ violating quantum field
theory with a DM candidate can yield any of the $T-$odd operators. Out
of those, the operators $\mathcal{O}_{12}, \, \mathcal{O}_{13}$ and
$\mathcal{O}_{14}$ are not generated in scenarios with $t-$channel
mediator with spin $\leq 1$;\footnote{$\mathcal{O}_{12}$ can be
  generated in a model with a spin$-1/2$ WIMP and spin$-0$ $s-$channel
  mediator, but the constraint from the nEDM forces this contribution
  to be negligible \cite{Drees:2019qzi}.} the latter two operators
anyway contribute little to the scattering cross section if all
operators have coefficients of similar size \cite{Kang:2018odb}. 
Therefore, we focus on $\mathcal{O}_{10}$ and $\mathcal{O}_{11}$ in 
the following.

Connecting a relativistic model for DM leading to specific DM--quark
and/or DM--gluon interactions with the NREFT in general involves two
steps. First, one integrates out the heavy mediator(s), where
``heavy'' here refers to all mediators $\phi$ with mass $m_\phi$ well
above $100 \MeV$, which is the maximal three--momentum exchange in
DM--nucleus scattering.\footnote{In order to include light mediators
  in the NREFT the differential cross section should be multiplied
  with $[m_\phi^2 / (m_\phi^2 + \vec{q}^2) ]^2$; this will lead to a
  softening of the recoil spectrum for $\vec{q}^2 \gtrsim m_\phi^2$.}
This results in relativistic but non--renormalizable four--field
DM--quark and/or DM--gluon operators. In the second step one takes the
non--relativistic limit of these four--field operators and matches them
onto the NREFT operators listed in
Table~\ref{table:operatorlist}. This leads to an effective Lagrangian
containing some (or all) of these operators, with coefficients
determined by the couplings and masses of the original relativistic
theory. The computation of DM--nucleus scattering rates from these
coefficients involves numerous nuclear ``response functions''; we
refer to ref.\cite{Anand:2013yka}, whose expressions we used in our
own numerical code.

As already mentioned, the NREFT is an expansion in powers of $v$ or
$\vec{q}$. If all coefficients in the NREFT are of comparable
magnitude, the total scattering rate typically receives the largest
contribution from the SI operator $\mathcal{O}_1$, which is of zeroth
order in the expansion. Moreover, this contribution is enhanced by $A^2$,
where $A$ is the nucleon number. For large momentum exchange, $|\vec{q}|
\gtrsim 1/r_N$ where $r_N$ is the radius of the target nucleus, the rate
is somewhat suppressed by a form factor, but even without form factor
this contribution to the scattering rate peaks at $|\vec{q}| \rightarrow
0$.

$\mathcal{O}_4$ is also of zeroth order in the expansion. However,
since the spins of the nucleons largely cancel in any given nucleus,
there is no $A^2$ enhancement; in fact, $\mathcal{O}_4$ does not
contribute at all if the target nucleus has no spin.

As already noted, for sufficiently heavy DM particle (and target
nucleus) the three--momentum exchange can reach
$\abs{\vec{q}}\sim 100$ MeV. In this high momentum exchange region of
phase space the operators that are linear in $\vec{q}$ and independent
of $\vec{v}^\perp$ are therefore only suppressed by a factor
$\sim 0.1$; these are the operators $\mathcal{O}_{10}$ and
$\mathcal{O}_{11}$ which are the focus of our study.\footnote{However,
  simplified models yielding $\mathcal{O}_{11}$ in the
  non--relativistic limit always seem to result in the coefficient
  $c_{11}$ containing an extra factor $m_N/m_{\rm DM}$
  \cite{Dent:2015zpa, Bishara:2017pfq, DelNobile:2018dfg,
    Drees:2019qzi}; see also eq.\eqref{eqn:model-iib-c11n} below. A
  discussion assuming $m_{\rm DM}$ independent Wilson coefficients
  therefore overestimates the importance of $\mathcal{O}_{11}$. Even
  with this caveat, $\mathcal{O}_{11}$ remains the potentially most
  important non--leading NREFT operator for
  $m_{\rm DM} \lesssim 100 \GeV$.} The contribution from
$\mathcal{O}_{11}$ is $A^2$ enhanced, up to a form factor, but
requires the DM particle to carry spin. The contribution from
$\mathcal{O}_{10}$ suffers similar cancellations as that from
$\mathcal{O}_4$, but can survive even for scalar DM particle.

Naively the contributions of operators involving $\vec{v}^\perp$
should be suppressed by a factor $v^2 \sim 10^{-6}$; indeed, compared
to the contribution from $\mathcal{O}_1$ this is almost true. However,
the richer structures can lead to nuclear response
$\propto \vec{S}_N \cdot \vec{L}_N$ \cite{Anand:2013yka}, $\vec{L}_N$
being the orbital angular momentum of a given nucleon; in this product
the contributions of paired nucleons do {\em not} cancel. As a result,
for heavy target nuclei and similar coefficients the contribution from
$\mathcal{O}_{12}$ often exceeds that from $\mathcal{O}_4$
\cite{Kang:2018odb}; however, as already noted $\mathcal{O}_{12}$ is
not generated in the models with neutral mediator that we consider in
this article. Similarly, for equal coefficients $\mathcal{O}_3$
typically contributes almost as much as $\mathcal{O}_{10}$ does
\cite{Kang:2018odb}, but $\mathcal{O}_3$ is not generated in leading
order when starting from a relativistic theory \cite{Dent:2015zpa}.
$\mathcal{O}_{10}$ and $\mathcal{O}_{11}$ are therefore the
potentially most important higher order NREFT operators that can be
generated from a relativistic QFT.

After these preliminaries, we are ready to introduce the simplified
models we consider in this analysis. Recall that, in the spirit of
ref.\cite{Dent:2015zpa}, we only impose invariance under
$SU(3)_C \times U(1)_{\rm em}$, not under the complete gauge group of
the SM.

\subsection{Model I}

Model I contains a complex spin--zero WIMP $S$ and a real spin--zero
mediator $\phi$; both are gauge singlets. We assume that the WIMP is
odd and the mediator and all the SM particles are even under a new
discrete symmetry $\mathbb{Z}_2$. This forbids dark matter decay, but
allows $\phi$ to couple to both $S$ and to SM quarks $q$. The most
general gauge invariant Lagrangian respecting the new $\mathbb{Z}_2$
symmetry and keeping the real and imaginary parts of $S$ degenerate
thus is:\footnote{Essentially identical results can be derived for a
  real spin--zero WIMP.}
\eq{ 	\label{eq:model-i-lagrangian}
  \mathcal{L}^{\text{I}}
  &= ( \partial_{\, \mu} S)^\dagger \, (\partial^{\, \mu} S )
  - m_S^2 \, S^\dagger S - \dfrac{\lambda_S}{2} \, (S^\dagger \, S)^2
  + \dfrac{1}{2} \partial_{\, \mu} \phi \, \partial^{\, \mu} \phi
  - \dfrac{1}{2} m_\phi^2 \, \phi^2 - \dfrac{m_\phi \, \mu_1}{3} \, \phi^3
  - \dfrac{\mu_2}{4} \, \phi^4 \nonumber \\
  & - g_1 \, m_S \, S^\dagger S \, \phi - \dfrac{g_2}{2} \, S^\dagger
  S \, \phi^2 -  h_1^{ij} \, \phi \, \bar{q}_{i}  q_{j}
  -i h_2^{ij} \, \phi \, \bar{q}_{i} \gamma^5 q_{j}\; .
}
$U(1)_{\text{em}}$ invariance implies that the mediator $\phi$ can
only couple to quarks with identical electric charge. Hence the quark
flavor indices $i$ and $j$ in the Yukawa coupling matrices are
restricted to the same quark type. However, non--vanishing couplings
of the mediator to quarks of different generations generate flavor
changing neutral currents (FCNC) processes at tree level. Experimental
constraints arising from meson mixing along with rare flavor changing
decays severely limit these flavor non--diagonal couplings. Therefore
we assume that the Yukawa coupling matrices in
eq.\eqref{eq:model-i-lagrangian} are diagonal in flavor space, in
particular $h_{1,2}^{q} \equiv h_{1,2}^{i i}$ where $q$ denotes the
quark flavor the mediator couples to.

In eq.(\ref{eq:model-i-lagrangian}) we have allowed both scalar and
pseudoscalar couplings of $\phi$ to the SM quarks. However, if
$g_1 h_1^q \neq 0$ there will be tree level contributions to
$\mathcal{O}_1$, which will then completely dominate WIMP--nucleus
scattering, in which case the nonleading operators in the NREFT would
not need to be considered. Moreover, $h_1^q h_2^q \neq 0$ for any
quark $q$ would yield an electric dipole moment of that quark at
one--loop order, leading to very strong constraints on the parameters
of the model. We therefore impose
\begin{equation} \label{eq:h1}
  h_1^q = 0 \ \forall q\,.
\end{equation}
We emphasize that this is completely ad hoc but necessary for 
considering the non-leading operators in the NREFT. In particular, 
it cannot be justified by any symmetry. In contrast, the requirement
$h_2^q = 0$, which would also remove the EDMs but allow
$\mathcal{O}_1$, could be justified by demanding $CP$
conservation. Nevertheless ``switching off'' $\mathcal{O}_1$ via
eq.(\ref{eq:h1}) is still less inelegant than requiring specific
relations between non--zero couplings, as we had to do in our earlier
analysis of models with charged $s-$channel mediators
\cite{Drees:2019qzi}. On the other hand, since eq.(\ref{eq:h1}) cannot
be enforced by a symmetry, we can already anticipate that higher order
contributions will generate an effective $h_1^q$ or, more generally,
lead to nonvanishing $\mathcal{O}_1$.

The matrix element for tree level $t-$channel scattering
$S(p_S) + q (p_q) \rightarrow S(p_{S}^\prime) + q(p_{q}^\prime )$ is
given by
\eq{ \label{eq:model-i-tree-o10-matrix-element}
 \mathcal{M}_{Sq \rightarrow Sq}^{ \text{I} } =
  -\dfrac{h_2^q \, g_1 \, m_S}{q^2-m_\phi^2} \, \bar{u}(p_q^\prime) \,
  i \gamma^5 \, u(p_q) \; .
}
For $\left| q^2 \right| \ll m^2_\phi$ we can ignore the $q^2$ term in
the $\phi$ propagator. The matrix element can then be matched on to a
relativistic effective operator $S^\dagger S \; \bar{q} \, i\gamma^5 \, q$,
which reduces to the operator $ \mathcal{O}_{10} $ in the
non--relativistic limit.

\begin{figure*}[t!]
\centering
\includegraphics[width=\linewidth]{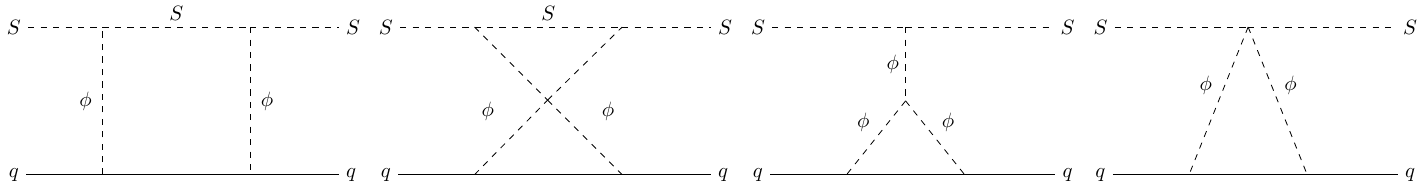}
\caption{One--loop box and triangle Feynman diagrams in Model I which
  give contributions to $\mathcal{O}_{1}$.}
\label{fig:Model_1-Skeleton}
\end{figure*}

For Model I, at the lowest order in perturbation theory, thus only the
operator $\mathcal{O}_{10}$ contributes to DM--nucleus
scattering. However, at the next--to--leading order the one--loop
Feynman diagrams shown in Fig.~\ref{fig:Model_1-Skeleton} can induce
contributions from the operator $\mathcal{O}_1$. We use the Dirac
equation and four--momentum conservation to write the resulting matrix
elements in a form that is symmetric in DM momenta; see
Appendix~\ref{sec:appendix-b-model-i} for details. The relativistic
effective Lagrangian for Model I derived in this manner can be written
as
\eq{ \label{eq:model-i-rel-eff-lagrangian}
  \mathcal{L}^{\text{I}}_{\text{eff}} \supset
  c^{q,d5}_{1,S} S^\dagger S \; \bar{q} q \,
  +\, c^{q,d5}_{10} S^\dagger S \; \bar{q} \, i\gamma^5 \, q \,
  +\, c_{1,V}^{q,d6}  i \left(S^\dagger \overleftrightarrow{\partial_\mu} S
  \right) \; \bar{q} \gamma^\mu q \, .
}
The Hermitean derivative on the complex scalars is defined as
$i S^\dagger \overleftrightarrow{\partial_\mu} S \equiv
\frac{i}{2}(S^\dagger \partial_\mu S - S \partial_\mu S^\dagger)$. The
subscripts $i$ on the quark-level Wilson coefficients
$c_i^{q,dj} \ (j=5,6)$ denote the NREFT operator that the
corresponding relativistic effective operator reduces; in case of
$\mathcal{O}_1$ we have distinguished the coefficient $c_{1,S}$ of the
product of two scalar currents from $c_{1,V}$ which multiplies the
product of two vector currents. Finally, the superscripts $d5$ and $d6$
refer to field operators with mass dimension $5$ and $6$, respectively.

{\renewcommand{\arraystretch}{1.25}
\begin{table}[t!]
\centering
\begin{tabular}{c c c c}
  & $ S^\dagger \Gamma_S S  \; \bar{q} \, \Gamma_q \,q $ &
  & $c^q_i$ \\ \hline \hline \rule{0pt}{5ex} Tree
  & $ c^{q,d5}_{10} S^\dagger S \; \bar{q} \, i\gamma^5 \, q$
  & $ \longrightarrow $ & $ \dfrac{h_2^q \, g_1}{m_\phi^2}$ \vspace*{1mm} \\
 \hline  \vspace*{-4mm} \\
  Box& $c_{1,V,B}^{q,d6}  i \left(S^\dagger \overleftrightarrow{\partial_\mu}
       S \right) \; \bar{q} \gamma^\mu q $ & $ \longrightarrow $
  & $ \dfrac{(h_2^q)^2 \, g_1^2 \, m_S^2}{16 \pi^2} \, M_1$ \\
  \rule{0pt}{5ex}& $ c^{q,d5}_{1,S,B} S^\dagger S \; \bar{q} q$&$\longrightarrow$
  &$\dfrac{(h_2^q)^2 \, g_1^2 \, m_S \, m_q}{16 \pi^2} \, M_2$
 \vspace*{1mm}  \\ \hline \rule{0pt}{5ex}Crossed
& $c_{1,V,C}^{q,d6} i\left(S^\dagger \overleftrightarrow{\partial_\mu} S \right)\;
   \bar{q} \gamma^\mu q$ & $\longrightarrow$
  &$-\dfrac{(h_2^q)^2 \, g_1^2 \, m_S^2}{16 \pi^2} \, M_3$ \\
  \rule{0pt}{5ex}Box& $ c^{q,d5}_{1,S,C} S^\dagger S \; \bar{q} q$
 & $ \longrightarrow $ &$\dfrac{(h_2^q)^2 \, g_1^2 \, m_S \, m_q}{16 \pi^2}
  \, M_4$ \vspace*{1mm} \\  \hline
  \rule{0pt}{5ex}Triangle 1 & $ c^{q,d5}_{1,S,T1} S^\dagger S \; \bar{q} q$
  & $\longrightarrow$ & $\dfrac{(h_2^q)^2 \, g_1 \, \mu_1}{16 \pi^2}
              \dfrac{m_q}{m_\phi} \, L_1$ \vspace*{1mm} \\  \hline
  \rule{0pt}{5ex}Triangle 2& $ c^{q,d5}_{1,S,T2} S^\dagger S \; \bar{q} q$
 & $ \longrightarrow $ & $ -\dfrac{(h_2^q)^2 \, g_2}{16 \pi^2} \dfrac{m_q}
                      {2 m_S} \, L_1$ \vspace*{1mm}\\  \hline
\end{tabular}
\caption{Non--relativistic reduction of relativistic effective
  operators in Model I. The middle column gives the relativistic
  four--field operators that appear in the matrix element for
  WIMP--nucleon scattering. The right column gives the corresponding
  quark--level Wilson coefficient $c_i^q$ of the relevant NREFT
  operators, where we have suppressed the dependence of the loop
  functions on dimensionless parameters, i.e. $M_i \equiv M_i(r,s)$
  and $L_1 \equiv L_1(r) $ with $r \equiv m_q/m_\phi$ and
  $s \equiv m_S/m_\phi$. The first line contributes to the coefficient
  of $\mathcal{O}_{10}$, all other lines describe contributions
  $\propto \mathcal{O}_1$.}
\label{table:coeff_model_i}
\end{table}
}

Table \ref{table:coeff_model_i} describes the matching of the
relativistic effective operators onto the NREFT operators in terms of
the parameters of Model I. The Wilson coefficients for the
dimension--5 operators $(S^\dagger S) (\bar{q} q)$ and
$(S^\dagger S) (\bar{q} i \gamma^5 q)$ have been divided by a factor
of $m_S$, i.e. these are coefficients of the dimension--6 operators
$m_S (S^\dagger S) (\bar q q)$ and
$m_S (S^\dagger S) (\bar q i \gamma^5 q)$; this ensures that the
expressions for all DM--nucleon cross sections contain the same factor
$\frac{\mu_{\chi N}^2}{\pi} \, (c_{i}^N)^2$ irrespective of the mass
dimension of the relativistic operator involved. The loop functions
$M_i(r,s)$ and $L_i(r)$ appearing in the box and triangle diagrams
have been expressed as functions of dimensionless parameters
$r \equiv m_q/m_\phi$ and $s \equiv m_S/m_\phi$. Analytical
expressions for these loop functions for Model I, and their
$m_q \rightarrow 0$ limits, can be found in
Appendix~\ref{sec:appendix-b-model-i-loop-functions}.

The dimension--5 scalar--scalar operator $(S^\dagger S) (\bar{q} q)$
as well as the dimension--6 vector--vector operator
$i \left(S^\dagger \overleftrightarrow{\partial_\mu} S \right) \;
\bar{q} \gamma^\mu q$ both reduce to the leading order SI operator
$\mathcal{O}_1$ in the non--relativistic limit. Only the box diagrams
give rise to both vector--vector and scalar--scalar relativistic
operators, whereas the two triangle diagrams yield only the
scalar--scalar operator. The first triangle diagram could be described
in terms of an effective, loop--generated coupling $h_1^q$, thereby
confirming our expectation that the choice $h_1^q = 0$ is not
technically natural; however, all four diagrams shown in
Fig.~\ref{fig:Model_1-Skeleton} contribute at the same order in
perturbation theory, and should therefore be included in a full NLO
treatment. The loop--generated Wilson coefficients are all suppressed
by the loop factor $1/16\pi^2$ compared to the tree level diagram. The
loop functions $M_i$ have mass dimension $-4$ while $L_1$ has mass
dimension $-2$, hence the quark--level Wilson coefficients $c_i^q$ all
have the same mass dimension $-2$. Since the scalar--scalar operator
violates chirality, the contributions to $c_{1,S}^{q,d5}$ are all
$\propto m_q$.

The quark bilinears in eq.\eqref{eq:model-i-rel-eff-lagrangian} and
Table~\ref{table:coeff_model_i} must be promoted to nucleon bilinears
in order to describe DM--nucleon scattering. The quark--level Wilson
coefficients therefore have to be combined with corresponding nucleon
embedding factors in order to derive the Wilson coefficients at the
nucleonic level \cite{Jungman:1995df, Fitzpatrick:2012ix}. Including
contributions from the box and the triangle diagrams, the nucleonic
Wilson coefficient $c^N_1$ of the NREFT operator $ \mathcal{O}_1$
becomes:
\eq{ \label{eq:model-i-coef-c1N}
  c_1^{N} |^{\text{I}} = \dfrac{1}{16 \, \pi^2} &\Big[ \, g_1^2 \, m_S^2
  \Big\{ \sum_{u,d} (h_2^q)^2 \mathcal{N}^{N}_q ( M_1(r_q, s) - M_3(r_q, s) )
  \Big\} \nonumber \\
  &+ m_N \left( \sum_{u,d,s} (h_2^q)^2 f^N_{Tq} + \dfrac{2}{27}
    f^N_{TG} \sum_{c,b,t} (h_2^q)^2 \right) \nonumber \\
  & \hspace*{5mm} \times \Big\{ g_1^2 m_S \, \Big( M_2(r_q,s) + M_4(r_q,s) \Big)
  + \left(
  \dfrac{g_1 \mu_1}{m_\phi} - \dfrac{g_2}{2 m_S}\right) L_1(r_q) \Big\} \Big]\,.
}
Promoting the vector quark bilinear
$\mathcal{N}^N_q \equiv \bra{\bar{N}} \bar{q} \gamma^\mu q \ket{N}$ to
the vector nucleon bilinear yields the number of valence quarks of
flavor $q$ in the nucleon $N$, i.e.
$\mathcal{N}^p_u = \mathcal{N}^n_d = 2, \; \mathcal{N}^p_d =
\mathcal{N}^n_u = 1$. The contribution of light quarks to the nucleon
mass (scalar nucleon bilinear) is given by
\eq{
\label{eq:fnTq}
q=u,d,s: \, \bra{\bar{N}} m_q \, \bar{q} q \ket{N} = m_N f^N_{Tq}\,.
}
The heavy quarks contribute to the nucleon mass via the trace anomaly of
the energy--momentum tensor \cite{Shifman:1978zn}:
\eq{
\label{eq:fnTG}
q = c,b,t: \, \bra{\bar{N}} m_q \, \bar{q} q \ket{N} = \dfrac{2}{27} m_N
f^N_{TG} = \dfrac{2}{27} m_N \Big( 1 - \sum\limits_{q'=u,d,s} f^N_{Tq'} \Big)
\; .
}

At first sight one might think that the contribution from the product
of two vector currents dominates, since the corresponding nucleonic
matrix elements $\mathcal{N}^N_q$ are large. However, in the limit
$r_q \rightarrow 0$ the difference $M_1(r_q,s) - M_3(r_q,s)$ vanishes
$\propto r_q^2$. In contrast, the sum
$M_2(r_q,s) + M_4(r_q,s) \rightarrow 2 M_2(0,s)$ remains finite for
massless quarks. As a result, the first term in
eq.\eqref{eq:model-i-coef-c1N} typically contributes much less than
the remaining terms, which originate from the product of scalar
currents.

For the tree--level scattering contribution reducing to
$\mathcal{O}_{10}$, the nucleonic Wilson coefficient in the limit of
vanishing momentum transfer is \cite{Fitzpatrick:2012ix}
\eq{ \label{eq:model-i-coef-c10N}
  c^{N}_{10}|^{\text{I}} = \dfrac{g_1}{m_\phi^2} \, \Big(
  \sum\limits_{u,d,s} h_2^{q} \, \Delta \tilde{q}^N - \Delta \tilde{G}^N
  \sum\limits_{c,b,t} \dfrac{h_2^{q}}{m_q} \Big)
}
where $\Delta \tilde{q}^N$ and $\Delta \tilde{G}^N$ are the light and
heavy quark contributions to the nucleon level pseudoscalar bilinear,
respectively. The latter hadronic matrix element is due to the QCD
chiral anomaly:
\eq{ \label{eq:model-i-DeltaG}
  q = c,b,t: \, \bra{\bar{N}} \partial_\mu(\bar{q} \gamma^\mu \gamma^5 q)
  \ket{N} = 2 m_q \bra{\bar{N}} \bar{q} i \gamma^5 q \ket{N}
  + \dfrac{\alpha_s}{4 \pi} \bra{\bar{N}} G^{a \mu \nu} \tilde{G}^a_{\mu \nu}
  \ket{N} \; ,
}
where $G$ and $\tilde G$ are the gluonic field strength tensor and its
dual, respectively. The left--hand side can be set to zero since
heavy quarks have no significant dynamics in the nucleon. Throughout
our calculations, we take the numerical values of the coefficients
that appear when quark bilinears are promoted to nucleon bilinears as
given in the Appendix of Ref.~\cite{Dent:2015zpa}.

\subsection{Model II}

In Model II we replace the complex scalar WIMP by a spin$-1/2$ gauge
singlet Dirac fermion $\rchi$, again using a real spin--zero mediator
$\phi$. The couplings of the SM quarks $q$ to the mediator are as in
Model I. The WIMP $\rchi$ can also couple to the mediator $\phi$ via a
scalar ($\lambda_1$) and a pseudoscalar coupling ($\lambda_2$). The
renormalizable $SU(3)_C \times U(1)_{\text{em}}$ invariant Lagrangian
is thus given by
\eq{ \label{eq:L_model2}
  \mathcal{L}^{\text{IIa}}
  &= i \bar{\rchi} \, \slashed{D} \, \rchi - m_\rchi \, \bar{\rchi} \, \rchi
  + \dfrac{1}{2} \partial_{\, \mu} \phi \, \partial^{\, \mu} \phi
  - \dfrac{1}{2} m_\phi^2 \, \phi^2 + \dfrac{m_\phi \, \mu_1}{3} \, \phi^3
  - \dfrac{\mu_2}{4} \, \phi^4 \nonumber \\
  & - \lambda_1 \, \phi \, \bar{\rchi} \, \rchi
  - i \, \lambda_2 \, \phi \, \bar{\rchi} \, \gamma^5 \, \rchi
  -  h_1^{ij} \, \phi \, \bar{q}_{i} q_{j}
  - i h_2^{ij} \, \phi \, \bar{q}_{i} \gamma^5 q_{j} \; .
}
We assume that the DM particle is odd and the SM particles along with
the mediator are even under a new discrete $\mathbb{Z}_2$ symmetry in
order to prevent DM decay. In order to avoid potentially very large
new contributions to FCNC processes we again take flavor diagonal
quark couplings, i.e. $h_1^q \equiv h_1^{ij} \delta_{ij}$ and
$h_2^q \equiv h_2^{ij} \delta_{ij}$.

If all Yukawa couplings (of a given flavor) appearing in the
Lagrangian \eqref{eq:L_model2} are of similar magnitude, WIMP--nucleus
scattering on heavy target nuclei will be completely dominated by a
tree level contribution from $\mathcal{O}_1$, with coefficient
$\propto \lambda_1 h_1^q$. We thus have to set at least one of these
couplings to zero. On the other hand, setting both of them to zero
would also ``switch off'' the operators $\mathcal{O}_{10}$ and
$\mathcal{O}_{11}$ which we seek to generate. We thus consider two
variants of Model II:
\eq{ \label{eq:model2_def}
  {\rm Model\ IIa}: \ & h_1^q = 0 \  \forall q\,; \nonumber \\
  {\rm Model\ IIb}: \ & \lambda_1 = h_2^q = 0 \ \forall q\,.
}
Setting $h_1^q \cdot h_2^q = 0$ again ensures that no electric dipole
moments are generated at one--loop level. As before, neither of these
choices is protected by a symmetry, i.e.  they are not technically
natural. We therefore again expect that $\mathcal{O}_1$ will be
generated by radiative corrections. On the other hand, $CP$ is
violated if for some flavor $q$,
\eq{ \label{eq:model2_CPV}
 \lambda_1 h_2^q \neq 0 \ \ {\rm or} \ \ \lambda_2 h_1^q \neq 0 \,.
}
We will discuss both variants of Model II in turn.

\subsubsection{Model IIa}

The matrix element for tree--level DM--quark scattering,
$\rchi(p_\rchi) + q (p_q) \rightarrow \rchi(p_\rchi^\prime) +
q(p_q^\prime)$, via $\phi$ exchange in the $t-$channel is given by
\eq{ \label{eq:model-iia-tree-o10-matrix-element}
  \mathcal{M}^{\text{IIa}}_{\rchi q \rightarrow \rchi q} = -
  \dfrac{h_2^q}{q^2 - m_\phi^2} \; \bar{u}(p_\rchi^\prime) ( \lambda_1 + i
  \gamma^5 \lambda_2 ) u(p_\rchi) \, \bar{u}(p_q^\prime) i \gamma^5 u(p_q) \; .
}
Taking the vanishing momentum transfer limit, $q^2 \rightarrow 0$, in
the $\phi$ propagator, the matrix element can be matched onto the
relativistic effective operators
$\left( \bar{\rchi} \rchi \right) \left( \bar{q} \, i \gamma^5 \, q
\right)$ and
$\left( \bar{\rchi}\, i \gamma^5 \, \rchi \right) \left( \bar{q} \, i
  \gamma^5 \, q \right)$, which reduce to the operators
$\mathcal{O}_{10}$ and $\mathcal{O}_6$, respectively, in the
non--relativistic limit. Recall, however, from Table~1 that
$\mathcal{O}_6$ is doubly suppressed by momentum transfer $\vec{q}$,
therefore generating a cross section suppressed by
$\mathcal{O}(\vec{q}^{\; 2}/m_N^2) \sim 10^{-2} - 10^{-3}$ relative to
the contribution from $\mathcal{O}_{10}$. Hence, at the leading order
in perturbation theory, $\mathcal{O}_{10}$ dominates the scattering
matrix element in Model IIa.

\begin{figure*}[htb!]
\centering
\begin{subfigure}[b]{0.5\textwidth}
\centering
\includegraphics[width=\linewidth]{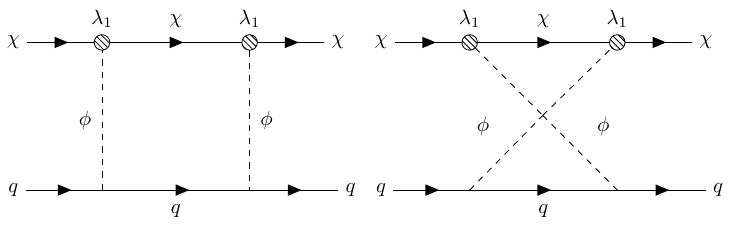}
\caption{Box and Crossed Box with $ \lambda_1 $}
\label{fig:Model_2-Box-Skeleton}
\end{subfigure}%
\begin{subfigure}[b]{0.5\textwidth}
\centering
\includegraphics[width=\linewidth]{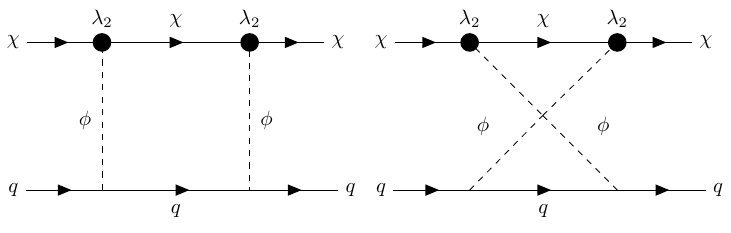}
\caption{Box and Crossed Box with $ \lambda_2 $}
\label{fig:Model_2-Box-Skeleton_1}
\end{subfigure}
\par\bigskip
\begin{subfigure}[b]{0.5\textwidth}
\centering
\includegraphics[width=0.6\linewidth]{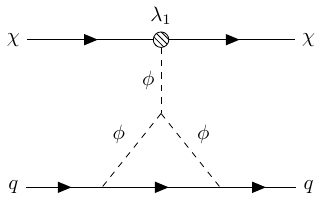}
\caption{Triangle}
\label{fig:Model_2-Triangle-Skeleton}
\end{subfigure}
\caption{One--loop box and triangle Feynman diagrams in Model IIa
  which contribute to $\mathcal{O}_1$. The shaded blob denotes the
  scalar DM--mediator coupling $\lambda_1$ while the dark blob
  denotes the pseudoscalar DM--mediator coupling $\lambda_2$.}
\label{fig:Model_2-Skeleton}
\end{figure*}

Although the operator $\mathcal{O}_1$ is not generated from leading
order Feynman diagrams, one--loop corrections to DM--nucleus
scattering again include contributions that reduce to $\mathcal{O}_1$
in the low energy limit; the corresponding Feynman diagrams are shown
in Fig.~\ref{fig:Model_2-Skeleton}.

In this model, there are two different box (and crossed box) diagrams:
one involves the scalar coupling $\lambda_1$ and the other involves
the pseudoscalar coupling $\lambda_2$. The triangle diagram involves
the scalar coupling $\lambda_1$ between the DM and the
mediator,\footnote{There is also a triangle diagram involving
  $\lambda_2$, as well as (crossed) box diagrams with one $\lambda_1$
  vertex and one $\lambda_2$ vertex; however, these diagrams do not
  contribute to $\mathcal{O}_1$. Moreover, the diagrams shown in
  Fig.~\ref{fig:Model_2-Skeleton} also generate other operators in the
  non--relativistic limit. We ignore these contributions, since they
  are of higher order in both the loop and NREFT expansions.} as well
as the cubic self--interaction of the mediator $\mu_1$. All diagrams
contain two vertices with the pseudoscalar coupling $h_2^q$ on the
quark line.

{\renewcommand{\arraystretch}{1.25}
\begin{table}[t!]
\centering
\begin{tabular}{c c c c}
  & $\bar{\rchi} \Gamma_\rchi \rchi  \; \bar{q} \, \Gamma_{q} \, q$ &
  & $c^q_i$ \\ \hline \hline
\rule{0pt}{5ex} Tree  & $c_{10}^{q,d6}  \bar{\rchi} \rchi \;
                        \bar{q} i \gamma^5 q $ & $\longrightarrow$
  & $ \dfrac{h_2^q \, \lambda_1}{m_\phi^2}$ \vspace*{1mm} \\ \hline
  \rule{0pt}{5ex}  Box$|_{\lambda_1}$
  & $c_{1,V,B1}^{q,d6}  \bar{\rchi} \gamma_{\mu} \rchi \;	\bar{q} \gamma^\mu q $
    & $\longrightarrow$ & $-\dfrac{\lambda_1^2 \, (h_2^q)^2}{16 \pi^2} \,
      N_1$ \\
  & $c_{1,S,B1}^{q,d6} \bar{\rchi} \rchi \; \bar{q} q$ & $\longrightarrow$
  & $-\dfrac{\lambda_1^2 \, (h_2^q)^2}{16 \pi^2} \, \left( 2m_\rchi m_q
 (N_2 - P_2) + m_\rchi^2 (N_3 - 2 P_1) + m_q^2 N_4 \right)$ \vspace*{1mm} \\
\hline  \vspace*{-4mm} \\
Crossed  & $c_{1,V,C1}^{q,d6} \bar{\rchi} \gamma_{\mu} \rchi \;
 \bar{q} \gamma^\mu q$ & $\longrightarrow$ & $\dfrac{\lambda_1^2 \, (h_2^q)^2}
  {16 \pi^2} \, N_5$ \\
  Box$|_{\lambda_1}$ & $c_{1,S,C1}^{q,d6} \bar{\rchi} \rchi \; \bar{q} q $
 & $\longrightarrow$
  & $\dfrac{\lambda_1^2 \, (h_2^q)^2}{16 \pi^2} \, \left( 2m_\rchi m_q
 (N_6 - P_4) + m_\rchi^2 (N_7 - 2 P_3) + m_q^2 N_8 \right)$ \vspace*{1mm} \\
\hline \vspace*{-4mm} \\
  Box$|_{\lambda_2}$ & $c_{1,V,B2}^{q,d6} \bar{\rchi} \gamma_{\mu} \rchi \;
                     \bar{q} \gamma^\mu q $ & $\longrightarrow$
 & $ -\dfrac{\lambda_2^2 \, (h_2^q)^2}{16 \pi^2} \, N_1$ \\
  & $c_{1,S,B2}^{q,d6} \bar{\rchi} \rchi \; \bar{q} q $ & $\longrightarrow$
  & $-\dfrac{\lambda_2^2 \, (h_2^q)^2}{16 \pi^2} \, \left( 2m_\rchi m_q N_2
    + m_\rchi^2 N_3 + m_q^2 N_4 \right)$ \vspace*{1mm} \\ 
\hline \vspace*{-4mm} \\
Crossed & $c_{1,V,C2}^{q,d6} \bar{\rchi} \gamma_{\mu} \rchi \;
          \bar{q} \gamma^\mu q $ & $\longrightarrow$
  & $ \dfrac{\lambda_2^2 \, (h_2^q)^2}{16 \pi^2} \, N_5$ \\
  Box$|_{\lambda_2}$ & $c_{1,S,C2}^{q,d6} \bar{\rchi} \rchi \; \bar{q} q $
   & $\longrightarrow$
  & $\dfrac{\lambda_2^2 \, (h_2^q)^2}{16 \pi^2} \, \left( 2m_\rchi m_q N_6
    + m_\rchi^2 N_7  + m_q^2 N_8 \right)$ \vspace*{1mm} \\ 
\hline \vspace*{-4mm} \\
  Triangle & $c_{1,S,T}^{q,d6} \bar{\rchi} \rchi \; \bar{q} q$&$\longrightarrow$
  & $-\dfrac{\lambda_1 \, \mu_1 \, (h_2^q)^2}{16 \pi^2} \,
    \dfrac{m_q}{m_\phi} R_1$ \vspace*{1mm} \\ \hline \vspace*{-4mm} \\
\end{tabular}
\caption{Non--relativistic reduction of effective operators in
  Model~IIa. In the expression for the quark--level Wilson
  coefficients, we have again suppressed the dependence of the loop
  functions on dimensionless parameters, i.e.
  $N_k \equiv N_k(r,s), \ P_l \equiv P_l(r,s)$ and
  $R_1 \equiv R_1(r)$, where $r \equiv m_q/m_\phi$ and
  $s \equiv m_\chi/m_\phi$. The first line contributes to the
  coefficient of $\mathcal{O}_{10}$, all other lines describe
  contributions $\propto \mathcal{O}_1$.}
\label{table:Model_IIa}
\end{table}
}

In order to simplify the loop calculation, we have used the Dirac
equation as well as $4-$momentum conservation, and took the limit of
vanishing momentum transfer. Further details can be found in
Appendix~\ref{sec:appendix-b-model-iia-iib}. Finally, we find the
following effective Lagrangian for Model IIa:
\eq{ \label{eq:model-ii-rel-eff-lagrangian}
  \mathcal{L}^{\text{IIa}}_{\text{eff}} \supset c_{10}^{q,d6}
  \left(\bar{\rchi} \rchi \right) \left( \bar{q} i \gamma^5 q \right)
  + c_{6}^{q,d6} \left(\bar{\rchi} i \gamma^5 \rchi \right)
  \left( \bar{q} i \gamma^5 q \right)
  + c_{1,S}^{q,d6}  \left(\bar{\rchi} \rchi \right) \left( \bar{q} q \right)
  + c_{1,V}^{q,d6}  \left(\bar{\rchi} \gamma_\mu \rchi \right)
  \left( \bar{q} \gamma^\mu q \right) \, ;
}
the notation is as in \eqref{eq:model-i-rel-eff-lagrangian}. For
completeness we have retained
$\left( \bar{\rchi} \, i \gamma^5 \, \rchi \right) \left( \bar{q} \, i
  \gamma^5 \, q \right)$ in the set of effective operators, although
it can safely be neglected as reasoned
earlier. Table~\ref{table:Model_IIa} displays the list of relativistic
effective operators and their matching to the NREFT operators in terms
of the parameters of Model~IIa. The quark--level Wilson coefficients
contain loop functions $N_k(r,s), \ k = 1, \dots, 8$,
$P_l(r,s), \ l = 1, \dots, 4$ and $R_1(r)$, which have been expressed
as functions of dimensionless parameters $r \equiv m_q/m_\phi$ and
$s \equiv m_\chi/m_\phi$. Analytic expressions for these loop functions
can be found in Appendix
\ref{sec:appendix-model-iia-iib-loop-functions}.

Collecting the results of Table~\ref{table:Model_IIa}, and embedding
the quark bilinears in nucleonic matrix elements, the Wilson
coefficient $c_1^N$ of the NREFT operator $\mathcal{O}_1$ is given by
\eq{ \label{eqn:model-iia-c1n}
  c_1^{N} |^{\text{IIa}} = -\dfrac{1}{16 \pi^2} &\biggr[ \,
  (\lambda_1^2 + \lambda_2^2) \Bigl( \sum\limits_{q=u,d} (h_2^q)^2
  \mathcal{N}^N_q (N_1(r,s) - N_5(r,s)) \Bigl) \nonumber \\
  &+ \, m_N \, \Bigl( \sum\limits_{q=u,d,s} (h_2^q)^2 f^N_{Tq}
  + \dfrac{2}{27} f^N_{TG} \sum\limits_{q=c,b,t} (h_2^q)^2  \Bigl)
  \nonumber \\
  &\hspace*{5mm} \times \Bigl\{ (\lambda_1^2 +\lambda_2^2) \Bigl( 2m_\rchi
  \bigl(N_2(r_q,s) - N_6(r_q,s) \bigl)
  + \, \dfrac{m_\rchi^2}{m_q} \, \bigl(N_3(r_q,s) - N_7(r_q,s) \bigl)
  \nonumber \\
  &\hspace*{8mm}+ \, m_q \, \bigl(N_4(r_q,s) - N_8(r_q,s) \bigl) \Bigl)
  \nonumber \\
  &\hspace*{8mm}- 2 m_\rchi \lambda_1^2 \, \Bigl( P_2(r_q,s) - P_4(r_q,s)
  + \dfrac{m_\rchi}{m_q} \bigl( P_1(r_q,s) - P_3(r_q,s) \bigl) \Bigl)
  \nonumber \\
  &\hspace*{8mm}+ \, \dfrac{ \lambda_1 \, \mu_1 }{m_\phi} \, R_1(r_q)
\Bigl\} 
\; \biggr] \; .
}
The nuclear bilinear coefficients $\mathcal{N}^N_q $, $f^T_{N q}$ and
$f^T_{N G}$ are the same as in eq.\eqref{eq:model-i-coef-c1N}. At
first sight $c_1^N|^{\text{IIa}}$ appears to be singular in the limit
$m_q \rightarrow 0$. Note that $f^T_{N_q}$ contains a factor of $m_q$,
see eq.\eqref{eq:fnTq}; hence contributions to $c_{1,S}^{q,d6}$ in
Table~\ref{table:Model_IIa} without an explicit factor $m_q$ appear
$\propto 1/m_q$ in eq.\eqref{eqn:model-iia-c1n}. However, they get
multiplied with the differences of loop functions
$N_3(r_q,s) - N_7(r_q,s)$ or $P_1(r_q,s) - P_3(r_q,s)$, which scale
like $r_q$ for $r_q \rightarrow 0$; these contributions therefore
approach a finite value as $m_q \rightarrow 0$. In contrast, the
difference $N_1(r_q,s) - N_5(r_q,s)$, which appears in the
contribution from the product of two vector currents, vanishes for
$r_q \rightarrow 0$. As a result, the dominant contributions to
$c_1^N|^{\text{IIa}}$ also originate from the product of two scalar
currents, as for Model I. Within this category, the contribution
$\propto N_4 - N_8$ vanishes as $r_q \rightarrow 0$ and is therefore
negligible for generation--independent couplings $h_2^q$, but all
other terms are very roughly comparable and must be taken into account.

The Wilson coefficient of the operator $\mathcal{O}_{10}$ arising from
the $t-$channel tree--level scattering diagram is
\eq{ \label{eqn:model-iia-coef-c10N}
  c^{N}_{10}|^{\text{IIa}} = \dfrac{\lambda_1}{m_\phi^2} \, \Big(
  \sum\limits_{q=u,d,s} h_2^{q} \, \Delta \tilde{q}^N - \Delta \tilde{G}^N
  \sum\limits_{q=c,b,t} \dfrac{h_2^{q}}{m_q} \Big) \; .
}

\subsubsection{Model IIb}

We next turn to Model IIb. It is also described by the Lagrangian of
eq.\eqref{eq:L_model2}, but we now postulate purely scalar Yukawa
couplings $h_1^q$ on the quark side and a purely pseudoscalar coupling
$\lambda_2$ on the dark matter side, see eq.\eqref{eq:model2_def}.
The tree--level matrix element for DM--quark scattering,
$\rchi (p_\rchi) + q(p_q) \rightarrow \rchi (p_\rchi^\prime) + q
(p_q^\prime)$, proceeding via $t-$channel $\phi$ exchange is then:
\eq{ \label{eq:model-iib-tree-o11-matrix-element}
  \mathcal{M}^{\text{IIb}}_{\rchi q \rightarrow \rchi q } =
  - \dfrac{h_1^q \, \lambda_2}{q^2 - m_\phi^2} \; \bar{u}(p_\rchi^\prime)
  i \gamma^5  u(p_\rchi) \, \bar{u}(p_q^\prime) u(p_q) \; .
}
This matrix element matches onto the dimension$-6$ effective operator
$\left( \bar{\rchi} i \gamma^5 \rchi \right) \left( \bar{q} q
\right)$, which reduces to the momentum--suppressed SI NREFT operator
$\mathcal{O}_{11}$ in the non--relativistic limit.

\begin{figure*}[htb!]
\centering
\includegraphics[width=0.8\linewidth]{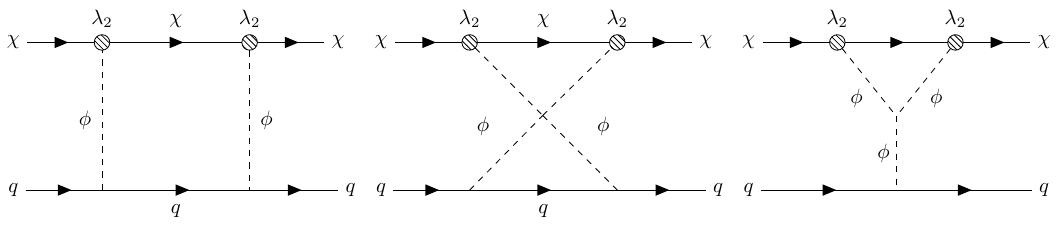}
\caption{One--loop (crossed) box and triangle Feynman diagrams in
  Model IIb which contribute to $\mathcal{O}_1$ in the
  non--relativistic limit.}
\label{fig:model_ii-b-skeleton}
\end{figure*}

Once again, the choice $\lambda_1 = 0$, which ensures the absence of
$\mathcal{O}_1$ at tree--level, is not protected by any symmetry. We
therefore again expect contributions $\propto \mathcal{O}_1$ to be
generated at the next order in perturbation theory. The relevant
Feynman diagrams are shown in Fig.~\ref{fig:model_ii-b-skeleton}.
Each diagram involves two factors of the pseudoscalar DM--mediator
coupling $\lambda_2$; the two $\gamma_5$ factors multiply to
unity. The triangle and (crossed) box diagrams in addition involve one
or two factors of the scalar DM--quark coupling $h_1^q$, respectively;
the former also involves the cubic self--coupling $\mu_1$ of the
mediator. The triangle diagram can be interpreted as generating a
scalar DM--mediator coupling. Explicit expressions for the resulting
amplitudes are provided in Appendix
\ref{sec:appendix-b-model-iia-iib}.

%\begin{figure*}[htb!]
%\centering
%\includegraphics[width=0.8\linewidth]{Skeleton-Model-IIb.pdf}
%\caption{One--loop (crossed) box and triangle Feynman diagrams in
%  Model IIb which contribute to $\mathcal{O}_1$ in the
%  non--relativistic limit.}
%\label{fig:model_ii-b-skeleton}
%\end{figure*}

When the mediator $\phi$ is integrated out, the following
relativistic effective Lagrangian describes DM--quark scattering:
\eq{ 	\label{eq:model-ii-b-rel-eff-lagrangian}
  \mathcal{L}^{\text{IIb}}_{\text{eff}} \supset c_{11}^{q,d6}
  \left(\bar{\rchi} i \gamma^5 \rchi \right) \left( \bar{q}  q \right)
  + c_{1,S}^{q,d6} \left( \bar{\rchi} \rchi \right) \left( \bar{q} q \right)
  + c_{1,V}^{q,d6} \left(\bar{\rchi} \gamma_\mu \rchi \right)
  \left( \bar{q} \gamma^\mu q \right)\, .
}
Table \ref{table:Model_IIb} displays the quark--level Wilson coefficients in
terms of the parameters of Model IIb. The loop functions $N_k$ and
$P_l$ are identical to the ones appearing in Model IIa. $S_1$ is
the only loop function not defined previously, and its analytic
expression along with that of the others can be found in
Appendix~\ref{sec:appendix-model-iia-iib-loop-functions}.

{\renewcommand{\arraystretch}{1.25}
\begin{table}[t!]
\centering
\begin{tabular}{c c c c}
  & $\bar{\rchi} \Gamma_\rchi \rchi \; \bar{N} \, \Gamma_N \,N$ & & $ c^q_i$
  \\ \hline \hline
\rule{0pt}{5ex} Tree & $ c_{11}^{q,d6} \bar{\rchi} i \gamma^5 \rchi \;
   \bar{q} \gamma^\mu q$ & $\longrightarrow$
  & $\dfrac{h_1^q \lambda_2 m_N}{m_\phi^2 m_\chi}$ \vspace*{1mm} \\ \hline
 \rule{0pt}{5ex}  Box & $c_{1,V,B}^{q,d6} \bar{\rchi} \gamma_\mu \rchi \;
  \bar{q} \gamma^\mu q$ &$\longrightarrow$&
 $-\dfrac{\lambda_2^2 \, h_1^{q \, 2}}{16 \pi^2} \, N_1$ \\
  & $c_{1,S,B}^{q,d6} \bar{\rchi} \rchi \;\bar{q} q$ &$\longrightarrow$
  & $-\dfrac{\lambda_2^2 \, h_1^{q \, 2}}{16 \pi^2} \, \left(
    2m_\rchi m_q (N_2 + P_1) + m_\rchi^2 N_3 + m_q^2 (N_4 + 2P_2) \right)$
\vspace*{1mm}  \\ \hline \vspace*{-4mm} \\
Crossed & $c_{1,V,C}^{q,d6} \bar{\rchi} \gamma_\mu \rchi \;
          \bar{q} \gamma^\mu q $ & $\longrightarrow$
   & $\dfrac{\lambda_2^2 \, h_1^{q \, 2}}{16 \pi^2} \, N_5$ \\
  Box & $c_{1,S,C}^{q,d6} \bar{\rchi} \rchi \; \bar{q} q$ & $\longrightarrow$
 & $\dfrac{\lambda_2^2 \, h_1^{q \, 2}}{16 \pi^2} \, \left(
   2m_\rchi m_q (N_6 - P_3) + m_\rchi^2 N_7+ m_q^2 (N_8 - 2P_4) \right)$
  \vspace*{1mm} \\ \hline \vspace*{-4mm} \\
Triangle & $c_{1,S,T}^{q,d6} \bar{\rchi} \rchi \; \bar{q} q$ & $\longrightarrow$
& $-\dfrac{h_1^q \, \mu_1 \, \lambda_2^2}{16 \pi^2} \, \dfrac{m_\rchi}{m_\phi}
       S_1$ \vspace*{1mm} \\ \hline
\end{tabular}
\caption{Non--relativistic reduction of effective operators in Model
  IIb. The arguments of the loop functions $N_k, \, P_l$ and $S_1$
  have again been suppressed; explicit expressions for these functions
  can be found in Appendix~\ref{sec:appendix-model-iia-iib-loop-functions}.}
\label{table:Model_IIb}
\end{table}
}

Collecting the results of Table~\ref{table:Model_IIb}, and inserting
the appropriate hadronic coefficients for the nucleonic matrix
elements of the quark bilinears, the Wilson coefficient of the
operator $\mathcal{O}_1$ is:
\eq{ \label{eq:model-ii-b-coef-c1n}
  c^N_1 |^{\text{IIb}} = -\dfrac{\lambda_2^2}{16 \pi^2} &\Biggl\{
  \sum\limits_{q=u,d} \mathcal{N}^N_q (h_1^q)^2
  \biggl( N_1(r_q,s) - N_5(r_q,s) \biggl) \nonumber \\
  &+ \, m_N \Bigl( \sum\limits_{q=u,d,s} f^N_{Tq}
  + \dfrac{2}{27} f^N_{TG} \sum\limits_{q=c,b,t} \Bigl) \nonumber \\
 & \hspace*{4mm} \times \Biggl[ (h_1^q)^2
 \Biggl( 2m_\rchi \biggl(N_2(r_q,s) - N_6(r_q,s)+ P_1(r_q,s) + P_3(r_q,s)
 \biggl) \nonumber \\
 & \hspace*{20mm} + \, m_q \biggl(N_4(r_q,s) - N_8(r_q,s) + 2 \Bigl( P_2(r_q,s)
 + P_4(r_q,s) \Bigl) \biggl) \nonumber \\ &\hspace*{20mm}
 + \dfrac{m_\rchi^2}{m_q} \, \Bigl( N_3(r_q,s) - N_7(r_q,s) \Bigl) \Biggl)
 \nonumber \\ & \hspace*{8mm} + \, \mu_1 h_1^q \, \dfrac{m_\rchi}{m_\phi}
 \, \dfrac{S_1(s)}{m_q} \Biggl] \Biggl\} \; ;
}
recall that $r_q = m_q/m_\phi$ and $s = m_\rchi / m_\phi$. As in case
of Model IIa, the combinations of loop functions multiplying
$\mathcal{N}^N_q$ and $m_\rchi^2/m_q$ vanish for $m_q \rightarrow 0$;
the term $\propto m_\chi^2/m_q$ thus yields a finite result in this
limit. However, the very last term in
eq.\eqref{eq:model-ii-b-coef-c1n} also contains an explicit $1/m_q$
factor; the loop function appearing in this term does not depend on
$m_q$ at all. Here the required chirality breaking on the quark line
is due to the {\em single} factor of $h_1^q$. We therefore expect this
term to dominate, unless the trilinear scalar coupling $\mu_1$ is for
some reason very small.

From eq.\eqref{eq:model-iib-tree-o11-matrix-element} the tree--level
contribution to the Wilson coefficient of $\mathcal{O}_{11}$ is:
\eq{ \label{eqn:model-iib-c11n}
  c_{11}^N |^{\text{IIb}} = \dfrac{\lambda_2} {m_\phi^2} \,
  \dfrac {m_N^2} {m_\chi} \,
  \Bigl( \sum\limits_{q=u,d,s} \dfrac{h_1^q f^N_{Tq}}{m_q}
  + \dfrac{2}{27} f^N_{TG} \sum\limits_{q=c,b,t} \dfrac{h_1^q}{m_q}\Bigl)
  \; =  \dfrac{\lambda_2 }{m_\phi^2} \, \dfrac {m_N^2} {m_\chi} \tilde{f}^N \; .
}
Here we have defined
$\tilde{f}^N \equiv \Bigl( \sum\limits_{q=u,d,s} \dfrac{h_1^q
  f^N_{Tq}}{m_q} + \dfrac{2}{27} f^N_{TG} \sum\limits_{q=c,b,t}
\dfrac{h_1^q}{m_q}\Bigl)$, and as usual neglected the $q^2$ term in
the $\phi$ propagator. As in the last term in
eq.\eqref{eq:model-ii-b-coef-c1n} the required chirality breaking on
the quark line is provided by $h_1^q$. Since the hadronic matrix
elements $f^N_{Tq}$ have been defined including an explicit factor of
$m_q$, the contributions in $\tilde{f}^N$ scale $\propto 1/m_q$,
leading to a large enhancement of the contribution of light quarks.
On the other hand, an extra factor $m_N/m_\chi$ appears since the
pseudoscalar DM current is
$\propto |\vec{q}|/m_\chi = (|\vec{q}|/m_N) \times (m_N/m_\chi)$. As
already noted in the discussion of Table~1, this appears to be quite
generic \cite{Dent:2015zpa, Bishara:2017pfq, DelNobile:2018dfg,
  Drees:2019qzi}.

\subsection{The Neutron EDM}

In all three cases we considered, the tree--level contribution to the
Wilson coefficient of $\mathcal{O}_{10}$ or $\mathcal{O}_{11}$ is
nonzero only if $CP$ is violated: in Model I one needs
$g_1 h_2^q \neq 0$, where $g_1$ is a scalar coupling while $h_2^q$ is
a pseudoscalar coupling; in Model IIa, $\lambda_1 h_2^q \neq 0$ is
required, where $\lambda_1$ is a scalar coupling; and in Model IIb,
$c_{11}^{d6} \propto \lambda_2 h_1^q$, where $\lambda_2$ is
pseudoscalar but $h_1^q$ is scalar. Since we assume the quark Yukawa
couplings to be flavor diagonal, the most sensitive probe of $CP$
violation is the EDM of the neutron (or of heavier nuclei). However,
$CP$ violation is a necessary condition for the generation of an EDM;
it is not by itself sufficient. In the case at hand, since we set
$h_1^q = 0$ in Models I and IIa and $h_2^q = 0$ in Model IIb, there is
no one--loop contribution to the neutron EDM, in contrast to the
models with charged mediator we considered in
ref.\cite{Drees:2019qzi}. In case of Model IIb, as far as the
one--loop diagram with a quark and a mediator $\phi$ in the loop is
concerned, the mediator can consistently be defined as being
$CP-$even, since only the coupling $h_1^q$ appears in the
diagram. Similarly, in Model I and Model IIa, in the one--loop diagram
with a quark--mediator loop the mediator can consistently be defined
as being $CP-$odd, since only the coupling $h_2^q$ appears.

Turning to higher loops, a $CP-$even mediator can have any self
coupling. Hence in Model IIb embellishing the one--loop diagram with
additional scalar vertices does not lead to $CP$ violation. In fact,
in this model the only coupling that is not consistent with
interpreting $\phi$ to be $CP-$even is $\lambda_2$, which couples
$\phi$ to the DM particle. However, this coupling cannot contribute to
electric dipole moments. It could appear in any diagram where the only
external particles are a through--going quark and a photon (or gluon)
only via a closed $\chi$ loop, which will either vanish (if an odd
number of $\phi$ legs is attached to it), or simply renormalize a
$CP-$even quantity like the $\phi$ $2-$point function. We thus
conclude that in Model IIb, {\em no} new contributions to the electric
dipole moments of SM particles are generated. The conceptually easiest
way to prove the existence of $CP$ violation in this model is via
$\rchi q$ scattering; in case of $2 \rightarrow 2$ scattering, spin
observables would have to be included in the construction of a
$CP-$odd quantity. While conceptually straightforward, experimentally
this seems prohibitively difficult; certainly there are no current
experimental constraints from such experiments.

\begin{figure}[t!]
\centering
\includegraphics[width=7.5cm]{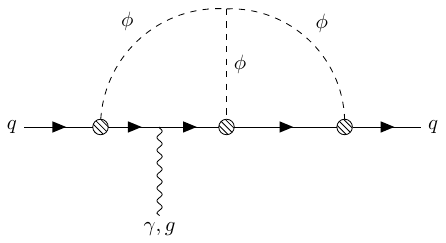}
\hfill
\includegraphics[width=7.5cm]{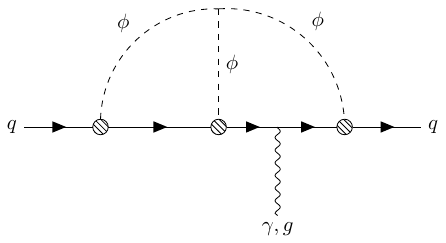}
\caption{Two--loop Feynman diagrams for quark EDMs and color--EDMs in
  Model \RNum{1} and \RNum{2}a. The blobs here indicate the insertion
  of the mediator pseudoscalar Yukawa coupling. }
\label{fig:two-loop-skeleton-diagram}
\end{figure}

The situation is very different in Models I and IIa, where in the
relevant one--loop diagram the mediator behaves like a
pseudoscalar. The reason is that a pseudoscalar cannot have a $\phi^3$
coupling. Hence, two--loop diagrams containing both the (pseudoscalar)
Yukawa coupling $h_2^q$ and the (scalar) trilinear coupling $\mu_1$,
see Fig.~\ref{fig:two-loop-skeleton-diagram}, can be expected to
generate a nonvanishing (chromo--)EDM for quark $q$. It should be
noted that here $\mu_1$ is relevant, not the couplings $\lambda_1$ or
$g_1$ appearing in the coefficient of $\mathcal{O}_{10}$. However, a
theory with $\mu_1 = 0$ but $\lambda_1 \neq 0$ is, strictly speaking,
not renormalizable, since a triangle diagram with $\rchi$ in the loop
will generate a {\em divergent} contribution to
$\mu_1 m_\phi \propto m_\rchi \lambda_1^3$; similarly, in Model I
there are divergent one--loop contributions $\propto g_1 g_2 m_S$ to
the $\phi^3$ vertex. We will come back to this point later. Because
only the trilinear self--coupling of the mediator and its pseudoscalar
Yukawa coupling to quarks are relevant here, the calculation of the
quark (chromo) EDM is exactly the same in these two models.

The quark EDM $d_q$ is calculated as the coefficient of a
dimension$-5$ $P-$ and $T-$odd interaction term
$(-i/2) \, \bar{q} \, \sigma_{\mu \nu} \gamma_5 \, q \, F^{\mu \nu}$
at zero momentum transfer. The same quark radiating a gluon instead a
photon leads to non--vanishing chromo EDM. These are calculated
similar to quark EDMs, by finding the coefficient of
$(-i/2) \, \bar{q} \, \sigma_{\mu \nu} t_a \gamma_5 \, q \, G^{\mu
  \nu}_a$ at zero momentum transfer.

The EDM operator breaks chirality, hence $d_q$ is proportional to an odd
number of chirality flips. These can come either from fermion masses
or from Yukawa couplings in the relevant Feynman diagrams. The
diagrams shown in Fig.~\ref{fig:two-loop-skeleton-diagram} contain
three Yukawa couplings, hence they can contribute even for
$m_q \rightarrow 0$. We provide the details of the computation of
these two--loop diagrams in Appendix
\ref{app:two-loop-calculations}. We calculate the quark EDM to be
\eq{ \label{eq:two-loop-qEDM}
  d_q = \dfrac{2 e \, Q_q \, (h_2^q)^3}{(16 \pi^2)^2} \, \mu_1 \, m_\phi
  \, \lim_{q^2 \rightarrow 0} \, \Bigl( \left[ \mathcal{X} \right] +
  \left[ \mathcal{Y} \right] \Bigl) \, ,
}
where $Q_q$ is the electric charge of quark $q$;
$\left[ \mathcal{X} \right]$ and $\left[ \mathcal{Y} \right]$ are loop
functions expressed as five dimensional integrals over five Feynman
parameters given in eqs.\eqref{eq:app:two-loop-X} and
\eqref{eq:app:two-loop-Y}. The color--EDM $\tilde{d}_q$ can be
obtained by replacing the external photon with a gluon. Hence
$\tilde{d}_q$ can be obtained from eq.\eqref{eq:two-loop-qEDM}
by replacing $eQ_q$ with the strong coupling $g_s$.

In order to calculate the value of the nEDM from $d_q$ and
$\tilde{d}_q$, we use
\eq{ \label{eq:nEDM_lattice_sumrules}
d_n = g^u_T d_u +  g^d_T d_d + g^s_T d_s + 1.1 \, e \,
( 0.5 \, \tilde{d}_u + \tilde{d}_d )\,.
}
Here the tensor charges $g^u_T = -0.233(28)$, $g^d_T = 0.774(66)$ and
$g^s_T = 0.009(8)$ have been calculated using lattice QCD
\cite{Bhattacharya:2015esa,Bhattacharya:2015wna} (see also
Refs.\cite{Gupta:2018lvp,Yamanaka:2018uud,Alexandrou:2017qyt}) at a
renormalization scale of $2 \GeV$. We are not aware of a reliable
lattice computation of the contribution of the chromo-EDMs to $d_n$;
we therefore employ a computation using QCD sum rules, again evaluated
at a renormalization scale of $2 \GeV$ \cite{Pospelov:2000bw},
although there is an $\mathcal{O}(50 \%)$ uncertainty in these results
\cite{Hisano:2012sc,Fuyuto:2012yf}. In this case the uncertainty in
the coefficients in Eq.(\ref{eq:nEDM_lattice_sumrules}) might shift
the boundary of the excluded region slightly, without affecting our
results qualitatively. Since in our numerical analyses we will assume
flavor universal quark Yukawa couplings, the contributions from the
(chromo--)EDMs of heavy quarks to the EDM of the neutron can safely be
neglected; in fact, since $g_T^s \ll g_T^d$ already the contribution
from the strange quark is essentially negligible for equal couplings.

%%%%%%%%%%%%%%%%%%%%%%%%%%%%%%%%%%%%%%%
\section{Results and Discussions} 
%%%%%%%%%%%%%%%%%%%%%%%%%%%%%%%%%%%%%%%
\label{sec:results-and-discussions}

In the previous section, we described the different contributions to
DM--nucleus scattering in Models I, IIa and IIb. These simplified
models were designed such there are no tree--level contributions to
the NREFT operator $\mathcal{O}_1$; instead, at tree--level only the
$P-$ and $T-$odd operators $\mathcal{O}_{10}$ (in Model I and IIa) or
$\mathcal{O}_{11}$ (in Model IIb) were generated. Recall that
$\mathcal{O}_1$ often gives the dominant contribution to DM--nucleus
scattering: it appears in leading order in the NREFT expansion, and
its contribution is coherently enhanced by $A^2$ if
$c_1^n \simeq c_1^p$, where $c_1^n$ and $c_1^p$ are the Wilson
coefficients accompanying $\mathcal{O}_1$ for neutron and proton
respectively. However, in the models considered here, $\mathcal{O}_1$
arises from box and triangle diagrams appearing at next--to--leading
order in perturbation theory. Therefore, it is not obvious \textit{a
  priori} which operator provides the dominant contribution to
DM--nucleus scattering.

In this section we compare the contributions from $\mathcal{O}_1$
quantitatively to the contributions from $\mathcal{O}_{10}$ or
$\mathcal{O}_{11}$. To that end, we compute the number of events due
to these operators for a recent XENON1T run \cite{XENON:2018voc} for a
variety of benchmark model parameters. Here we use the explicit
expressions for the loop functions contributing to the Wilson
coefficient of $\mathcal{O}_1$ given in Appendix~A. It should be noted
that the contribution from $\mathcal{O}_1$ does not interfere with
those from $\mathcal{O}_{10}$ and $\mathcal{O}_{11}$, due to the
different $CP$ properties of these operators. Hence the total
scattering rate is simply given by the sum of these
contributions.\footnote{Recall that we only extract the contribution
  to $\mathcal{O}_1$ from our loop diagrams. These diagrams will also
  contribute to additional NREFT operators; however, those
  contributions will be of higher order both in the loop and NREFT
  expansions, and can thus safely be neglected.}

Moreover, for Models I and IIa we check numerically if the nEDM
surpasses its experimental upper bound \cite{Abel:2020pzs,
  Pendlebury:2015lrz, Burghoff:2011xk}. We perform the integration of
the loop functions in eq.\eqref{eq:two-loop-qEDM} using the Monte
Carlo integration routine \texttt{SUAVE} of the \texttt{CUBA}
numerical library \cite{Hahn:2004fe}. The absolute numerical error
reported by the routine is around $\mathcal{O}(10^{-8})$ for the two
values of the mediator mass $m_\phi$ we consider.

\begin{figure*}[t!]
\centering
\includegraphics[width=1.15\linewidth]{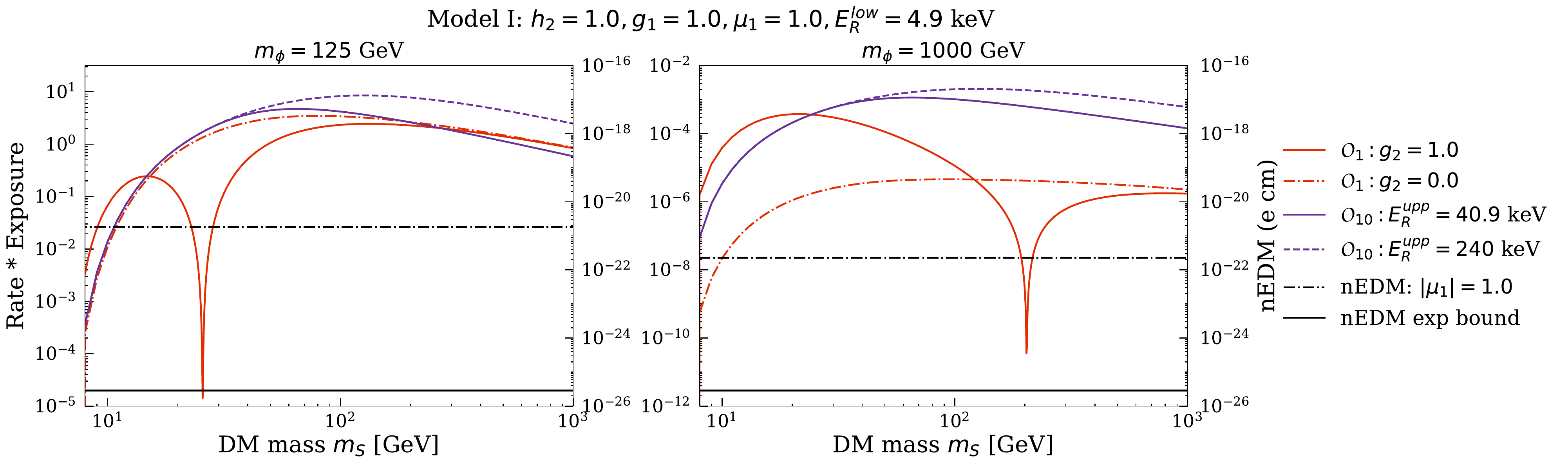}
\caption{The left $y-$axis shows the total number of scattering events
  in Model I with $\mu_1 = 1.0$ at XENON1T with a run time exposure of
  1 tonne--year as a function of the DM mass $m_S$ for two values of
  the mediator mass, $m_\phi = 125 \GeV$ (left frame) and
  $m_\phi = 1000 \GeV$ (right frame). The \textbf{violet} and
  \textbf{red} curves show the number of events due to
  $\mathcal{O}_{10}$ and $\mathcal{O}_1$, respectively. The
  \textbf{black} lines refer to the right $y-$axis; the \textbf{solid
    black} line is the experimental upper limit on nEDM at 90\%
  C.L., while the \textbf{dashed black} line is the predicted nEDM for
  the given choice of parameters.}
\label{fig:Model-I-Positive-Mu1}
\end{figure*}
\begin{figure*}[h!]
\centering
\includegraphics[width=1.15\linewidth]{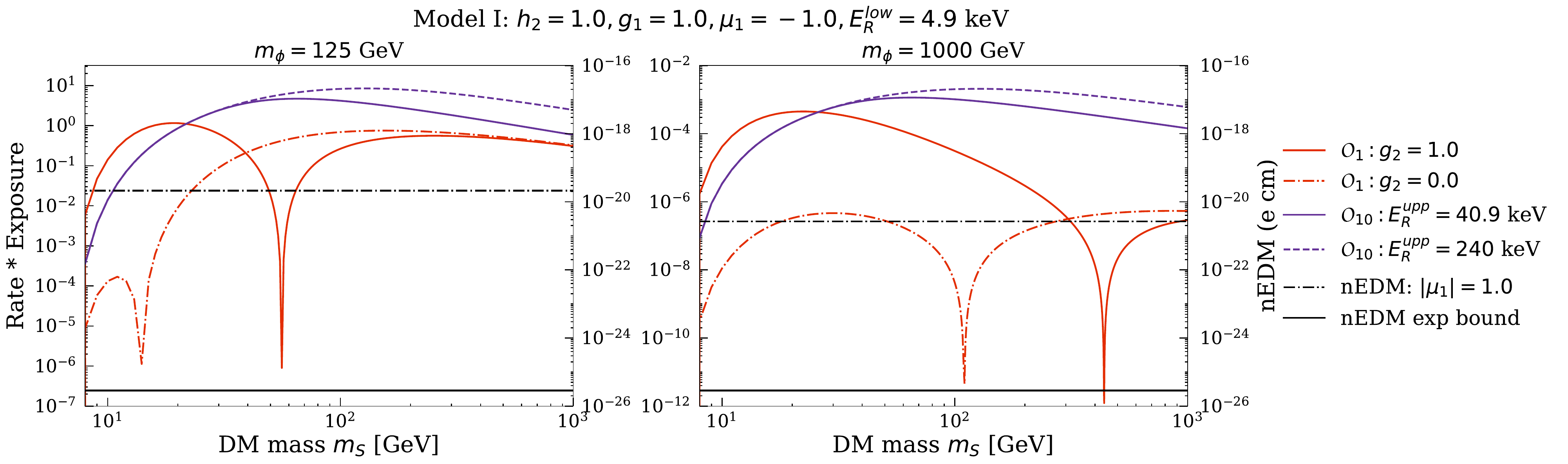}
\caption{As in Fig.~\ref{fig:Model-I-Positive-Mu1}, except for
  $\mu_1 = - 1.0$.}
\label{fig:Model-I-Negative-Mu1}
\end{figure*}

In order to compute the number of events for the 2018 run of XENON1T
\cite{XENON:2018voc}, we integrate the differential event rate,
computed using the the Mathematica code \texttt{dmformfactor}, which
is based on the formalism of ref.\cite{Anand:2013yka}, over the recoil
energy $E_R$ from $4.9$ keV to $40.9$ keV, in accordance with the
recoil energy region of interest (ROI) used in that run. In addition,
we include results (shown as dashed curves) where the upper end of
integration is set at $E_R^{\text{upp}} = 240 \keV$, as considered
in a dedicated effective field theory search by the XENON
experiment. We multiply the integrated rate with the 2018 XENON1T
runtime exposure of $278.8 \text{ days} \times 1.3\text{ tonne} = 1.0$
tonne-yr. We have further assumed a standard isothermal DM halo with
$\rho_\chi = 0.3 \GeV \text{cm}^{-3}$, $v_0 = 220$ km/s,
$v_e = 232$ km/s and $v_{\rm esc} = 544$ km/s. We calculate our event
rates using a weighted sum over the different isotopes occurring in
naturally abundant Xenon.

We choose the following sets of parameter values for Model I:
$h_2^q \equiv h_2 = g_1 = 1.0$ with $m_\phi = 125 \GeV$ or
$m_\phi = 1000 \GeV$. Note that we choose a rather large and flavor
universal Yukawa coupling of the quarks. For the given couplings, the
former (smaller) mediator mass value leads to a DM--nucleon cross
section from the operator $\mathcal{O}_{10}$ that lies just below the
sensitivity of the 2018 XENON1T run in the $\sigma_{\chi p} - m_S$
exclusion plane. The latter (larger) mediator mass corresponds to a
DM--nucleon cross section from the operator $\mathcal{O}_{10}$ that
lies just above the irreducible background from coherent
neutrino--nucleus scattering (the so--called ``neutrino floor''); it
will be very difficult to probe even smaller cross sections in direct
search experiments.

Figures \ref{fig:Model-I-Positive-Mu1} and
\ref{fig:Model-I-Negative-Mu1} depict the resulting number of events
from the operators $\mathcal{O}_{10}$ and $\mathcal{O}_1$ as a
function of the DM mass $m_S$ in Model I, for different values of the
trilinear mediator coupling $\mu_1$. Each figure contains two frames,
corresponding to the two values of $m_\phi$ we consider.

Since the tree--level contribution from $\mathcal{O}_{10}$ does not
depend on $\mu_1$, it remains the same in both figures. We note that
this contribution is enhanced by up to an order of magnitude once the
DM mass $m_S$ exceeds the mass of the target nuclei ($\simeq 125$ GeV
for Xenon) if the upper end of the ROI is increased from $40.9$ keV to
$240$ keV. This is not surprising, since $\mathcal{O}_{10}$ predicts a
recoil energy spectrum that peaks at sizable energy, due to the
explicit factor of $\vec{q}$ that appears in the definition of this
operator. Note that at $m_S < m_{\rm Xe}$, the maximal recoil energy
is $\mathcal{O}(v^2 m_S^2/m_{\rm Xe})$ rather than
$\mathcal{O}(v^2 m_{\rm Xe})$ for $m_S > m_{\rm Xe}$; hence this
enhancement only occurs at larger DM masses.

Figure~\ref{fig:Model-I-Positive-Mu1} is for $\mu_1 = 1.0$. We see
that the contribution from $\mathcal{O}_{10}$ generally exceeds that
from $\mathcal{O}_1$, except at low DM mass where the low maximal
recoil energy leads to low values of $|\vec{q}|$. We also note that
for $g_2 = 0$, i.e. in the absence of the last diagram in
Fig.~\ref{fig:Model_1-Skeleton}, the $\mathcal{O}_1$ contribution,
shown by the dashed red lines, drops faster with increasing $m_\phi$
than the (tree--level) contribution from $\mathcal{O}_{10}$ does. For
a positive coupling $g_2 = 1.0$ the contributions from the (crossed)
box and $g_2$ triangle diagrams have opposite signs, leading to a
vanishing $\mathcal{O}_1$ contribution for $m_S \simeq 0.2 m_\phi$.

In order to understand the comparison of the tree--level
$\mathcal{O}_{10}$ contributions with the loop suppressed
$\mathcal{O}_1$ contributions semi--quantitatively, we estimate
the ratio of the number of events from the two contributions
$N_{\mathcal{O}_1}/N_{\mathcal{O}_{10}}$ as:
\eq{ \label{eqn:Model-I-estimate}
  \left. \dfrac{ N_{\mathcal{O}_1} }{ N_{\mathcal{O}_{10}} }
  \right\rvert_{\text{I}} = \left. \dfrac{ R_{\mathcal{O}_1} }
    { R_{\mathcal{O}_{10}} } \right\rvert_{\text{I}} \;
  \sim \dfrac{ (c_1^{N} |^{\text{I}})^2 }{ (c_{10}^{N} |^{\text{I}})^2 }
  \dfrac{A^2}{\langle S_{\rm Xe} \rangle^2} \;
  \dfrac{m_N^2}{\vec{q}^{\; 2}} \; .
}
Here $R_{\mathcal{O}_i}$ denotes the scattering rate due to operator
$\mathcal{O}_i$ integrated over the recoil energy window. The factor
$(\vec{q}^{\; 2}/m_N^2)^{-1}$ is due to the momentum suppression of
$\mathcal{O}_{10}$. In this estimate, we have assumed the nuclear
response functions to be independent of the recoil energy, and
approximated the ratio of the nuclear response functions as
$A^2/\langle S_N\rangle^2$ since the SI response (from
$\mathcal{O}_1$) is coherently enhanced by $A^2$ while the SD
response (from $\mathcal{O}_{10}$) is suppressed by the spin
expectation value squared $ \langle S_{\rm Xe}\rangle^2 $. We
estimate $c_1^N|^{\textrm{I}}$ using the vanishing quark mass limit of
eq.\eqref{eq:model-i-coef-c1N}:
\eq{ \label{eqn:Model-I-c1n-masless-quark-limit}
  c_1^N|^{\textrm{I}}_{m_q \rightarrow 0} = \dfrac{h_2^2}{16 \pi^2} m_N f_T^N
  \left\{ g_1^2 m_S (M_2 + M_4)|_{r \rightarrow 0}
    + \left( \dfrac{g_1 \mu_1}{m_\phi} - \dfrac{g_2}{2 m_S}\right)
    L_1(r)|_{r \rightarrow 0}\right\} \; .
}
For small $s=m_S/m_\phi$ and $g_2 = 1.0$, the last (triangle) term in
eq.\eqref{eqn:Model-I-c1n-masless-quark-limit} dominates:
\eq{ \label{eqn:Model-I-c1n-g2-approx}
  c_1^N|^{\textrm{I}} \approx \dfrac{h_2^2}{16 \pi^2} \dfrac{g_2}{4} m_N f^N
  \dfrac{1}{m_S} \dfrac{1}{m_\phi^2} \, ,
}
where we used $L_1(0) = -1/2m_\phi^2$ and defined
$f^N \equiv \left( \sum_{u,d,s} f^N_{Tq} + \dfrac{2}{27} f^N_{TG}
  \sum_{c,b,t} \right)$. Inserting
eq.\eqref{eqn:Model-I-c1n-g2-approx} and
eq.\eqref{eq:model-i-coef-c10N} for $c_{10}^N|^{\textrm{I}}$ in
eq.~\eqref{eqn:Model-I-estimate}, we obtain for
$ m_S < m_\phi = 1000 \GeV$:
\eq{ \label{eq:Model-I-ratio-events-g2-approx}
  \left. \dfrac{ N_{\mathcal{O}_1} }{ N_{\mathcal{O}_{10}} }
  \right\rvert_{\text{I}} \approx \left( \dfrac{h_2}{16 \pi^2}
    \dfrac{g_2}{4 g_1} \dfrac{f^N_T}{\tilde{\Delta}^N}
    \dfrac{m_N}{m_S}  \right)^2 \dfrac{A}{\langle S_N \rangle^2}
  \dfrac{m_N}{2 E_R} \; ,
}
where we defined
$\tilde{\Delta}^N = \Big( \sum\limits_{q=u,d,s} \Delta\tilde{q}^N -
\Delta \tilde{G}^N \sum\limits_{q=c,b,t} \dfrac{1}{m_q} \Big)$ and
used $m_{\rm Xe} = A m_N$. Plugging in $h_2 = g_1 = g_2 = 1.0$,
$E_R = 30 \keV$ for Xenon ($A=131$) at $m_S = 30 \GeV$ results in a
ratio of $0.8$, which agrees well with the solid red curve for
$\mathcal{O}_1$ being marginally below the purple curve for
$\mathcal{O}_{10}$ in the right frame of
Figure~\ref{fig:Model-I-Positive-Mu1}.

For $m_\phi = 125 \GeV$ and not too large $m_S$
$c_1^N|^{\textrm{I}}$ is instead dominated by the scattering
contribution from the $g_1$ triangle diagram since $g_1 \mu_1/m_\phi$
is less suppressed than $g_2/2 m_S$. In that case we should replace
$g_2 / (2 m_S)$ by $g_1 \mu_1/m_\phi$ in
eq.\eqref{eq:Model-I-ratio-events-g2-approx}. Again using
$E_R = 30 \keV$ and $h_2 = g_1 = \mu_1 = 1.0$, the estimated
contribution from $\mathcal{O}_1$ is then $0.2$ times that from
$\mathcal{O}_{10}$. This is in good agreement with the left panel of
Figure \ref{fig:Model-I-Positive-Mu1} for $m_S \geq 100 \GeV$, above
the accidental cancellation around $m_S \simeq 30 \GeV$.

For $m_S > m_\phi$ and unit couplings, the box diagrams instead
dominate, in which case
\eq{ \label{eq:Model-I-c1n-box-approx}
  c_1^N|^{\textrm{I}} \approx \dfrac{h_2^2 g_1^2}{16 \pi^2}
  f_T^N m_N m_S \left.(M_2 + M_4)\right\rvert_{r \rightarrow 0} \; .
}
The ratio of the number of scattering events then becomes
\eq{ \label{eq:Model-I-ratio-events-box-approx}
  \left. \dfrac{ N_{\mathcal{O}_1} }{ N_{\mathcal{O}_{10}}
    }\right\rvert_{\text{I}} \approx \left( \dfrac{h_2 g_1}{16 \pi^2}
    \dfrac{f^N_T}{\tilde{\Delta}^N}  m_N \, m_S \, m_\phi^2 \,
    \left(M_2 + M_4 \right)|_{r \rightarrow 0} \right)^2
  \dfrac{A}{\langle S_N \rangle^2} \dfrac{m_N}{2 E_R} \; .
}
Using $E_R = 30 \keV$ and $h_2 = g_1 = \mu_1 = 1.0$, for
$m_S = 1000 \GeV$ and $m_\phi = 125 \GeV$ the estimated number of
events for $\mathcal{O}_1$ is 2.8 times that for
$\mathcal{O}_{10}$. The left panel of
Fig.~\ref{fig:Model-I-Positive-Mu1} shows that this is a slight
overestimate even if the smaller ROI is used. For the larger maximal
recoil energy, $E_R^{\textrm{upp}} = 240 \keV$, $\mathcal{O}_{10}$
still contributes roughly three times more scattering events than
$\mathcal{O}_1$. Using the same couplings, for
$m_S = m_\phi = 1000 \GeV$, the estimated contribution from
$\mathcal{O}_1$ is smaller by a factor $2.5 \times 10^{-2}$ relative
to that from $\mathcal{O}_{10}$. This rough estimate matches the ratio
between the $\mathcal{O}_1$ and $\mathcal{O}_{10}$ contributions in
the right frame of Figure~\ref{fig:Model-I-Positive-Mu1}.

In Fig.~\ref{fig:Model-I-Negative-Mu1} we have changed the sign of
$\mu_1$ while keeping $g_1$ and $g_2$ positive (or zero).  Both
triangle diagrams therefore now contribute with opposite signs than
the (crossed) box diagrams. As a result, the coefficient of
$\mathcal{O}_1$ now vanishes at $m_S \simeq 0.5 m_\phi$ for
$g_2 = 1.0$, and at $m_S \simeq 0.1 m_\phi$ for $g_2 = 0$.

As before, for $g_2 = 1.0$ and small DM masses the Wilson coefficient
of $\mathcal{O}_1$ can be approximated by the triangle diagram
involving the $g_2$ coupling. From
eq.\eqref{eq:Model-I-ratio-events-g2-approx} for $m_S = 100 \GeV$,
$m_\phi = 1000 \GeV$ and $h_2 = g_1 = 1.0$ we estimate
$7.2 \times 10^{-2}$ as ratio of the $\mathcal{O}_1$ and
$\mathcal{O}_{10}$ contributions, in rough agreement with the right
panel of Fig.~\ref{fig:Model-I-Negative-Mu1}. For $m_S > m_\phi$, the
box diagrams again dominate. Since they do not depend on $\mu_1$ our
earlier estimates still apply. However, since the triangle diagrams
aren't quite negligible even at $m_S = 8 m_\phi$, the largest ratio
covered in Fig.~\ref{fig:Model-I-Negative-Mu1}, this somewhat
overestimates the importance of $\mathcal{O}_1$.

Before turning to the neutron EDM, we discuss results for DM--Xenon
scattering for Model IIa. We again chose a flavor--universal Yukawa
coupling $h_2^q \equiv h_2 = 1$, and the same two values of mediator
mass as for Model I. The coupling $\lambda_1$ appearing in $c_{10}$ is
set to 1, and we show results for the pseudoscalar mediator--WIMP
coupling $\lambda_2 = 1$ or $0$.

\begin{figure}[h!]
	\centering
	\includegraphics[width=1.15\linewidth]{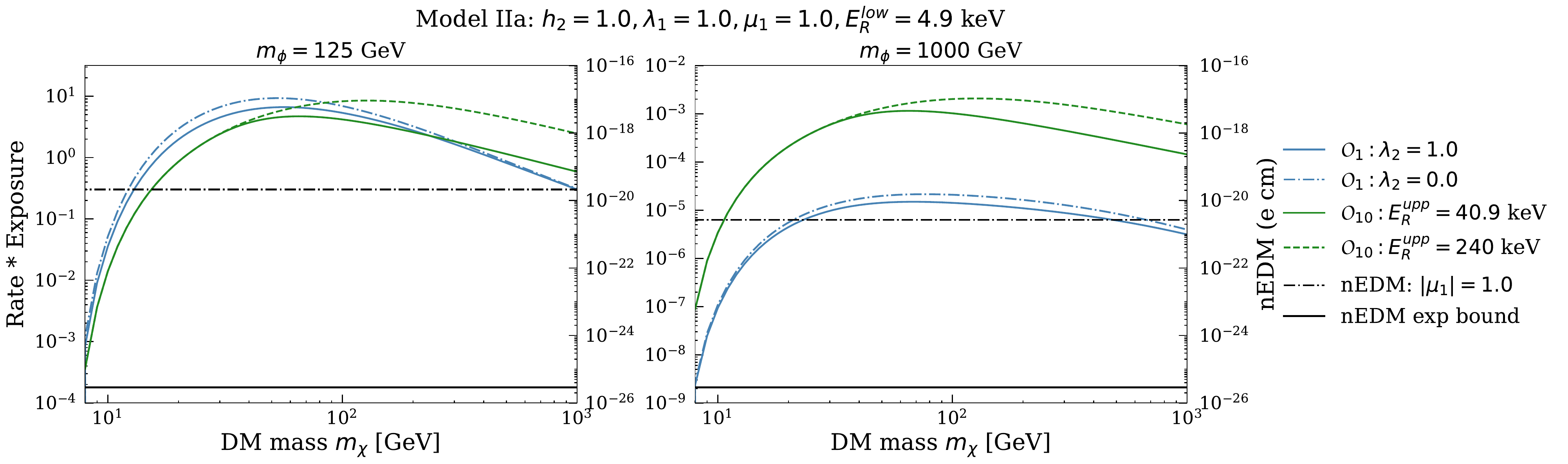}
	\caption{The left $y-$axes show the total number of scattering events
		in Model IIa with $\mu_1 = 1.0$ at XENON1T with a runtime exposure
		of 1 tonne-year as a function of the DM mass $m_\chi$ for two
		values of the mediator mass, $m_\phi = 125 \GeV$ (left frame) and
		$m_\phi = 1000 \GeV$ (right frame). The \textbf{green} and
		\textbf{blue} curves show the number of events due to
		$\mathcal{O}_{10}$ and $\mathcal{O}_1$ respectively. The \textbf{black}
		lines refer to the right $y-$axes;
		\textbf{solid black} line is the experimental upper limit on the nEDM, and
		the \textbf{dashed black} line is value of nEDM predicted by Model
		IIa.}
	\label{fig:Model-IIa-Positive-Mu1}
\end{figure}

Figure \ref{fig:Model-IIa-Positive-Mu1} shows the number of scattering
events from $\mathcal{O}_{10}$ and $\mathcal{O}_1$ for $\mu_1 =
1.0$. We again find that for the large mediator mass
($m_\phi = 1000 \GeV$, right frame) the tree--level contribution
from $\mathcal{O}_{10}$ dominates by approximately two orders of
magnitude for the entire range of $m_\chi$ shown.\footnote{Due to the
  lower cut on the recoil energy and the upper limit on the WIMP
  velocity related to the galactic escape velocity, the entire
  scattering rate vanishes for WIMP masses below $5$ GeV.} On the
other hand, for $m_\phi = 125 \GeV$ the two contributions are
roughly comparable, with $\mathcal{O}_1$ dominating at small WIMP
masses and $\mathcal{O}_{10}$ dominating for larger masses, in
particular if the upper cut on the recoil energy is relaxed to $240$
keV.

\begin{figure}[!h]
	\centering
	\includegraphics[width=1.15\linewidth]{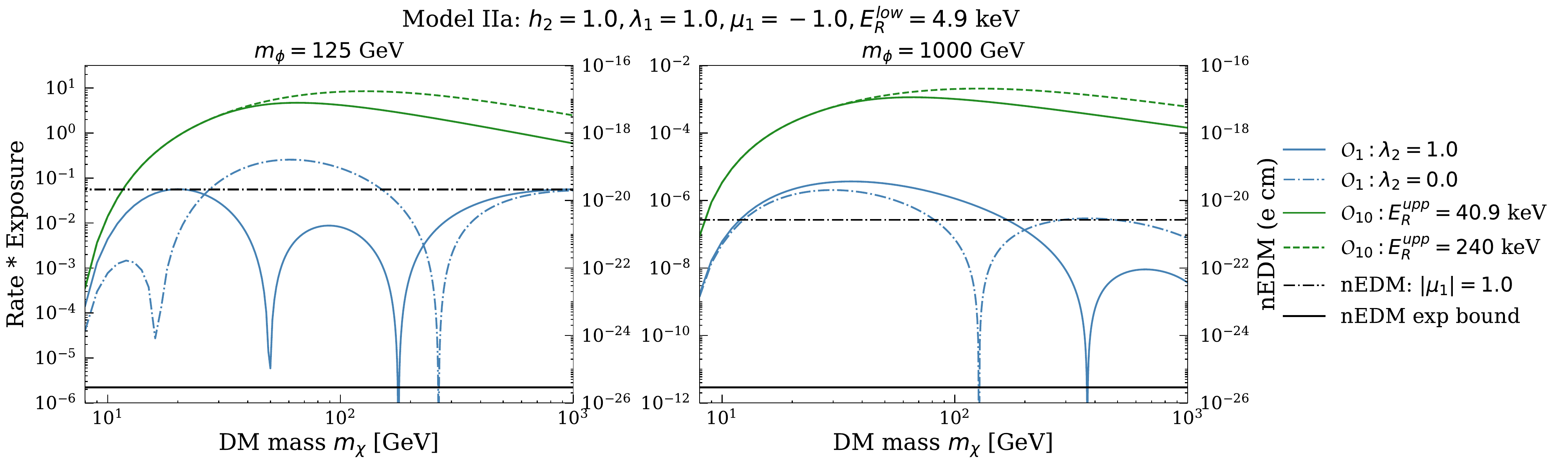}
	\caption{As in Fig.\ref{fig:Model-IIa-Positive-Mu1}, but for
		$\mu_1 = -1.0$.}
	\label{fig:Model-IIa-Negative-Mu1}
\end{figure}

The main qualitative difference to the results of Model I is that
$c_1$ remains nonzero over the entire range of DM mass shown. The
terms from the scalar--scalar current
$\propto \lambda_1^2 + \lambda_2^2$ in eq.\eqref{eqn:model-iia-c1n},
third line, contribute with opposite sign from those
$\propto \lambda_1^2$ (in the fifth line), which have the same sign as
the contribution from the triangle diagram (the last term). This last
(triangle) term dominates both for $m_\chi \ll m_\phi$ and for
$m_\chi \gg m_\phi$, but is slightly smaller than the total
contribution from box diagrams for $m_\chi \sim m_\phi$. As a result,
when the (relative) sign between these contributions is flipped by
choosing $\mu_1 = -1$, as in Fig.~\ref{fig:Model-IIa-Negative-Mu1},
$c_1$ vanishes at the two values of $s = m_\chi/m_\phi$ where the
total box contribution has the same magnitude as the contribution from
the triangle diagram. Since this cancellation happens at fixed values
of $s$, it occurs at larger $m_\chi$ when $m_\phi$ is increased (right
frame).

The upshot of this discussion is that both in Model I and in Model IIa
the contribution from the non--leading NREFT operator
$\mathcal{O}_{10}$ can indeed dominate the loop suppressed
contribution from the leading operator $\mathcal{O}_1$, especially for
large mediator masses. However, we saw at the end of the previous
Chapter that in these models an nEDM is produced by two--loop diagrams
involving a quark and the mediator $\phi$, see
Fig.~\ref{fig:two-loop-skeleton-diagram}. Its $90 \%$
C.L. experimental upper bound is depicted by the solid black curves in
Figs~\ref{fig:Model-I-Positive-Mu1} to
\ref{fig:Model-IIa-Negative-Mu1} which refer to the right $y-$axes,
while the nEDM predicted by the models for the given choice of
parameters is shown by the dashed black curves. We estimate the
numerical value of the nEDM from the dominant contribution from the
down quark cEDM $\tilde{d}_d$. Using eqs.\eqref{eq:two-loop-qEDM} and
\eqref{eq:nEDM_lattice_sumrules}, we obtain for $h_2 = \mu_1 = 1.0$:
\eq{ \label{eq:nEDM_I}
  |d_n| \approx \dfrac{2 \, g_S }{(16 \pi^2)^2} \, \left(
    \dfrac{m_\phi}{{\rm GeV}} \right) \,
  \dfrac{5.28 \times 10^{-2}} {(m_\phi/{\rm GeV})^2} \; \textrm{e} \cdot
  {\rm GeV}^{-1}
  \approx \dfrac{1.58 \times 10^{-19}} {(m_\phi/{\rm GeV})} \, \textrm{e}
  \cdot \cm \;,
}
where we used $g_S = 1.9$ for the strong coupling at scale of a
few\GeV. The loop integrals in eq.\eqref{eq:two-loop-qEDM} only depend
on the quark mass $m_q$ and the mediator mass $m_\phi$. On dimensional
grounds, they can be written as $1/m_\phi^2$ times a function of the
dimensionless ratio $m_q/m_\phi$. For the down quark we can safely set
$m_q \rightarrow 0$ for the values of $m_\phi$ we consider; the loop
integral then evaluates to $5.28 \times
10^{-2}/m_\phi^2$. Eq.\eqref{eq:nEDM_I} therefore predicts
$|d_n| \sim {\cal O}(10^{-21}) \; e \cdot \cm$ for
$m_\phi = 125 \GeV$. Increasing $m_\phi$ to $1$ TeV reduces the
nEDM value by a factor of $8$; recall that we parameterize the
trilinear scalar coupling as $\mu_1 m_\phi$, so for fixed $|\mu_1|$,
$d_n \propto 1/m_\phi$. This agrees with the dashed black lines in the
panels of Figs.~\ref{fig:Model-I-Positive-Mu1} to
\ref{fig:Model-IIa-Negative-Mu1}.

Evidently for this set of couplings and $m_\phi = 125$ GeV, the
predicted nEDM exceeds the upper bound by about $5$ orders of
magnitude; increasing $m_\phi$ to 1 TeV still leads to a discrepancy
by about $4$ orders of magnitude. Clearly these sets of parameters,
which led to in principle observable effects from $\mathcal{O}_{10}$,
are not realistic.

We saw in eq.\eqref{eq:two-loop-qEDM} that the predicted nEDM scales
like $h_2^3 \, \mu_1$. In order to suppress the produced nEDM below
the upper limit, one thus has to reduce this product of couplings by
$\mathcal{O}(10^{-4}-10^{-5})$. This could be achieved by reducing the
Yukawa coupling $h_2$ by a factor of at least $30$ while keeping
$|\mu_1| = 1.0$ the same. However, this would reduce the contribution
from $\mathcal{O}_{10}$ to the scattering rate by a factor of $10^3$;
the contribution from $\mathcal{O}_1$ would even be reduced by a
factor of $10^6$. The resulting cross sections lie below their
corresponding neutrino floor(s); this part of parameter space of
Models I and IIa is thus beyond the sensitivity of direct search
experiments.

\begin{figure*}[t!]
\centering
\includegraphics[width=1.\linewidth]{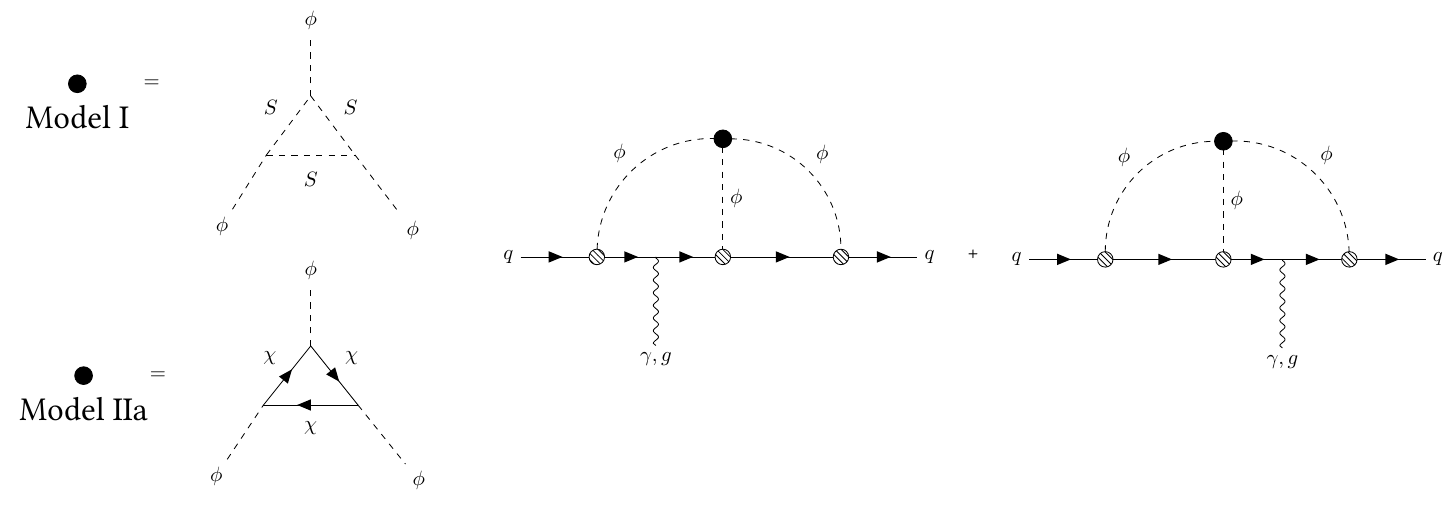}
\caption{One--loop radiative corrections to the three--point function
  of the mediator $\langle \phi^3 \rangle$ given by triangle diagrams
  in Model I and Model IIa; such diagrams induce a non--vanishing
  effective $\mu_1$ even if this trilinear coupling is initially set
  to zero. The quark EDM and cEDMs are then induced at the three--loop
  level. }
\label{fig:two-loop-constraint}
\end{figure*}

Alternatively one can reduce $\mu_1$ by a factor $\geq 10^4$. This
effectively removes one of the triangle diagrams contributing to
$\mathcal{O}_1$, but the contribution from $\mathcal{O}_{10}$ as well
as the box and -- for Model I -- the second triangle diagram for
$\mathcal{O}_1$ remain unchanged. The contribution from
$\mathcal{O}_{10}$ would then still dominate over that
$\mathcal{O}_1$, except for small $m_S$ as discussed above. However,
setting $\mu_1 = 0$ does not increase the symmetry of the
theory. Therefore setting $\mu_1$ to zero only suppresses it at the lowest
order in perturbation theory. Radiative corrections
will in general induce a non--zero value of
this cubic self--coupling.

As shown in Fig.~\ref{fig:two-loop-constraint}, this happens at
one--loop level in both Model I and Model IIa, via triangle diagrams
with the DM particle running in the loop. Crucially, in both cases the
DM--mediator coupling used here is the same coupling that appears in
$c_{10}$, i.e. one cannot ``switch off'' these triangle diagrams
without simultaneously setting the contribution from
$\mathcal{O}_{10}$ to zero. In case of Model IIa, the diagram shown in
Fig.~\ref{fig:two-loop-constraint} is (logarithmically) divergent,
i.e. setting $\mu_1 = 0$ leads, strictly speaking, to a
non--renormalizable theory. The triangle diagram shown for Model I is
convergent. If the quartic coupling $g_2$ is nonzero, there is also a
divergent ``bubble'' diagram involving one $g_2$ vertex and one $g_1$
vertex, but $g_2$ does not contribute to $c_{10}$ at tree--level, so
in the following discussion we will set $g_2 = 0$.

Inserting these one--loop triangle diagrams into the upper vertex of
the EDM diagrams results in three--loop diagrams.  Rather than
performing the challenging full three--loop calculation, we use our
earlier two--loop result for $d_q$ and insert a lower bound on
$\mu_1$ due to the triangle diagrams:
\eq{ \label{eq:mu1bound}
  \textrm{Model I:} \quad &\mu_1 \gtrsim \dfrac{g_1^3}{16 \, \pi^2} \,
  \dfrac{m_S}{2 m_\phi} \; ,\\
  \textrm{Model IIa: } \quad &\mu_1 \gtrsim \dfrac{\lambda_1^3}{16 \,
    \pi^2} \, \dfrac{2m_\chi}{m_\phi} \; .
}
The relative factor of $4$ accounts for the four degrees of freedom running
in the loop in Model IIa; computationally it results from the Dirac trace
appearing in the evaluation of the fermionic triangle diagram.
In the next step, we convert this lower
bound on $\mu_1$ into a lower bound on the quark EDM (and cEDM) using
eq.\eqref{eq:two-loop-qEDM}:
\eq{ \label{eq:dn_est}
  \textrm{Model I: } \quad & d_q \gtrsim \dfrac{e \, Q_q}{(16 \pi^2)^3} \,
  (h_2 \, g_1)^3 \, m_S \, \lim_{q^2 \rightarrow 0} \,
  \Bigl( \left[ \mathcal{X} \right] + \left[ \mathcal{Y} \right] \Bigl) \, ;
  \\
  \textrm{Model IIa: } \quad &d_q \gtrsim 4 \,\dfrac{e \, Q_q}{(16
    \pi^2)^3} \, (h_2 \, \lambda_1)^3 \, m_\chi \, \lim_{q^2
    \rightarrow 0} \, \Bigl( \left[ \mathcal{X} \right] + \left[
    \mathcal{Y} \right] \Bigl) \, .
}
This leads directly to a lower bound on the nEDM, which can be
translated into upper bounds on $\abs{h_2^3 \, g_1^3}$ and
$\abs{h_2^3 \, \lambda_1^3}$ in Model I and Model IIa, respectively,
by requiring our theoretical lower bound on $d_n$ not exceed the
stringent experimental upper bound.

Note that the products of couplings which are bounded by the nEDM also
appear in the coefficient $c_{10}$, see
eqs.\eqref{eq:model-i-coef-c10N} and \eqref{eqn:model-iia-coef-c10N},
although with different powers: the bound on the nEDM scales
cubically with the product of couplings whereas $c_{10}$ scales
linearly, i.e. the contribution from $\mathcal{O}_{10}$ to the DM
scattering rate scales quadratically. The same is true for the
``DM--nucleon scattering cross section'' due to $\mathcal{O}_{10}$,
defined by:
\eq{
  \sigma_{\mathcal{O}_{10}}|^{\textrm{I}} =  \dfrac{3 \mu_{S \, N}^2}{\pi}
  \left( c^{N}_{10}|^{\text{I}} \right)^2 \quad \text{and}
  \quad \sigma_{\mathcal{O}_{10}}|^{\textrm{IIa}} =  \dfrac{3 \mu_{\chi \, N}^2}
  {\pi} \left( c^{N}_{10}|^{\text{IIa}} \right)^2 \; .
}
The upper bounds on $\abs{h_2 \, g_1}$ and $\abs{h_2\, \lambda_1}$
from the nEDM will thus lead to upper bounds on this cross section.

\begin{figure*}[t!]
\centering
\includegraphics[width=0.48\linewidth]{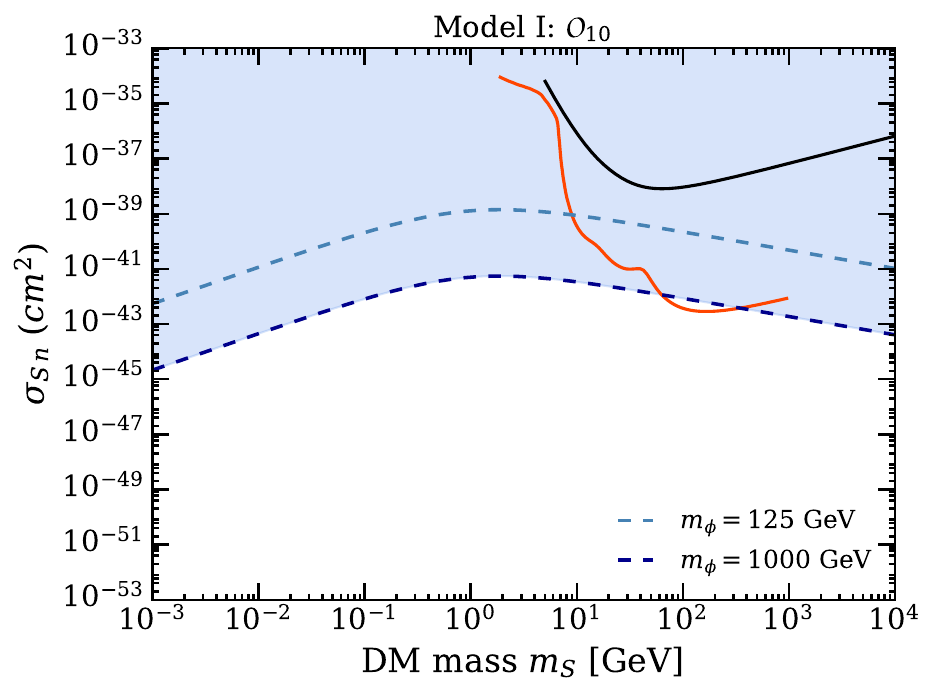}
\includegraphics[width=0.48\linewidth]{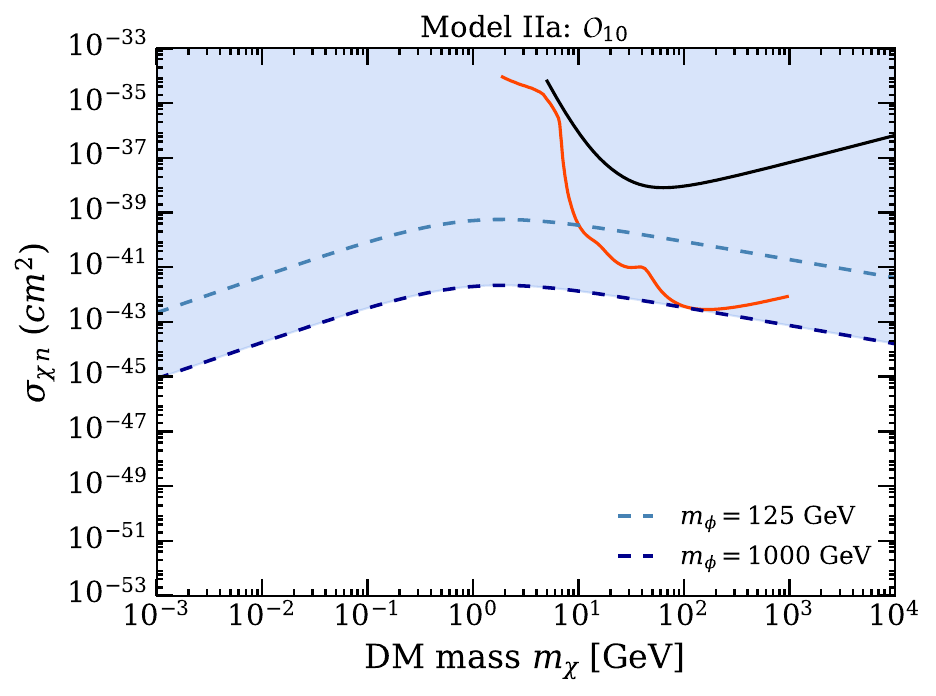}
\caption{The \textbf{black} curves show the XENON1T upper bound on the
  DM--nucleon cross section from $\mathcal{O}_{10}$ in Model I (left
  frame) and Model IIa (right frame) as a function of DM mass. The
  indirect constraint from the nEDM is shown by the \textbf{dashed
    light blue} curve for $m_\phi = 125 \GeV$ and the \textbf{dashed
    dark blue} curve for $m_\phi = 1000 \GeV$; the shaded region is
  thus excluded by the nEDM bound for $m_\phi = 1000 \GeV$. The
  \textbf{red} curve denotes the neutrino floor for $\mathcal{O}_{10}$
  and has been taken from Ref.~\cite{Dent:2016iht}.}
\label{fig:model-i-iia-nEDM-sigma-constraint}
\end{figure*}

This is illustrated in
Figs.~\ref{fig:model-i-iia-nEDM-sigma-constraint}, which show excluded
regions in the plane spanned by $\sigma_{\mathcal{O}_{10}}$ and the DM
mass, for Model I (left) and Model IIa (right). The dashed blue curves
depict the nEDM--derived $90 \%$ c.l. upper limit on the cross section
for our two standard choices of the mediator mass $m_\phi$.  This
bound becomes weakest at DM mass $m_{\rm DM} \simeq 1 \GeV$,
i.e. close to the nucleon mass; here $m_{\rm DM} = m_S \ (m_\chi)$ for
Model I (Model IIa). For fixed couplings our estimated $d_n$ of
eq.\eqref{eq:dn_est} increases linearly with $m_{\rm DM}$, while
$\sigma_{\mathcal{O}_{10}} \propto m^2_{\rm DM}$ for
$m_{\rm DM} \ll 1 \GeV$ but becomes independent of it for large DM
mass. As a result, taking into account the different powers of
couplings involved in the two quantities, the $d_n-$derived upper
bound on $\sigma_{\mathcal{O}_{10}}$ grows like $m_{\rm DM}^{4/3}$ for
$m_{\rm DM} \ll 1 \GeV$, but declines like $m_{\rm DM}^{-2/3}$ for
$m_{\rm DM} \gg 1 \GeV$. Moreover,
$\sigma_{\mathcal{O}_{10}} \propto m_\phi^{-4}$ while our estimated
bound on $d_n \propto m_\phi^{-2}$; the $d_n-$derived upper bound on
$\sigma_{\mathcal{O}_{10}}$ therefore scales $\propto m_\phi^{-8/3}$
for all DM masses, i.e. the bound becomes more stringent for larger
mediator mass. Finally, the bound is stronger in Model IIa by a factor
of $4^{2/3} \simeq 2.5$ due to the relative factor of $4$ between the
radiatively generated value of $\mu_1$ given in
eqs.\eqref{eq:mu1bound}.

The red curves in Fig.~\ref{fig:model-i-iia-nEDM-sigma-constraint}
show the irreducible background level from coherent neutrino--nucleus
scattering (``neutrino floor'') as estimated in
Ref.~\cite{Dent:2016iht}. We see that the indirect constraint is
around five orders of magnitude below the present XENON1T sensitivity,
and for most DM masses well below the neutrino floor for
$m_\phi = 1000 \GeV$. For this value of the mediator mass the current
constraints on $d_n$ therefore imply that the interactions due to
$\mathcal{O}_{10}$ are essentially unobservable; recall that for this
large mediator mass, $\mathcal{O}_1$ contributes even less to the
scattering rate.

For $m_\phi = 125 \GeV$, the nEDM constraint still lies well below the
current sensitivity. On the other hand, for
$m_{\rm DM} \gtrsim 7 \GeV$ it is up to two orders of magnitude above
the neutrino floor. Our analysis can therefore not completely exclude
the possibility that future Xenon experiments might become sensitive
to contributions from $\mathcal{O}_{10}$, if the upper bound on $d_n$
remains unchanged. It should be noted, however, that saturating the
bound on $d_n$ requires relatively large couplings. Setting
$g_1 = \lambda_1 = 1$ and $m_\phi = 125 \GeV$, the estimate
\eqref{eq:dn_est} is saturated for Yukawa coupling
$h_2 \simeq 0.13 [m_\phi/(\kappa m_{\rm DM})]^{1/3}$, where
$\kappa = 1/2 \ (2)$ in Model I (Model IIa); reducing the size of the
DM--mediator coupling would require even larger $h_2$. Given our
assumption of flavor--universal Yukawa couplings, experiments at LEP,
the Tevatron and the LHC should be able to set quite stringent bounds
on $h_2$ for $m_\phi \lesssim 125 \GeV$.

\begin{figure*}[t!] 
\centering
\includegraphics[width=1.15\linewidth]{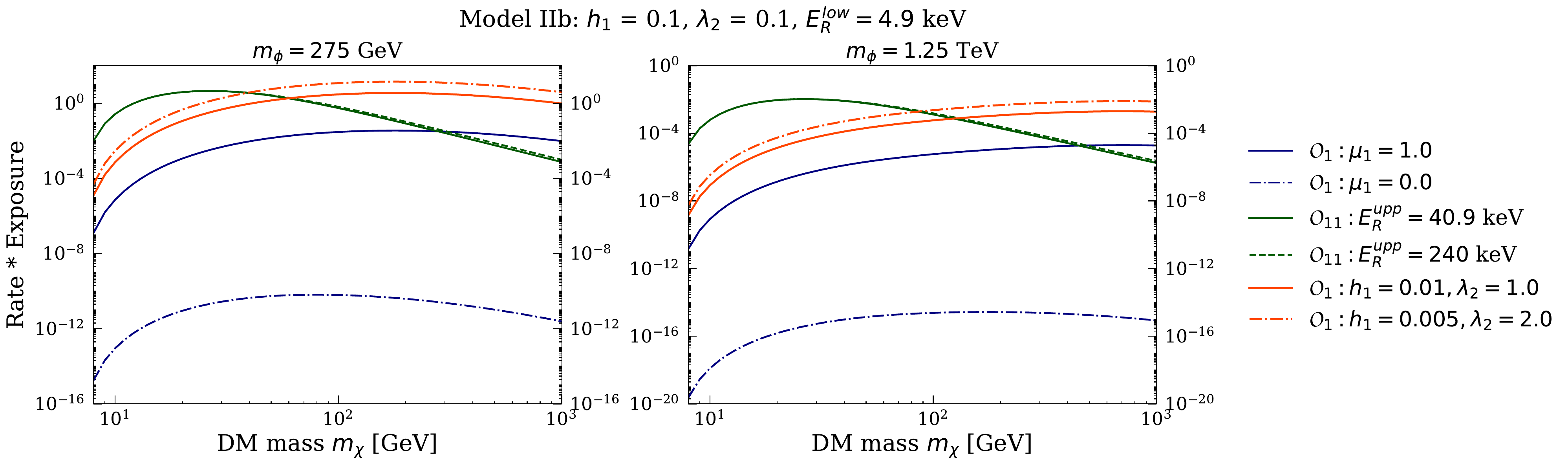}
\caption{Total number of scattering events in Model IIb at XENON1T
  with a runtime exposure of 1 tonne--year as a function of the DM
  mass $m_\chi$ for $h_1 = \lambda_2 = 0.1$, with $m_\phi = 275\GeV$
  (left) and $m_\phi = 1.25\TeV$ (right). The \textbf{green} curves
  show the number of events due to $\mathcal{O}_{11}$ and the
  \textbf{blue} curves show the number of events due to
  $\mathcal{O}_1$. The \textbf{red} curves also show contributions
  from $\mathcal{O}_1$ with the same product $h_1 \lambda_2$, and
  hence the same contribution from $\mathcal{O}_{11}$, but larger
  $\lambda_2$.}
\label{fig:model-iib-set1}
\end{figure*}

We finally discuss numerical results for Model IIb. We recall that
this model had been constructed to generate the operator
$\mathcal{O}_{11}$ at tree--level, which is independent of the spin of
the target nucleus. This required a scalar Yukawa coupling of the
mediator to quarks, and a pseudoscalar coupling to the DM
fermion. While this assignment quite manifestly again violates $CP$,
it does not generate new contributions to the neutron EDM. However,
at one--loop contributions to the leading spin--independent operator
$\mathcal{O}_1$ are generated also in this model.

\begin{figure*}[t!]
\centering
\includegraphics[width=1.15\linewidth]{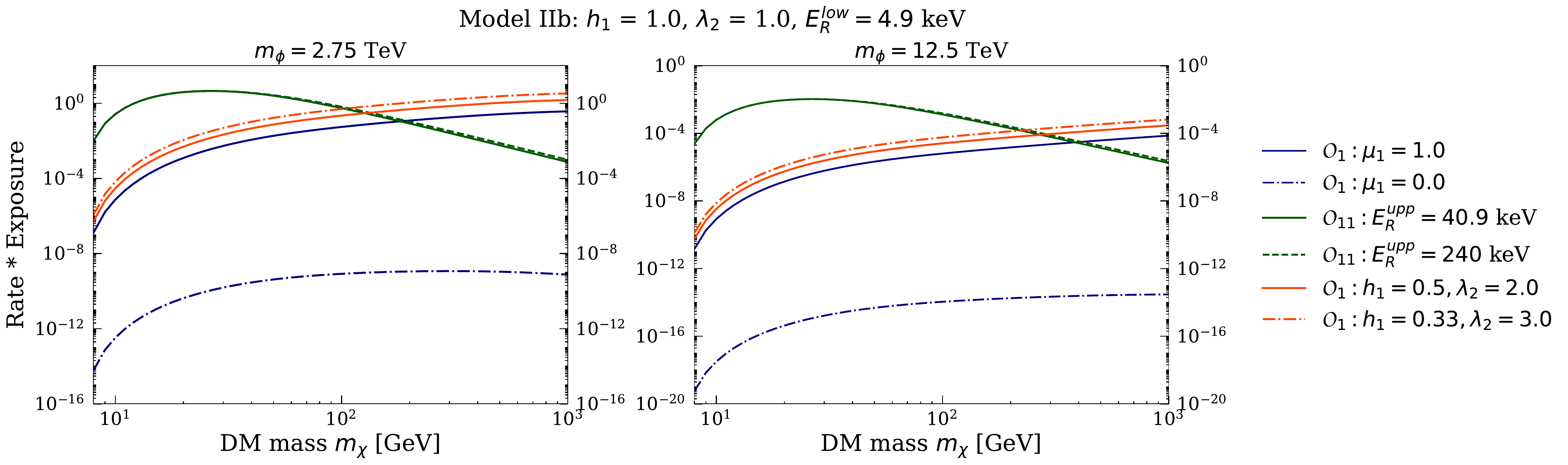}
\caption{As in Fig.~\ref{fig:model-iib-set1}, but with
  $\lambda_2 = h_1 = 1$ and correspondingly larger mediator masses:
  $m_\phi = 2.75 \TeV$ (left) and $m_\phi = 12.5 \TeV$ (right).}
\label{fig:model-iib-set2}
\end{figure*}

In Figs.~\ref{fig:model-iib-set1} and \ref{fig:model-iib-set2}, we
display the number of scattering events due to $\mathcal{O}_1$ and
$\mathcal{O}_{11}$ for $1.0$ tonne--year exposure of the XENON1T
experiment as a function of the DM mass $m_\chi$. In each figure the
mediator mass in the left frame is chosen such that the contribution
from $\mathcal{O}_{11}$ saturates the XENON1T constraint for
$m_\chi \simeq 30$ GeV, while in the right frame this contribution is
barely above the neutrino floor for
$10\GeV \lesssim m_\chi \lesssim 50\GeV$ (and slightly below it for
larger $m_\chi$). We again assume flavor--universal Yukawa couplings
$h_1^q \equiv h_1$. The event rates due to $\mathcal{O}_{11}$ are
shown by the green lines, while the blue curves show the event rate
due to $\mathcal{O}_1$ for $h_1 = \lambda_2 = 0.1$, with ($\mu_1 = 1$,
solid) or without ($\mu_1 = 0$, dashed) the triangle diagram. The red
curves also show contribution due to $\mathcal{O}_1$ with $\mu_1 = 1$,
but for $\lambda_2 > h_1$, keeping the product $\lambda_2 \cdot h_1$,
and hence the contribution from $\mathcal{O}_{11}$, constant.

Evidently the loop--induced contribution from $\mathcal{O}_1$ can
only be competitive if the triangle diagram is not suppressed. This is
in accord with our discussion of eq.\eqref{eq:model-ii-b-coef-c1n},
which showed that this contribution is expected to dominate if $\mu_1$
is sizable. We reiterate that there are logarithmically divergent
one--loop contributions of order $h_1^3 m_q/(16 \pi^2m_\phi)$ to this
coupling from quark triangle diagrams, hence there is no reason to
assume that $\mu_1$ is very small.

We also see that in all cases the tree--level contribution from
$\mathcal{O}_{11}$ drops quickly for $m_\chi \gtrsim 50\GeV$. This is
partly due to the reduced flux of DM particles, which scales
$\propto 1/m_\chi$, but mostly because $c_{11} \propto 1/m_\chi$, as
shown in eq.\eqref{eqn:model-iib-c11n}. In contrast, the loop--induced
contribution from $\mathcal{O}_1$ in many cases keeps increasing with
increasing $m_\chi$ over the entire range shown. This is because for
$m_\chi \ll m_\phi$, the triangle loop function satisfies
$s S_1(s) \simeq m_\chi / (2 m_\phi^3)$, hence this contribution to
the scattering cross section scales $\propto m_\chi^2$ for
$m_\chi \ll m_\phi$. As a result, for $\mu_1 = 1$ we always find that
the contribution from $\mathcal{O}_{11}$ dominates for small $m_\chi$,
while that from $\mathcal{O}_1$ is dominant at large $m_\chi$; the DM
mass where the two contributions are equal depends on the values of
the other parameters. The behavior of the loop function $S_1$ also
explains why the contribution due to $\mathcal{O}_1$ drops faster with
increasing $m_\phi$ than that due to $\mathcal{O}_{11}$, as long as
$m_\phi > m_\chi$.

We note that the number of events from $\mathcal{O}_{11}$ remains
essentially unchanged when the maximal recoil energy is increased from
$40.9 \keV$ to $240 \keV$. As in case of $\mathcal{O}_{10}$, the
scattering rate due to $\mathcal{O}_{11}$ is peaked at non--zero
recoil energies, due to the $\vec{q}$ factor in the definition of this
operator. However, unlike $\mathcal{O}_{10}$, $\mathcal{O}_{11}$ leads
to coherent scattering on the entire nucleus. The rate is thus
proportional to the square of the spin--independent elastic form
factor of Xenon, which is quite soft. It is this coherent enhancement
which leads to a much larger scattering rate from $\mathcal{O}_{11}$
than from $\mathcal{O}_{10}$, for similar Wilson
coefficients. However, the product of recoil energy $E_R$ (which is
$\propto \vec{q}^2$) and squared form factor already peaks at
$E_R \simeq 20 \keV$; the convolution with the DM velocity
distribution further suppresses the rate at large $E_R$.

Semi--quantitatively, the ratio of the two contributions can be
estimated as follows:
\eq{
  \left. \dfrac{ N_{\mathcal{O}_1} }{
      N_{\mathcal{O}_{11}} }\right\rvert_{\text{IIb}} =
  \left. \dfrac{ R_{\mathcal{O}_1} }{ R_{\mathcal{O}_{11}}
    }\right\rvert_{\text{IIb}} \; \sim \dfrac{ (c_1^{N}
    |^{\text{IIb}})^2 }{ (c_{11}^{N} |^{\text{IIb}})^2 }
  \; \dfrac{m_N^2}{\vec{q}^{\; 2}} \; .
}
The nuclear response is the same for both contributions and therefore
does not appear in the ratio. The factor $(q^2/m_N^2)^{-1}$ is due to
the momentum transfer dependence of $\mathcal{O}_{11}$. We only retain
the leading triangle contribution to $c^N_1 |^{\text{IIb}}$, see
eq.\eqref{eq:model-ii-b-coef-c1n},
\eq{
  c^N_1 |^{\text{IIb}} \approx -\dfrac{\lambda_2^2}{16 \,
 \pi^2} \, \mu_1 \, m_N \, \tilde{f}^N \, \dfrac{m_\chi}{m_\phi} \, S_1 \; .
}
Using $c^N_{11} |^{\text{IIb}}$ from
eq.\eqref{eqn:model-iib-c11n}, the ratio of events is thus given by
\eq{ \label{eq:r_2b}
  \left. \dfrac{ N_{\mathcal{O}_1} }{ N_{\mathcal{O}_{11}}
    }\right\rvert_{\text{IIb}} \approx \left( \dfrac{\lambda_2 \,
      \mu_1}{16 \, \pi^2} \dfrac {m_\chi^2 m_\phi} {m_N}
     S_1(s) \right)^2 \dfrac{m_N}{2 A E_R} \; .
}
For example, for $\lambda_2 = \mu_1 = 1.0,\; m_\phi = 2.75 \TeV$ and
$m_\chi = 200 \GeV$, and taking $E_R = 20 \keV$ as typical recoil
energy in order to account for the soft form factor for coherent
scattering, gives $0.45$ for the ratio of event numbers, in rough
agreement with the results shown in the left frame of
Fig.~\ref{fig:model-iib-set2}. For the same couplings but increasing
$m_\phi$ to $12.5 \TeV$, eq.\eqref{eq:r_2b} predicts equal event rates
for $m_\chi = 520 \GeV$, quite close to the intersection point between
the green and blue lines in the right frame of
Fig.~\ref{fig:model-iib-set2}. Moreover, eq.\eqref{eq:r_2b} also
explains why increasing $\lambda_2$ while keeping
$\lambda_2 \cdot h_1$ fixed (red curves) increases the contribution
from $\mathcal{O}_1$. In fact, for $m_\phi \lesssim 1 \TeV$ reducing
$h_1$ should help to avoid possible constraints on the model from
searches at the LHC.

%%%%%%%%%%%%%%%%%%%%%%%%%%%%%%%%%%%%%%%
\section{Summary and Conclusions} 
\label{sec:conclusions}

In this paper we explored the detection prospects involving $P-$ and
$T-$odd operators arising in the NREFT formalism of WIMP--nucleon
scattering. These operators appear at next--to--leading order in an
expansion in WIMP velocity $v$ and momentum transfer $\vec{q}$. Since
these quantities accompanying the $P-$ and $T-$odd operators are quite
small, these operators are expected to be insignificant relative to
the leading order operators {\em if} the corresponding Wilson
coefficients are of similar magnitude. Hence the additional operators
can make significant contributions only when the coefficient of the
leading spin--independent operator $\mathcal{O}_1$ is strongly
suppressed or vanishes entirely. This typically requires ad hoc
choices of couplings in a relativistic QFT, i.e. one can generally not
find a symmetry that suppresses the contribution from $\mathcal{O}_1$
without also suppressing the Wilson coefficients of the additional
operators.

Crucially, the $P-$ and $T-$odd NREFT operators can only occur in the
low energy limit of a QFT that violates the $CP$ symmetry. This can
lead to stringent constraints on the theory, in particular from
electric dipole moments.

We addressed these concerns in the framework of three simplified
models with uncharged $t-$channel mediators, taken from
ref.~\cite{Dent:2015zpa}. These models extend the SM by a real scalar
mediator particle $\phi$ which does not carry electric or color
charge, and a DM particle which has spin $0$ (Model I) or spin $1/2$
(Model IIa and IIb). These models can generate flavor changing neutral
currents already at tree--level unless the Yukawa couplings of the
mediator are diagonal in the quark mass basis. This can easily be
ensured if these new couplings are flavor--universal, which we
therefore assumed in our numerical examples.

The couplings in these models are chosen such that at the lowest order
in perturbation theory, only the $P-$ and $T-$odd operators
$\mathcal{O}_{10}$ and $ \mathcal{O}_{11} $ arise in the
non--relativistic limit. In particular, $\phi$ must not have scalar
couplings to both quarks and the DM particle. In Models I and IIa, the
quark couplings are pseudoscalar while the DM couplings are scalar,
while in Model IIb the quark couplings are scalar but the DM couplings
are pseudoscalar. Note that there is no symmetry that forbids scalar
quark couplings in Models I and IIa, or scalar DM couplings in Model
IIb. It is therefore not surprising that at the next order in
perturbation theory, one--loop box and triangle diagrams do induce the
canonical SI interactions described by the operator $\mathcal{O}_1$ in
these models. We compared the tree--level interactions giving rise to
$\mathcal{O}_{10}$ or $\mathcal{O}_{11}$ with the one--loop suppressed
interactions yielding $\mathcal{O}_1$. To that end we computed the
total number of events due to the two types of interactions for a
Xenon target.

In the case of Model I and Model IIa and assuming large couplings in
order to generate detectable event rates, we found that the
contributions from $\mathcal{O}_{10}$ can be roughly comparable to
those from $\mathcal{O}_1$ for mediator mass $m_\phi = 125 \GeV$,
but for heavier mediator the contributions from $\mathcal{O}_{10}$
clearly dominates. It thus appears as if $\mathcal{O}_{10}$ could
indeed be the most relevant NREFT operator in these models.

However, we pointed out that the quark--mediator interactions in both
models produce two--loop contributions to $d_n$, the electric dipole
moment of the neutron (nEDM). These contributions scale linearly with
the cubic self--interaction of the mediator $\mu_1$. For parameter
choices that lead to detectable event rates from $\mathcal{O}_{10}$
and $\mu_1$ of order unity (in units of $m_\phi$), the predicted $d_n$
is several orders of magnitude larger than the upper limit reported by
experiments. Even if we set $\mu_1 = 0$, non--vanishing trilinear
self--interactions are generated at one--loop level by couplings that
also appear in the Wilson coefficient of $\mathcal{O}_{10}$.
Estimating a lower bound on $\mu_1$ from these loop diagrams, we find
that the resulting upper bound on the WIMP--nucleus scattering rate is
still well below current sensitivity, and often even below the
irreducible background (``neutrino floor''), especially for large
$m_\phi$ where $\mathcal{O}_{10}$ potentially dominates the scattering
rate. As in case of models with charged, $s-$channel mediator, where
$d_n$ is generated already at one--loop level \cite{Drees:2019qzi}, it
is thus essential to consider the $d_n$ constraint when considering
prospects for detecting WIMP--nucleus scattering due to
$\mathcal{O}_{10}$. We expect this result to hold 
true for a simplified model containing a spin$-1$ WIMP and a neutral 
spin$-0$ mediator with pseudoscalar coupling to quarks, which would 
also give rise to ${\mathcal O}_{10}$. Since the WIMP--mediator coupling 
is again CP--even, electric dipole moments for quarks will be induced 
radiatively as well in this case.

In Model IIb, which generates the spin--independent NREFT operator
$\mathcal{O}_{11}$, we again found that the loop--induced
contributions from $\mathcal{O}_1$ can be larger than the tree--level
contribution due to $\mathcal{O}_{11}$, {\em if} the trilinear scalar
coupling $\mu_1$ is not suppressed. We emphasized that in this model
quark loops generate a logarithmically divergent one--loop
contribution to $\mu_1$. For $\mu_1 = 1$, $\mathcal{O}_1$ typically
dominates for large DM masses, while $\mathcal{O}_{11}$ is dominant
for small masses, the cross--over point depending on the values of the
other parameters. In this model no new contributions to $d_n$ are
generated, and therefore this operator is not subject to the stringent
constraints of the neutron EDM, even though it is also $P-$ and
$T-$odd. At least at the level of a simplified model one can therefore
engineer a scenario where the non--leading operator $\mathcal{O}_{11}$
dominates the WIMP--nucleon scattering rate.

However, it is by no means clear whether this remains true in the
framework of UV--complete theories. Recall that we only require our
Lagrangians to be invariant under $SU(3)_C \times U(1)_{\rm em}$; we
did not enforce invariance under the electroweak gauge symmetry. In
particular, the $\phi \bar q q$ couplings are not
$SU(2) \times U(1)_Y$ invariant if $\phi$ is a singlet. The simplest
choice would be to identify $\phi$ with the Higgs boson of the SM,
which resides in a doublet of $SU(2)$. However, the $\phi \bar q q$
couplings are then known to be very small. Moreover, the WIMP can then
also not be a gauge singlet, and would thus have additional (gauge)
interactions leading to additional constraints as well as new
contributions to WIMP--nucleon scattering. Alternatively one can
couple a singlet WIMP to a singlet scalar which mixes with the SM
Higgs boson; however, in such a scenario the WIMP--nucleon scattering
rate would be suppressed even further by the mediator--Higgs mixing
angle, which has to be rather small in order not to distort the
properties of the physical $125 \GeV$ particle too much.

We remind the reader that we assumed universal flavor--diagonal
couplings of the mediator to the quarks. This simplifies the model
building, since these couplings are then flavor--diagonal in any
basis. On the other hand, it might appear more natural to assume that
the new couplings increase with increasing quark mass, just as the
Yukawa couplings of the SM do. In this case tree--level FCNC are
avoided if the matrices of new Yukawa couplings commute with the
Yukawa coupling matrices of the SM; from the model building point of
view it is not clear why this should be the case. Moreover, keeping
the couplings to the top quark $\lesssim 1$ would then require very
small couplings to the first generation. In such a scenario the
contributions from heavy quarks might well dominate the DM scattering
cross section, as well as -- in Models I and IIa -- the electric
dipole moment of the neutron. In order to yield detectable event
rates, the couplings to third generation quarks would have to be
larger, and/or the mediator lighter, than in our numerical examples.
It seems rather unlikely to us that this would lead to qualitatively
different conclusions in Models I and IIa with unsuppressed trilinear
self--coupling of the mediator, given the very large discrepancy
between the predicted nEDM and its experimental upper bound that we
found in our numerical examples; however, we have not performed an
explicit computation to check this.\footnote{The cEDMs of heavy quarks
  generate light quark cEDMs, as well as an electron EDM, at the
  three--loop level \cite{Ema:2022wxd}. Moreover, the contribution of
  the gluonic Weinberg operator, which will generically be generated
  at three--loop level in Models I and IIa, would probably also have
  to be taken into account \cite{Yamanaka:2020kjo}.}

In summary, the results presented in this paper as well as
ref.\cite{Drees:2019qzi} strongly indicate that the current
experimental upper bound on the electric dipole moment of the neutron
excludes the possibility that the operator $\mathcal{O}_{10}$ can make
contributions to WIMP--nucleus scattering to which current or
near--future experiments are sensitive. While no such strong statement
can be made for the operator $\mathcal{O}_{11}$ generated by the
exchange of a neutral mediator in the $t-$channel, it is currently
unclear whether such a model can be constructed that respects the full
gauge symmetry of the SM and leads to detectable WIMP--nucleon
scattering being dominated by $\mathcal{O}_{11}$.

\acknowledgments

We acknowledge the use of TikZ--Feynman \cite{Ellis:2016jkw} for
creating Feynman diagrams and Package--X \cite{Patel:2016fam} for
verifying parts of our loop computations. We thank Claude Duhr for 
discussions about the two loop diagram. RM was partially supported 
by the Bonn Cologne Graduate School of Physics and Astronomy.

%%%%%%%%%%%%%%%%%%%%%%%%%%%%%%%%%%%%%%%
\appendix

\section{1-loop calculations}
\label{sec:one-loop-calculations}

In this appendix we provide details of the calculations of the
one--loop box and triangle diagrams appearing in Model I, IIa and IIb.

\subsection{Model I Matrix Element}
\label{sec:appendix-b-model-i}

\begin{figure}[h]
\centering
$\vcenter{\hbox{\includegraphics[width=6.5cm]{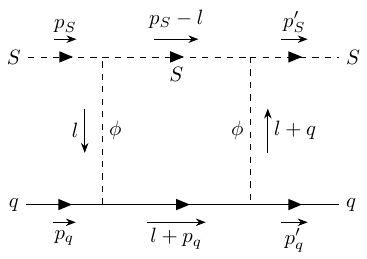}}}$
 $\vcenter{\hbox{\includegraphics[width=6.5cm]{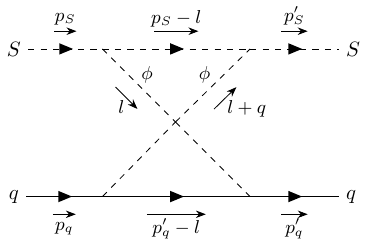}}}$
 \caption{One--loop box and crossed box diagrams contributing to
   $\mathcal{O}_1$ in Model \RNum{1}.}
 \label{fig:model1_boxes}
\end{figure}

The box diagram shown in the left Fig.~\ref{fig:model1_boxes} gives
the following contribution to the matrix element for DM--quark
scattering:
\eq{
  i \mathcal{M}^{\mathrm{\RNum{1}}}_{1} =  (-i g_1 m_S)^2 &\bigintssss
  \dfrac {\Diff4l} {(2\pi)^4} \dfrac {i} {(p_S-l)^2 - m_S^2}
  \dfrac {i} {l^2 - m_\phi^2} \dfrac {i} {(l+q)^2 - m_\phi^2}
  \dfrac {i} {(p_q + l)^2 - m_q^2} \nonumber \\
 &\times \bar{u}(p_q^{\prime}) \, h_2^q \gamma^5 \,
  (\slashed{p}_q + \slashed{l} + m_q) \, h_2^q \gamma^5 \, u(p_q) \, ;
}
the crossed box diagram shown in the right Fig.~\ref{fig:model1_boxes}
contributes:
\eq{
  i \mathcal{M}^{\mathrm{\RNum{1}}}_{2} =  (-i g_1 m_S)^2 &\bigintssss
  \dfrac {\Diff4l} {(2\pi)^4} \dfrac {i} {(p_S-l)^2 - m_S^2}
  \dfrac {i} {l^2 - m_\phi^2} \dfrac {i} {(l+q)^2 - m_\phi^2}
  \dfrac {i} {(p_q^{\prime} - l)^2 - m_q^2} \nonumber \\
  & \times \bar{u}(p_q^{\prime}) h_2^q \gamma^5 \,
  (\slashed{p}_q^{\prime} - \slashed{l} + m_q) h_2^q \gamma^5 \, u(p_q) \, .
}
After simplifying the numerator by commuting the two $\gamma^5$
matrices and using the Dirac equation, we obtain for the two diagrams:
\eq{
  i \mathcal{M}^{\mathrm{\RNum{1}}}_{1} &= g_1^2 \, (h_2^q)^2 \, m_S^2 \;
  \mathrm{M}^{\mathrm{\RNum{1}}}_{1, \mu} \; \bar{u}(p_q^{\prime})
  \gamma^\mu u(p_q) \quad  \text{and} \\
  i \mathcal{M}^{\mathrm{\RNum{1}}}_{2} &= - \, g_1^2 \, (h_2^q)^2 \, m_S^2 \;
  \mathrm{M}^{\mathrm{\RNum{1}}}_{2, \mu} \; \bar{u}(p_q^{\prime})
  \gamma^\mu u(p_q) \; .
}
Here $\mathrm{M}^{\mathrm{\RNum{1}}}_{1, \mu}$ and
$\mathrm{M}^{\mathrm{\RNum{1}}}_{2, \mu}$ are loop integrals:
\eq{
  \mathrm{M}^{\mathrm{\RNum{1}}}_{1, \mu} &= \bigintssss
  \dfrac {\Diff4l} {(2\pi)^4} \, \dfrac {l_\mu } {[(p_S-l)^2 - m_S^2] \;
    [l^2 - m_\phi^2] \; [(l+q)^2 - m_\phi^2] \; [(p_q + l)^2 - m_q^2]} \; , \\
  \mathrm{M}^{\mathrm{\RNum{1}}}_{2, \mu} &= \bigintssss
  \dfrac{\Diff4l}{(2\pi)^4} \, \dfrac{ l_\mu }{[(p_S-l)^2 - m_S^2] \;
    [l^2 - m_\phi^2] \; [(l+q)^2 - m_\phi^2] \; [(p_q^\prime - l)^2 -
    m_q^2]} \; .
}
After Feynman parametrization, these loop integrals can be expressed
in the vanishing momentum transfer limit ($q \rightarrow 0$, i.e.
$p_S \rightarrow p'_S$ and $p_q \rightarrow p'_q$) in terms of loop
functions $M_i(r=m_q/m_\phi, s=m_S/m_\phi), \ i = 1, \dots, 4$, which are
given in Appendix~\ref{sec:appendix-b-model-i-loop-functions}. The
contributions from the box diagram and crossed box diagrams can
then finally be written as:
\eq{
  \mathcal{M}^{\mathrm{\RNum{1}}}_{1} &= \dfrac {(h_2^q)^2 \, g_1^2 \, m_S^2}
  {16 \pi^2} \, \left[ \left( \dfrac {p_{S, \mu} + p^{\prime}_{S, \mu}} {2}
    \right) \, M_1 \, \bar{u}(p_q^{\prime}) \gamma^\mu u(p_q) + m_q \, M_2 \,
    \bar{u}(p_q^{\prime}) \, u(p_q) \, \right] \; , \\
  \mathcal{M}^{\mathrm{\RNum{1}}}_{2} = - \, &\dfrac{(h_2^q)^2 \, g_1^2 \,
    m_S^2}{16 \pi^2} \, \left[ \left( \dfrac{p_{S, \mu} +
        p^{\prime}_{S, \mu}}{2} \right) \, M_3 \,
    \bar{u}(p_q^{\prime}) \gamma^\mu u(p_q) - m_q \, M_4 \,
    \bar{u}(p_q^{\prime}) \, u(p_q) \, \right] \; .
}
Even though $p_S = p'_S$ for $q=0$ we've written these contributions
in terms of the symmetric sum $p_S + p'_S$, which facilitates matching
onto the effective Lagrangian of
eq.\eqref{eq:model-i-rel-eff-lagrangian} after integrating out the
mediator. 

\begin{figure}[h]
\centering
\includegraphics[width=6.5cm]{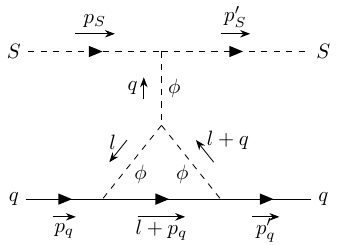}
\includegraphics[width=6.5cm]{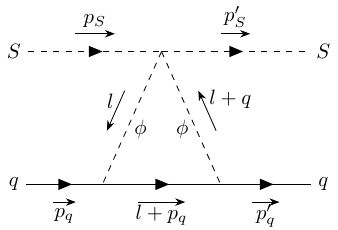}
\caption{One--loop triangle diagrams contributing to $\mathcal{O}_1$
  in Model \RNum{1}.}
\label{fig:model1_tri}
\end{figure}

Fig.~\ref{fig:model1_tri} shows contributions to WIMP--quark scattering
in Model I from triangle diagrams. The matrix element for the left
diagram reads:
\eq{
  i \mathcal{M}^{\mathrm{\RNum{1}}}_{\triangle_1} = (-i m_\phi \, \mu_1) \,
  (-i g_1 m_S) \dfrac {i} {q^2 - m_\phi^2} &\bigintssss 
  \dfrac {\Diff4l} {(2\pi)^4} \dfrac {i} {l^2 - m_\phi^2}
  \dfrac {i} {(l+q)^2 - m_\phi^2} \dfrac {i} {(p_q + l)^2 - m_q^2} \nonumber \\
  &\times \bar{u}(p_q^{\prime}) \, h_2^q \gamma^5 \,
  (\slashed{p}_q + \slashed{l} + m_q) \, h_2^q \gamma^5 \, u(p_q)\;;
}
the right triangle diagram contributes:
\eq{i \mathcal{M}^{\mathrm{\RNum{1}}}_{\triangle_2} = \left(
    \dfrac {-i \, g_2} {2} \right) \, &\bigintssss \dfrac {\Diff4l} {(2\pi)^4}
 \dfrac {i} {l^2 - m_\phi^2} \, \dfrac {i} {(l+q)^2 - m_\phi^2}
  \dfrac {i} {(p_q + l)^2 - m_q^2} \nonumber \\
  & \times  \bar{u}(p_q^{\prime}) \, h_2^q \gamma^5 \,
  (\slashed{p}_q + \slashed{l} + m_q) \, h_2^q \gamma^5 \, u(p_q)\;.
}
Again commuting the two $\gamma^5$ matrices and using the Dirac
equation, we obtain for these two diagrams:
\eq{
  i \mathcal{M}^{\mathrm{\RNum{1}}}_{\triangle_1} &= g_1 \mu_1 (h_2^q)^2 \,
  \dfrac {m_S \, m_\phi} {q^2 - m_\phi^2} \;
  \mathrm{M}^{\mathrm{\RNum{1}}}_{3, \mu} \; \bar{u}(p_q^{\prime})
  \gamma^\mu u(p_q) \quad  \text{and} \\
  i \mathcal{M}^{\mathrm{\RNum{1}}}_{\triangle_2} &= \dfrac{g_2}{2} (h_2^q)^2 \;
  \mathrm{M}^{\mathrm{\RNum{1}}}_{3, \mu} \; \bar{u}(p_q^{\prime})
  \gamma^\mu u(p_q) \;.
}
The loop integral $\mathrm{M}^{\mathrm{\RNum{1}}}_{3, \mu} $ is:
\eq{
  \mathrm{M}^{\mathrm{\RNum{1}}}_{3, \mu} =  \bigintssss
  \dfrac {\Diff4l} {(2\pi)^4} \dfrac {l_\mu} {[l^2 - m_\phi^2] \;
    [(l+q)^2 - m_\phi^2] \; [(p_q + l)^2 - m_q^2]} \; . 
}
In the limit $q \rightarrow 0$ it can be expressed in terms of the
loop function $L_1$, whose analytic expression is given in
Appendix~\ref{sec:appendix-b-model-i-loop-functions}. The contribution
of the two triangle diagrams to the matrix element can then finally be
written as
\eq{
  \mathcal{M}^{\mathrm{\RNum{1}}}_{\triangle_1} &= \dfrac
  {g_1 \mu_1 (h_2^q)^2} {16 \, \pi^2} \, \dfrac {m_q m_S} {m_\phi} \,
  L_1 (m_q^2, m_\phi^2) \,  \; \bar{u}(p_q^{\prime}) \, u(p_q) \; , \\
  \mathcal{M}^{\mathrm{\RNum{1}}}_{\triangle_2} &= -\dfrac {g_2 (h_2^q)^2}
  {16 \, \pi^2} \, \dfrac{m_q }{2} \, L_1(m_q^2, m_\phi^2) \,  \;
  \bar{u}(p_q^{\prime}) \, u(p_q) \; .
}

\subsubsection{Model I Loop Functions}
\label{sec:appendix-b-model-i-loop-functions}

We first define the function $\text{L}(x)$ of the real variable $x$ as:
\eq{ \label{eq:Ldef}
 \text{L}(x) = \left\{ \begin{array}{lcl}
                   \sqrt{1-4x^2} \cdot \ln \left( \frac { 1 + \sqrt{1-4x^2} }
                   {2|x|} \right) \ \ & {\rm for} \ \ & |x| \leq 0.5 \\
                   -\sqrt{4x^2-1} \cdot \arctan \left( \sqrt{4x^2 - 1} \right)
                   \ \
         & {\rm for} \ \ & |x| \geq 0.5 \end{array} \right. \ .
   }
In terms of this function, the loop functions $M_i, \ i = 1, \dots, 4$ and
$L_1$ can be written as:
\eq{
  M_1 = \dfrac {1} {3 \, m_\phi^4} \; &\Bigg[ \dfrac {1} {s \, (r+s)}
  - \dfrac {(1 + 2 r^2) }{2 \, r^2 \, (r+s)^2} \; \text{L}(r)
  + \dfrac {s-2r} {4 \,r^2 \, s^3} \; \ln(1/s^2) \nonumber \\
  &- \dfrac {\ln(r^2/s^2)} {4 \,r^2 \, (r+s)^2} \;
  + \dfrac {2 r + 3 s - 4 r s^2 - 6 s^3 + 8 r s^4 }
  {2 \, s^3 \, (r+s)^2 \, (1 - 4s^2)} \; \text{L}(s) \Bigg] \; ; \\
  M_2 = \dfrac {1} {3 \, m_\phi^4} \; &\Bigg[ -\dfrac {1} {r \, (r+s)}
  + \dfrac { 1 + 2s^2 } {2s^2 \; (r+s)^2} \; \text{L}(s)
  - \dfrac {r-2s} {4 \,r^3 \, s^2} \; \ln( 1/s^2 ) \nonumber \\ 
  &- \dfrac { (3r+2s) \; \ln(r^2/s^2)} {4 \,r^3 \, (r+s)^2} \;
  + \dfrac {2s + 3r - 4r^2s - 6r^3 + 8r^4 s }
  {2 \, r^3 \, (4r^2 -1)\, (r+s)^2} \; \text{L}(r) \Bigg] \; ; \\
  M_3 = \dfrac {1} {3 \, m_\phi^4} \; &\Bigg[ \dfrac {1} {s \, (s-r)}
  - \dfrac {1 + 2 r^2 } {2 \, r^2 \, (r-s)^2} \; \text{L}(r)
  + \dfrac {s+2r} {4 \,r^2 \, s^3} \; \ln(1/s^2) \nonumber \\
  &- \dfrac {\ln(r^2/s^2)} {4 \,r^2 \, (r-s)^2} \;
  - \dfrac {2 r - 3 s - 4 r s^2 + 6 s^3 + 8 r s^4 }
  {2 \, s^3 \, (r-s)^2 \, (1- 4s^2)} \; \text{L}(s) \Bigg]\; ; \\
  M_4 = \dfrac {1} {3 \, m_\phi^4} \; &\Bigg[ -\dfrac {1} {r \, (r-s)}
  + \dfrac {1 + 2s^2} {2s^2 \; (r-s)^2} \; \text{L}(s)
  - \dfrac {r+2s} {4 \,r^3 \, s^2} \; \ln(1/s^2) \nonumber \\ 
  &- \dfrac { (3r-2s)\; \ln(r^2/s^2)} {4 \,r^3 \, (r-s)^2} \;
  + \dfrac{ 2s - 3r - 4r^2s + 6r^3 + 8r^4 s }
  {2 \, r^3 \, (1- 4r^2)\, (r-s)^2} \; \text{L}(r) \Bigg] \; ; \\
  L_1 = \dfrac{1}{m_\phi^2} \, &\left[ \dfrac{1}{r^2}
    + \dfrac {r^2-1} {2 r^4} \, \ln(1/r^2)
    + \dfrac {1-3r^2} {r^4 \, (1-4r^2)}\, \text{L}(r) \right]\; .
}
We note that $M_3(r,s) = M_1(-r,s) = M_1(r,-s)$ and $M_4(r,s) = M_2(-r,s)
= M_2(r,-s)$, i.e. the functions $M_i$ remain invariant when both
arguments change their sign. Of course, in our application only positive
arguments are physical, since $r = m_q/m_\phi$ and $s = m_S/m_\phi$.

In our examples we assume flavor--universal Yukawa couplings. In this
case the contribution from the light quarks will dominate the
WIMP--nucleon scattering matrix elements. We therefore also give the
\textbf{massless quark limits of the loop functions}:
\eq{
  M_1, M_3 &\xrightarrow{ r \to 0 } \dfrac {1} {2 m_\phi^4} \dfrac {1} {s^2}
  \left[ 1 - \dfrac {1} {2s^2} \ln\left( \dfrac{1}{s^2} \right)
    + \dfrac {1-2s^2} {s^2\, (1-4s^2)} \text{L}(s) \right]\; ; \\
  M_2, M_4 &\xrightarrow{ r \to 0 } \dfrac {1} {6 m_\phi^4} \dfrac {1} {s^2}
  \left[ 1 - \dfrac {1} {2s^2} \ln\left( \dfrac{1}{s^2} \right)
    + \dfrac {1+2s^2} {s^2} \text{L}(s) \right] \; ; \\
L_1 &\xrightarrow{ r \to 0 } -\dfrac {1} {2 m_\phi^2} \; .
}

\subsection{Model IIa and IIb Matrix Elements}
\label{sec:appendix-b-model-iia-iib}

\begin{figure}[t!]
\centering
$\vcenter{\hbox{\includegraphics[width=7.5cm]{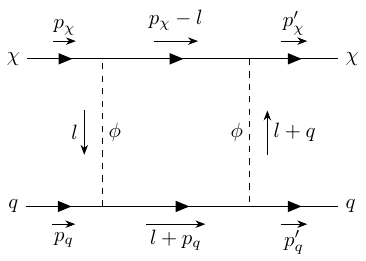}}}$
$\vcenter{\hbox{\includegraphics[width=7.5cm]{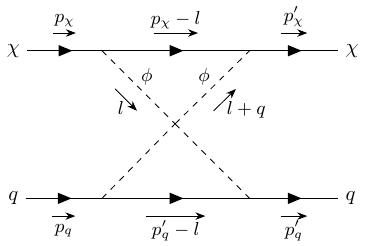}}}$
\caption{One--loop box diagrams contributing to $\mathcal{O}_1$ in Model
  \RNum{2}a or Model \RNum{2}b.}
\label{fig:model2_boxes}
\end{figure}

We now turn to the models with fermionic WIMP $\chi$. We again begin
with the box and crossed box diagrams shown in
Fig.~\ref{fig:model2_boxes}. The contribution to the matrix element
for DM--quark scattering from the box diagram in Model IIa, involving
the scalar DM--mediator coupling $\lambda_1$, is
\eq{
  i \mathcal{M}^{\mathrm{\RNum{2}a}}_{1} = (-i \lambda_1)^2 &\bigintssss
  \dfrac {\Diff4l} {(2\pi)^4} \dfrac {i} { (p_\rchi - l)^2 - m_\rchi^2}
  \dfrac {i} {l^2 - m_\phi^2} \dfrac{i} {(l+q)^2 - m_\phi^2}
  \dfrac{i}{ (p_q + l)^2 - m_q^2 }  \nonumber \\
  &\times  \bar{u}(p_\rchi^{\prime}) (\slashed{p}_\rchi - \slashed{l} + m_\rchi)
  u(p_\rchi) \bar{u}(p_q^{\prime}) h_2^q \gamma^5
  (\slashed{p}_q + \slashed{l} + m_q) h_2^q \gamma^5 u(p_q) \;;
}
the crossed box diagram contributes:
\eq{
  i \mathcal{M}^{\mathrm{\RNum{2}a}}_{2} = (-i \lambda_1)^2 &\bigintssss
  \dfrac {\Diff4l} {(2\pi)^4} \dfrac {i} {(p_\rchi - l)^2 - m_\rchi^2}
  \dfrac {i} {l^2 - m_\phi^2} \dfrac {i} {(l+q)^2 - m_\phi^2}
  \dfrac {i} { (p_q^\prime - l)^2 - m_q^2 } \nonumber \\
  &\times \bar{u}(p_\rchi^{\prime}) (\slashed{p}_\rchi - \slashed{l} + m_\rchi)
  u(p_\rchi) \bar{u}(p_q^{\prime}) h_2^q \gamma^5
  (\slashed{p}_q^{\prime} - \slashed{l} + m_q) h_2^q \gamma^5 u(p_q) \;.
}
After simplifying the numerator by commuting the two $\gamma^5$
matrices and using the Dirac equation for the external quarks and
WIMPs, we obtain for the two diagrams:
\eq{
  i \mathcal{M}^{\mathrm{\RNum{2}a}}_{1} = - \lambda_1^2 \, (h_2^q)^2 \;
  & \Bigr( \mathrm{N}^{\mathrm{\RNum{2}a}}_{1, \mu \nu} \, \left[
    \bar{u}(p_q^{\prime}) \gamma^\mu u(p_q) \right] \left[
    \bar{u}(p_\rchi^{\prime}) \gamma^\nu u(p_\rchi) \right] \nonumber \\
  &- 2 m_\rchi \mathrm{N}^{\mathrm{\RNum{2}a}}_{2, \mu} \left[
    \bar{u}(p_q^{\prime}) \gamma^\mu u(p_q) \right] \left[
    \bar{u}(p_\rchi^{\prime}) u(p_\rchi) \right] \Bigr) \; ;\\
  i \mathcal{M}^{\mathrm{\RNum{2}a}}_{2} = \lambda_1^2 \, (h_2^q)^2 \;
  & \Bigr( \mathrm{N}^{\mathrm{\RNum{2}a}}_{3, \mu \nu} \left[
    \bar{u}(p_q^{\prime}) \gamma^\mu u(p_q) \right]  \left[
    \bar{u}(p_\rchi^{\prime}) \gamma^\nu u(p_\rchi) \right] \nonumber \\
  &- 2 m_\rchi \mathrm{N}^{\mathrm{\RNum{2}a}}_{4, \mu} \left[
    \bar{u}(p_q^{\prime}) \gamma^\mu u(p_q) \right] \left[
    \bar{u}(p_\rchi^{\prime}) u(p_\rchi) \right] \Bigr) \;.
}
Here $\mathrm{N}^{\mathrm{\RNum{2}a}}_{1, \mu \nu}$,
$ \mathrm{N}^{\mathrm{\RNum{2}a}}_{3, \mu \nu} $,
$ \mathrm{N}^{\mathrm{\RNum{2}a}}_{3, \mu} $ and
$ \mathrm{N}^{\mathrm{\RNum{2}a}}_{4, \mu} $ are loop integrals:
\eq{
  \mathrm{N}^{\mathrm{\RNum{2}a}}_{1, \mu \nu} &=  \bigintssss
  \dfrac {\Diff4l} {(2\pi)^4} \, \dfrac {l_\mu  \, l_\nu}
  { [(p_\rchi-l)^2 - m_\rchi^2] \; [l^2 - m_\phi^2] \; [(l+q)^2 - m_\phi^2]
    \; [(p_q + l)^2 - m_q^2]} \; ; \label{eqn:M-IIa-Loop-Integral-1} \\
  \mathrm{N}^{\mathrm{\RNum{2}a}}_{2, \mu} &= \bigintssss
  \dfrac {\Diff4l} {(2\pi)^4} \, \dfrac { l_\mu} {[(p_\rchi-l)^2 - m_\rchi^2]
    \; [l^2 - m_\phi^2] \; [(l+q)^2 - m_\phi^2] \; [(p_q + l)^2 - m_q^2]} \;
  ; \label{eqn:M-IIa-Loop-Integral-2} \\
  \mathrm{N}^{\mathrm{\RNum{2}a}}_{3, \mu \nu} &=  \bigintssss
  \dfrac {\Diff4l} {(2\pi)^4} \, \dfrac {l_\mu \, l_\nu}
  { [(p_\rchi-l)^2 - m_\rchi^2] \; [l^2 - m_\phi^2] \; [(l+q)^2 - m_\phi^2]
    \; [(p_q^\prime - l)^2 - m_q^2]} \; ; \label{eqn:M-IIa-Loop-Integral-3} \\
  \mathrm{N}^{\mathrm{\RNum{2}a}}_{4, \mu} &= \bigintssss
  \dfrac {\Diff4l} {(2\pi)^4} \, \dfrac {l_\mu} { [(p_\rchi-l)^2 - m_\rchi^2]
    \; [l^2 - m_\phi^2] \; [(l+q)^2 - m_\phi^2] \;
    [(p_q^\prime - l)^2 - m_q^2]} \label{eqn:M-IIa-Loop-Integral-4} \; .
}
After Feynman parametrization and taking the limit $q \rightarrow 0$,
the loop integrals in eqs.\eqref{eqn:M-IIa-Loop-Integral-1} to
\eqref{eqn:M-IIa-Loop-Integral-4} can be expressed in terms of loop
functions $N_i, \ i = 1, \dots, 8$ and $P_j, \ j = 1, \dots, 4$ whose
analytic expressions can be found in Appendix
\ref{sec:appendix-model-iia-iib-loop-functions}. Both sets of
functions depend on $r=m_q/m_\phi$ and $s = m_\chi/m_\phi$. The
contributions from the box and crossed box diagrams can then be
written as:
\eq{
  \mathcal{M}^{\mathrm{\RNum{2}a}}_{1}= -\, \dfrac {\lambda_1^2 \, (h_2^q)^2}
  {16 \, \pi^2} \, &\Big\{ N_1 \, \left[\bar{u}(p_\rchi^{\prime}) \,
    \gamma^{\mu} \, \bar{u}(p_\rchi) \right] \;
  \left[ \bar{u}(p_q^{\prime}) \, \gamma_{\mu} \,u(p_q) \right] \nonumber \\
  &+ \Big( 2 m_\rchi m_q \, ( N_2 - P_2 ) + m_\rchi^2 \, ( N_3 - 2 P_1 )
  + m_q^2 \, N_4 \Big) \nonumber \\
  & \hspace*{5mm} \times \left[\bar{u}(p_\rchi^{\prime}) \,
    \bar{u}(p_\rchi) \right] \;  \left[ \bar{u}(p_q^{\prime}) \,u(p_q) \right]
  \Big\}\; ; 
}
\eq{
  \mathcal{M}^{\mathrm{\RNum{2}a}}_{2} = \dfrac {\lambda_1^2 \, (h_2^q)^2}
  {16 \, \pi^2} \, &\Big\{ N_5 \, \left[\bar{u}(p_\rchi^{\prime}) \,
    \gamma^{\mu} \, \bar{u}(p_\rchi) \right] \;  \left[ \bar{u}(p_q^{\prime})
    \, \gamma_{\mu} \,u(p_q) \right] \nonumber  \\
  &+ \Big( 2 m_\rchi m_q \, (N_6 - P_4) + m_\rchi^2 \, (N_7 - 2 P_3)
  + m_q^2 \, N_8 \Big) \nonumber \\
  & \hspace*{5mm} \times \left[\bar{u}(p_\rchi^{\prime}) \, \bar{u}(p_\rchi)
  \right] \;  \left[ \bar{u}(p_q^{\prime}) \,u(p_q) \right] \Big\} \;.
}

The diagrams of Fig.~\ref{fig:model2_boxes} also contribute when both
DM--mediator couplings are $\lambda_2$. The contribution from the box
diagram reads:
\eq{
  i \mathcal{M}^{\mathrm{\RNum{2}a}}_{3} = \lambda_2^2 (h_2^q)^2 &\bigintssss
  \dfrac {\Diff4l} {(2\pi)^4} \dfrac{i} { (p_\rchi - l)^2 - m_\rchi^2}
  \dfrac {i} {l^2 - m_\phi^2} \dfrac {i} {(l+q)^2 - m_\phi^2}
  \dfrac {i} {(p_q + l)^2 - m_q^2} \nonumber \\
  &\times \bar{u}(p_\rchi^{\prime}) \gamma^5
  ( \slashed{p}_\rchi - \slashed{l} + m_\rchi) \gamma^5 u(p_\rchi)
  \bar{u}(p_q^{\prime}) \gamma^5 (\slashed{p}_q + \slashed{l} + m_q)
  \gamma^5 u(p_q)\;.
}
The crossed box contributes:
\eq{
  i \mathcal{M}^{\mathrm{\RNum{2}a}}_{4} = \lambda_2^2 (h_2^q)^2 &\bigintssss
  \dfrac {\Diff4l} {(2\pi)^4} \dfrac {i} {(p_\rchi - l)^2 - m_\rchi^2}
  \dfrac {i} {l^2 - m_\phi^2} \dfrac {i} {(l+q)^2 - m_\phi^2}
  \dfrac {i} { (p_q^\prime - l)^2 - m_q^2 } \nonumber \\
  &\times \bar{u}(p_\rchi^{\prime}) \gamma^5 (\slashed{p}_\rchi - \slashed{l}
  + m_\rchi) \gamma^5 u(p_\rchi) \bar{u}(p_q^{\prime}) \gamma^5
  (\slashed{p}_q^{\prime} - \slashed{l} + m_q) \gamma^5 u(p_q)\;.
}
After commuting the two $\gamma^5$ matrices and using the Dirac
equation, these simplify to
\eq{
  i \mathcal{M}^{\mathrm{\RNum{2}a}}_{3} &= -\lambda_2^2 \, (h_2^q)^2 \;
  \mathrm{N}^{\mathrm{\RNum{2}a}}_{1, \mu \nu} \, \left[
    \bar{u}(p_q^{\prime}) \gamma^\mu u(p_q) \right]  \left[
    \bar{u}(p_\rchi^{\prime}) \gamma^\nu u(p_\rchi) \right]\;; \\
  i \mathcal{M}^{\mathrm{\RNum{2}a}}_{4} &= \lambda_2^2 \, (h_2^q)^2 \;
  \mathrm{N}^{\mathrm{\RNum{2}a}}_{3, \mu \nu} \, \left[ \bar{u}(p_q^{\prime})
    \gamma^\mu u(p_q) \right] \left[ \bar{u}(p_\rchi^{\prime}) \gamma^\nu
    u(p_\rchi) \right]\;,
}
where $\mathrm{N}^{\mathrm{\RNum{2}a}}_{1, \mu \nu}$ and
$\mathrm{N}^{\mathrm{\RNum{2}a}}_{3, \mu \nu} $ are loop integrals
defined in eqs.\eqref{eqn:M-IIa-Loop-Integral-1} and
\eqref{eqn:M-IIa-Loop-Integral-3}. In the $q \rightarrow 0$ limit
we finally obtain:
\eq{
  \mathcal{M}^{\mathrm{\RNum{2}a}}_{3} = -\, \dfrac{\lambda_2^2 \, (h_2^q)^2}
  {16 \, \pi^2} \, &\Big\{ N_1 \, \left[\bar{u}(p_\rchi^{\prime}) \,
    \gamma^{\mu} \, \bar{u}(p_\rchi) \right] \;  \left[ \bar{u}(p_q^{\prime})
    \, \gamma_{\mu} \,u(p_q) \right] \\
  &+ (2 m_\rchi m_q \, N_2 + m_\rchi^2 \, N_3 + m_q^2 \, N_4 ) \, \left[
    \bar{u}(p_\rchi^{\prime}) \, \bar{u}(p_\rchi) \right] \;  \left[
    \bar{u}(p_q^{\prime}) \,u(p_q) \right] \Big\} \nonumber \;; \\
  \mathcal{M}^{\mathrm{\RNum{2}a}}_{4} = \dfrac{\lambda_2^2 \, (h_2^q)^2}
  {16 \, \pi^2} \, &\Big\{ N_5 \, \left[\bar{u}(p_\rchi^{\prime}) \,
    \gamma^{\mu} \, \bar{u}(p_\rchi) \right] \;  \left[ \bar{u}(p_q^{\prime})
    \, \gamma_{\mu} \,u(p_q) \right]  \\
  &+ (2 \, m_\rchi \, m_q \, N_6 + m_\rchi^2 \, N_7 + m_q^2 \, N_8 ) \, \left[
    \bar{u}(p_\rchi^{\prime}) \, \bar{u}(p_\rchi) \right] \;  \left[
    \bar{u}(p_q^{\prime}) \,u(p_q) \right] \Big\} \nonumber \;.
}
The functions $N_i$ already appeared in the contributions
$\propto \lambda_1^2$; they are defined in Appendix
\ref{sec:appendix-model-iia-iib-loop-functions}. This completes the
contribution from the diagrams of Fig~\ref{fig:model2_boxes}, since
diagrams involving one scalar and one pseudoscalar coupling on the
WIMP line do not contribute to $\mathcal{O}_1$.

\begin{figure}[t!]
\centering
\includegraphics[width=6.5cm]{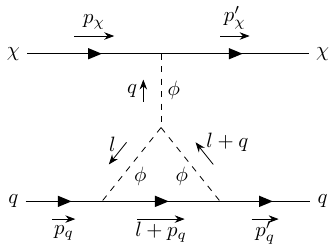}
\caption{One--loop triangle diagram contributing to $\mathcal{O}_1$ in
  Model \RNum{2}a involving the couplings $\lambda_1$, $\mu_1$ and
  $h_2^q$.}
\label{fig:model2_tri}
\end{figure}

There is only one triangle diagram contributing to $\mathcal{O}_1$ at
one--loop order in Model IIa, as shown in Fig.~\ref{fig:model2_tri};
note that only the diagram involving the scalar DM--mediator coupling
$\lambda_1$ contributes to $\mathcal{O}_1$. Its contribution to the
DM--quark scattering matrix element is given by:
\eq{
  i \mathcal{M}^{\mathrm{\RNum{2}a}}_{\triangle} = \dfrac
  { i(-i \lambda_1) (-i m_\phi \mu_1) } {q^2 - m_\phi^2}  \, &\bigintssss
  \dfrac {\Diff4l} {(2\pi)^4} \dfrac {i} {l^2 - m_\phi^2}
  \dfrac {i} {(l+q)^2 - m_\phi^2} \dfrac {i} {(l+q)^2 - m_\phi^2} \\
  &\times \left[ \bar{u}(p_\rchi^{\prime}) \, u(p_\rchi)\right]
  \bar{u}(p_q^{\prime}) \, h_2^q \gamma^5 \, ( \slashed{p}_q + \slashed{l}
  + m_q ) \, h_2^q \gamma^5 \, u(p_q) \;.
}
After simplifying the numerator as before, we get:
\eq{
  \mathcal{M}^{\mathrm{\RNum{2}a}}_{\triangle} = -
  \dfrac {\lambda_1 \, (h_2^q)^2 \, \mu_1} {16 \, \pi^2} \left(
    \dfrac {m_q} {m_\phi} \right) R_1 \left[ \bar{u}(p_\rchi^\prime) \,
    u(p_\rchi) \right] \left[ \bar{u}(p_q^\prime) \, u(p_q)  \right]\;.
}
Here a new loop function $R_1$ appears, which is also defined in Appendix
\ref{sec:appendix-model-iia-iib-loop-functions}.

The calculations for Model IIb are very similar. The (crossed) box
diagrams look exactly the same as in Model IIa, but now we have
scalar Yukawa couplings on the quark line and pseudoscalar couplings
on the WIMP line. We only list the final results. For the box and
crossed box we obtain:
\eq{
  \mathcal{M}^{\mathrm{\RNum{2}b}}_1 = - \, \dfrac{\lambda_2^2 \, (h_1^q)^2}
  {16 \, \pi^2} \, &\Big[ N_1 \, \left[\bar{u}(p_\rchi^{\prime}) \,
    \gamma^{\mu} \, \bar{u}(p_\rchi) \right] \;  \left[ \bar{u}(p_q^{\prime})
    \, \gamma_{\mu} \,u(p_q) \right] \nonumber \\
  &+ \big(2 m_\rchi m_q \, (N_2 + P_1) + m_\rchi^2 \, N_3 + m_q^2 \,
  (N_4 + 2 P_2 ) \big) \nonumber \\
  & \hspace*{5mm} \times \left[\bar{u}(p_\rchi^{\prime}) \, \bar{u}(p_\rchi)
  \right] \;  \left[ \bar{u}(p_q^{\prime}) \,u(p_q) \right] \Big] \\
  \mathcal{M}^{\mathrm{\RNum{2}b}}_2 = \dfrac{\lambda_2^2 \, (h_1^q)^2}
  {16 \, \pi^2} \, &\Big[ N_5 \, \left[\bar{u}(p_\rchi^{\prime}) \,
    \gamma^{\mu} \, \bar{u}(p_\rchi) \right] \;  \left[ \bar{u}(p_q^{\prime})
    \, \gamma_{\mu} \,u(p_q) \right] \nonumber \\
  &+ \big( 2 \, m_\rchi \, m_q \, (N_6-P_3) + m_\rchi^2 \, N_7 + m_q^2 \,
  (N_8-2P_4) \big) \nonumber \\ & \hspace*{5mm} \times
  \left[\bar{u}(p_\rchi^{\prime}) \, \bar{u}(p_\rchi) \right] \;
  \left[ \bar{u}(p_q^{\prime}) \,u(p_q) \right] \Big] \;.
}
The functions $N_i$ and $P_k$ are the same as in Model~IIa, and are
defined in Appendix \ref{sec:appendix-model-iia-iib-loop-functions}.
The Model IIb triangle diagram shown in
Fig.~\ref{fig:model_ii-b-skeleton} yields:
\eq{
  \mathcal{M}^{\mathrm{\RNum{2}b}}_{\triangle} = -\dfrac
  {h_1^q \, \lambda_2^2 \, \mu_1} {16 \, \pi^2} \left( \dfrac {m_\rchi}
    {m_\phi} \right) S_1 \left[ \bar{u}(p_\rchi^\prime) \, u(p_\rchi) \right]
  \left[ \bar{u}(p_q^\prime) \, u(p_q)  \right] \;.
}
Here a new loop function $S_1$ appears; it is also defined in
Appendix \ref{sec:appendix-model-iia-iib-loop-functions}.

\subsubsection{Model IIa and IIb Loop Functions}
\label{sec:appendix-model-iia-iib-loop-functions}

These functions are again expressed in terms of the function $\text{L}(x)$
defined in eq.\eqref{eq:Ldef}.

\eq{
  N_1 = -\dfrac{1}{6 \, m_\phi^2} &\Bigg[ \dfrac {1} {r s} -
  \dfrac {r^2 - 1} {r^3(r+s)} \text{L}(r)
  + \dfrac{rs - s^2 + r^2(3s^2 - 1)}{2 \, r^3 s^3} \ln\Big(
  \dfrac{1}{s^2} \Big) \nonumber \\
  &+ \dfrac {1-3r^2} {2 \, r^3(r+s) } \ln\Big( \dfrac{r^2}{s^2} \Big)
  - \dfrac {s^2-1} {s^3(r+s)} \text{L}(s)  \Bigg] \;;
}
\eq{
  N_2 = \dfrac{1}{30 \, m_\phi^4} &\Bigg[
  \dfrac {3r^2 + 2rs + 3s^2 - 8r^2s^2} {r^2s^2(r+s)^2}
  + \dfrac {5r - 5r^3+3s+r^2s+8r^4s} {r^4(r+s)^3} \text{L}(r) \nonumber \\
  &+ \dfrac {-3r^2+4rs-3s^2+5r^2s^2} {2r^4s^4} \ln\Big( \dfrac{1}{s^2} \Big)
  + \dfrac {5r-15r^3+3s-5r^2s} {2r^4(r+s)^3} \ln\Big( \dfrac{r^2}{s^2} \Big)
  \nonumber \\
&+ \dfrac {3r+5s+rs^2-5s^3+8rs^4} {s^4(r+s)^3}  \text{L}(s) \Bigg] \;;
}
\eq{
  N_3 = \dfrac{1}{15 \, m_\phi^4} &\Bigg[
  \dfrac {6r^2 + 9rs + s^2 + 4r^2s^2} {rs^3(r+s)^2}
  + \dfrac {1 - 3r^2-4r^4} {r^3(r+s)^3} \text{L}(r) \nonumber \\
  &+ \dfrac {-6r^2+3rs-s^2+5r^2s^2} {2r^3s^5} \ln\Big( \dfrac{1}{s^2} \Big)
  + \dfrac {1-5r^2} {2r^3(r+s)^3} \ln\Big( \dfrac{r^2}{s^2} \Big) \\
  &+ \dfrac {6r^2+15rs+10s^2-17r^2s^2-45rs^3-35s^4-2r^2s^4+10s^6+16r^2s^6}
  {s^5(r+s)^3(1-4s^2)} \text{L}(s) \Bigg]\;;  \nonumber
}
\eq{
  N_4 = \dfrac{1}{15 \, m_\phi^4} &\Bigg[
  \dfrac {r^2 + 9rs + 6s^2 + 4r^2s^2} {r^3s(r+s)^2}
  + \dfrac {-r^2+3rs-6s^2+5r^2s^2} {2r^5s^3} \ln\Big( \dfrac{1}{r^2} \Big)
  \nonumber \\
  &+ \dfrac {10r^2-35r^4+10r^6+15rs-45r^3s+6s^2-17r^2s^2-2r^4s^2+16r^6s^2}
  {r^5(1-4r^2)(r+s)^3} \text{L}(r) \nonumber \\
  &+ \dfrac {-1+5s^2} {2s^3(r+s)^3} \ln\Big( \dfrac{r^2}{s^2} \Big)
  + \dfrac {1-3s^2-4s^4} {s^3(r+s)^3} \text{L}(s) \Bigg]\;;
}
\eq{
  N_5 = \dfrac{1}{6 \, m_\phi^2} &\Bigg[ \dfrac{1}{r s}
  - \dfrac {1- r^2} {r^3(r-s)} \text{L}(r)
  + \dfrac {-r^2 - rs -s^2 + 3r^2s^2} {2\,r^3 s^3} \ln\Big( \dfrac{1}{s^2}
  \Big) \nonumber \\
  &+ \dfrac {-1+3r^2} {2 \, r^3(r-s)} \ln\Big( \dfrac{r^2}{s^2} \Big)
  - \dfrac {1-s^2} {s^3(-r+s)} \text{L}(s) \Bigg] \;;
}
\eq{
  N_6 = \dfrac{1}{30 \, m_\phi^4} &\Bigg[
  -\dfrac {3r^2 - 2rs + 3s^2 - 8r^2s^2} {r^2s^2(r-s)^2}
  - \dfrac {5r - 5r^3-3s-r^2s-8r^4s} {r^4(r-s)^3} \text{L}(r) \nonumber \\
  &+ \dfrac {3r^2+4rs+3s^2-5r^2s^2} {2r^4s^4} \ln\Big( \dfrac{1}{s^2} \Big)
  + \dfrac {-5r+15r^3+3s-5r^2s} {2r^4(r-s)^3} \ln\Big( \dfrac{r^2}{s^2} \Big)
  \nonumber \\
&- \dfrac {3r-5s+rs^2+5s^3+8rs^4} {s^4(r-s)^3} \text{L}(s) \Bigg] \;;
}
\eq{
  N_7 = \dfrac{1}{15 \, m_\phi^4} &\Bigg[
  -\dfrac {6r^2 - 9rs + s^2 + 4r^2s^2} {rs^3(r-s)^2}
  + \dfrac {1 - 3r^2-4r^4} {r^3(r-s)^3} \text{L}(r) \nonumber \\
  &+ \dfrac {6r^2+3rs+s^2-5r^2s^2} {2r^3s^5} \ln\Big( \dfrac{1}{s^2} \Big)
  + \dfrac {1-5r^2} {2r^3(r-s)^3} \ln\Big( \dfrac{r^2}{s^2} \Big) \\
  &- \dfrac {6r^2-15rs+10s^2-17r^2s^2+45rs^3-35s^4-2r^2s^4+10s^6+16r^2s^6}
  {s^5(r-s)^3(1-4s^2)} \text{L}(s) \Bigg] \;;  \nonumber
}
\eq{
  N_8 = \dfrac{1}{15 \, m_\phi^4} &\Bigg[
  -\dfrac {r^2 - 9rs + 6s^2 + 4r^2s^2} {r^3s(r-s)^2}
  + \dfrac {r^2+3rs+6s^2-5r^2s^2} {2r^5s^3} \ln\Big( \dfrac{1}{r^2} \Big)
  \nonumber \\
  &+ \dfrac {10r^2-35r^4+10r^6+15rs+45r^3s+6s^2-17r^2s^2-2r^4s^2+16r^6s^2}
  {r^5(1-4r^2)(r-s)^3} \text{L}(r) \nonumber \\
  &+ \dfrac {-1+5s^2} {2s^3(-r+s)^3} \ln\Big( \dfrac{r^2}{s^2} \Big)
  + \dfrac {1-3s^2-4s^4} {s^3(-r+s)^3} \text{L}(s) \Bigg] \;;
}
\eq{
  P_1 = \dfrac{1}{3 \, m_\phi^4} &\Bigg[ \dfrac {1} {s(r+s)}
  - \dfrac {1+2r^2} {2r^2(r+s)^2} \text{L}(r)
  + \dfrac {-2r+s} {4r^2s^3} \ln\Big( \dfrac{1}{s^2} \Big) \nonumber \\
  &- \dfrac {1} {4r^2(r+s)^2} \ln\Big( \dfrac{r^2}{s^2} \Big)
  + \dfrac {2r+3s-4rs^2-6s^3+8rs^4} {2s^3(r+s)^2(1-4s^2)} \text{L}(s)
  \Bigg] \;;
}
\eq{
  P_2 = \dfrac{1}{3 \, m_\phi^4} &\Bigg[ -\dfrac {1} {r(r+s)}
  + \dfrac {-r+2s} {4r^3s^2} \ln\Big( \dfrac{1}{r^2} \Big)
  + \dfrac {1+2s^2} {2s^2(r+s)^2} \text{L}(s)\nonumber \\
  &- \dfrac {1} {4s^2(r+s)^2} \ln\Big( \dfrac{r^2}{s^2} \Big)
  - \dfrac {3r-6r^3+2s-4r^2s+8r^4s} {2r^3(r+s)^2(1-4r^2)} \text{L}(r)
  \Bigg] \;;
}
\eq{
  P_3 = \dfrac{1}{3 \, m_\phi^4} &\Bigg[ -\dfrac {1} {s(r-s)}
  - \dfrac {1+2r^2} {2r^2(r-s)^2} \text{L}(r)
  + \dfrac {2r+s} {4r^2s^3} \ln\Big( \dfrac{1}{s^2} \Big) \nonumber \\
  &- \dfrac {1} {4r^2(r-s)^2} \ln\Big( \dfrac{r^2}{s^2} \Big)
  - \dfrac {2r-3s-4rs^2+6s^3+8rs^4} {2s^3(r-s)^2(1-4s^2)} \text{L}(s)
  \Bigg] \;;
}
\eq{
  P_4 = \dfrac{1}{3 \, m_\phi^4} &\Bigg[ \dfrac {1} {r(r-s)}
  + \dfrac {r+2s} {4r^3s^2} \ln\Big( \dfrac{1}{r^2} \Big)
  - \dfrac {1+2s^2} {2s^2(-r+s)^2} \text{L}(s) \nonumber \\
  &+ \dfrac {1} {4s^2(r-s)^2} \ln\Big( \dfrac{r^2}{s^2} \Big)
  + \dfrac {3r-6r^3-2s+4r^2s-8r^4s} {2r^3(r-s)^2(1-4r^2)} \text{L}(r)
  \Bigg] \;;
}
\eq{
  R_1 = \dfrac{1}{m_\phi^2} \left[ -\dfrac {1} {r^2}
    - \dfrac {-1+r^2} {2r^4} \ln\left( \dfrac{1}{r^2} \right)
    - \dfrac {1-3r^2} {r^4 \, (1-4r^2)} \, \text{L}(r) \right] \;;
}
\eq{
  S_1 = \dfrac {1} {m_\phi^2} \left[ -\dfrac {1} {s^2}
    - \dfrac {-1+s^2} {2s^4} \ln\left( \dfrac {1}{s^2} \right)
    - \dfrac {1-3s^2} {s^4 \, (1-4s^2)} \, \text{L}(s) \right] \;.
}
The functions appearing in the evaluation of the (crossed) box
diagrams are again invariant under a simultaneous sign change of both
arguments. In addition, they are pairwise related: $N_5(r,s) = N_1(-r,s),\
N_6(r,s) = -N_2(-r,s), \ N_7(r,s) = N_3(-r,s), \ N_8(r,s) = N_4(-r,s), \
P_3(r,s) = P_1(-r,s)$ and $P_4(r,s) = -P_2(-r,s)$.

We also again provide the \textbf{massless quark limits of the loop
  functions}. Since $N_3, \; N_7, \; P_1$ and $P_3$ get multiplied
with $1/m_q$, we keep terms up to linear in $r$ in these functions:
\eq{
  P_1, P_3 &\xrightarrow{ r \to 0 } \dfrac {1} {2 m_\phi^4} \left[
    \dfrac {1} {s^2} - \dfrac {1} {2 s^4} \ln\left( \dfrac{1}{s^2} \right)
    + \dfrac {1-2s^2} {s^4} \text{L}(s) \right] \nonumber \\
&\hspace*{4mm} \pm \dfrac {r} {3 s m_\phi^4} \Big[ - \dfrac {2} {s^2}
    + \dfrac {1} {s^4} \ln \left( \dfrac {1}{s^2} \right)
    + \dfrac {4s^4 + 4s^2 - 2} {s^4(1-4s^2)} \text{L}(s) \Big] \;; \\    
  P_2, - P_4 &\xrightarrow{ r \to 0 } \dfrac {1} {6 m_\phi^4} \left[
    \dfrac {1} {s^2} - \dfrac {1} {2 s^4} \ln\left( \dfrac{1}{s^2}\right)
    + \dfrac{1+2s^2} {s^4} \text{L}(s) \right] \;; \\
  N_1, \; N_5 &\xrightarrow{r \rightarrow 0} -\dfrac {1} {6 m_\phi^4} \Big[
  \dfrac {1} {s^2} - \dfrac {1-3s^2} {2s^4} \ln\Big( \dfrac{1}{s^2} \Big)
  + \dfrac {1-s^2} {s^4} \text{L}(s) \Big] \;; \\
  N_2, - N_6 &\xrightarrow{r \rightarrow 0} \dfrac {1} {6 m_\phi^4} \Big[
  \dfrac {1} {s^4} - \dfrac {3} {2 s^2}
  + \dfrac {-1+3s^2} {2s^6} \ln\Big( \dfrac{1}{s^2} \Big)
  + \dfrac {1-s^2} {s^6} \text{L}(s) \Big] \;; \\
  N_3, \; N_7 &\xrightarrow{r \rightarrow 0} \dfrac {1} {3 m_\phi^4} \Big[
  \dfrac {2} {s^4} + \dfrac {3s^2-2} {2s^6} \ln\Big( \dfrac{1}{s^2} \Big)
  + \dfrac {2-7s^2+2s^4} {s^6\, (1-4s^2) } \text{L}(s) \Big] \nonumber \\
&\hspace*{4mm}   \mp \dfrac {r} {2 s^3 m_\phi^4} \Big[ 1 - \dfrac {2} {s^2}
    - \dfrac {2 s^2 - 1} {s^4} \ln \left( \dfrac {1}{s^2} \right)
    - \dfrac {2} {s^4} \left( \dfrac {1-4s^2+2s^4} {1-4s^2} \right) \text{L}(s) \Big] \;; \\
  N_4, \; N_8 &\xrightarrow{r \rightarrow 0} \dfrac {1} {15 m_\phi^4} \Big[
  \dfrac {1} {s^4} - \dfrac {7} {2s^2}
  - \dfrac {1-5s^2} {2s^6} \, \ln\Big(\dfrac{1}{s^2}\Big)
  + \dfrac {1-3s^2-4s^4} {s^6} \text{L}(s) \Big] \;; \\
R_1 &\xrightarrow{ r \to 0 } \dfrac{1}{2 m_\phi^2}  \;.
}
In eq.(A.61) the $+$ sign in the second line refers to $P_1$ and the $-$ sign
to $P_3$; similarly, in the second line of eq.(A.65) the $+$ sign refers
to $N_3$ and the $-$ sign to $N_7$.

%\clearpage
\section{2--Loop Calculations}
\label{app:two-loop-calculations}

In this appendix we provide details of the calculation of the two--loop 
diagrams contributing to the neutron EDM.

\begin{figure}[h]
\centering
\includegraphics[width=7.5cm]{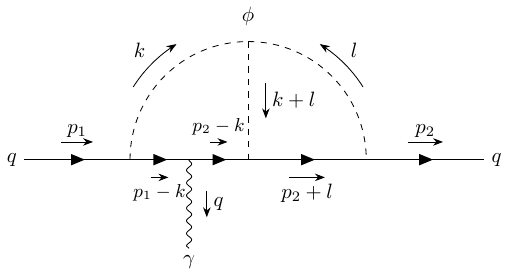}
\hfill
\includegraphics[width=7.5cm]{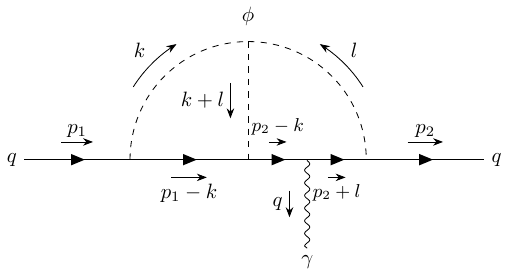}
\caption{Two--loop Feynman diagrams generating quark EDMs and color--EDMs in
  Model \RNum{1} and \RNum{2}a.}
\label{app:two-loop-diagram}
\end{figure}

The matrix elements for diagram 1 (left) and diagram 2 (right) of
Fig.\ref{app:two-loop-diagram} generating EDMs in Model I and IIa can be
written as:
\eq{ \label{eq:dn_matel}
  i \mathcal{M}_{1} &= - \dfrac {i} {(16 \, \pi^2)^2} \, 2 \, eQ_q \, (h_2^q)^3
  \, \mu_1 \, m_\phi \, \epsilon^\star_\mu (q)  \left[ u(p_2) \,
    \mathcal{X}^\mu \, \gamma^5 \, u(p_1) \right] \, ; \\
  i \mathcal{M}_{2} &= - \dfrac {i} {(16 \, \pi^2)^2} \, 2 \, eQ_q \, (h_2^q)^3
  \, \mu_1 \, m_\phi \, \epsilon^\star_\mu (q)  \left[ u(p_2) \,
    \mathcal{Y}^\mu \, \gamma^5 \, u(p_1) \right] \, .
}
Here $Q_q$ is the electric charge of the quark and the loop functions
$\mathcal{X}^\mu = \sum_{i}^4 \mathcal{X}^{\mu}_{i}$ and
$\mathcal{Y}^{\mu} = \sum_{i}^4 \mathcal{Y}^{\mu}_{i}$ are described
below.
\eq{
  \mathcal{X}_1^{\mu} &= \int_{X} \dfrac {-2x_1} {\xi^4}  \, p_2^\mu
    (t_3 + x_1 z_1 t_1) \left[ \dfrac {-3} {\Delta^\prime}
    + \dfrac {1} {\xi^2} \, \dfrac {2(p_1 \cdot p_2) t_2 (t_3 + x_1 z_1 t_1)
      + m_q^2 \, (t_2^2 + (t_3 + x_1 z_1 t_1 )^2 ) } {\Delta^{\prime \, 2}}
  \right] ; \label{eqn:Two-Loop-X-1}
\\
\mathcal{X}_2^{\mu} &= \int_{X} \dfrac {2x_1} {\xi^2} \; p_2^\mu \left[
  \dfrac {-1} {\Delta^\prime} + \dfrac {2} {\xi^2} \,
  \dfrac{ \left( t_2 \, (p_1 \cdot p_2) + m_q^2 \, (t_3 + x_1 z_1 t_1 ) \right)
    (t_3 + x_1 z_1 t_1)} {\Delta^{\prime \, 2}} \right] ;
\label{eqn:Two-Loop-X-3} \\
\mathcal{X}_3^{\mu} &= \int_{X} \dfrac {-2z_1 m_q^2} {\xi^4} \,
\dfrac{ \left( t_3 + x_1 z_1 t_1 - t_2 \right) }
{ \Delta^{\prime \, 2} } \, \left( t_2 \, p_1^\mu + (t_3 + x_1 z_1 t_1) \,
  p_2^\mu \right) ; \label{eqn:Two-Loop-X-4} \\
\mathcal{X}_4^{\mu} &= \int_{X} \dfrac {4z_1 m_q^2} {\xi^2}  \,
\dfrac { (t_3 + x_1 z_1 t_1) } { \Delta^{\prime \, 2} } \, p_2^\mu\,.
\label{eqn:Two-Loop-X-5}
}
Here $t_{1,2,3}$, $x_1$ and $z_1$ are the five Feynman parameters required
to describe diagram 1. The five dimensional measure $\int_X$ is
\eq{
\int_{X} \equiv  \int\displaylimits_{0}^{1} dt_1 \int\displaylimits_{0}^{1-t_1}
dt_2 \int\displaylimits_{0}^{1-t_1-t_2} dt_3 \int\displaylimits_{0}^{1} dx_1
\int\displaylimits_{0}^{1-x_1} dz_1, \quad \xi =
\left[ 1 - t_1 (1-x_1(1-x_1)) \right]^{1/2} .
}
Finally, the denominator $\Delta'$ is given by:
\eq{
  \Delta^\prime = - m_\phi^2 (z_1t_1 + t_2 + t_3 - 1) + z_1^2 m_q^2 t_1
  + \left( \dfrac{x_1 z_1 p_2 t_1 + p_1 t_2 + p_2 t_3}{\xi} \right)^2 \, .
}

The loop functions $\mathcal{Y}_i^\mu$ for the second diagram are
obtained by the following replacements of Feynman parameters in the
$\mathcal{X}_i^\mu$ in eqs.\eqref{eqn:Two-Loop-X-1} to
\eqref{eqn:Two-Loop-X-5}: $t_1 \rightarrow u_1$,
$t_2 \rightarrow u_3$, $t_3 \rightarrow u_2$, $x_1 \rightarrow x_2$
and $z_1 \rightarrow z_2$, as well as the replacement
$p_2^\mu \leftrightarrow p_1^\mu$. This yields:
\eq{	
  \mathcal{Y}_1^{\mu} &= \int_{Y} \dfrac {-2x_2} {\chi^4}  p_1^\mu \left(
    u_2 + x_2 z_2 u_1 \right) \left[ \dfrac{-3} {\Delta^{\prime\prime}}
    + \dfrac {1} {\chi^2} \dfrac {2(p_1 \cdot p_2) u_3 (u_2 + x_2 z_2 u_1)
      + m_q^2 (u_3^2 + (u_2 + x_2 z_2 u_1 )^2 ) }
    {\Delta^{\prime\prime \, 2}}\right] ;  \label{eqn:Two-Loop-Y-1} \\
  \mathcal{Y}_2^{\mu} &= \int_Y \dfrac {2x_2} {\chi^2} \; p_1^\mu \left[
    \dfrac {-1} {\Delta^{\prime\prime}} + \dfrac {2} {\chi^2} \,
    \dfrac { \left( u_3 \, (p_1 \cdot p_2) + m_q^2 \, (u_2 + x_2 z_2 u_1 )
      \right) (u_2 + x_2 z_2 u_1)} {\Delta^{\prime \, 2}}\right] ;
  \label{eqn:Two-Loop-Y-3} \\
  \mathcal{Y}_3^\mu &= \int_Y \dfrac {2z_2 m_q^2} {\chi^4} \, \dfrac{
    \left( u_3 - x_2 z_2 u_1 - u_2\right) } { \Delta^{\prime\prime \, 2} }
  \left( u_3 \, p_2^\mu + (u_2 + x_2 z_2 u_1) \, p_1^\mu \right) ;
  \label{eqn:Two-Loop-Y-4} \\
  \mathcal{Y}_4^\mu &= \int_Y \dfrac {4z_2 m_q^2} {\chi^2}  \,
  \dfrac { (u_2 + x_2 z_2 u_1)} { \Delta^{\prime\prime \, 2} } \, p_1^\mu \, .
  \label{eqn:Two-Loop-Y-5} 
}
The integration measure $\int_Y$ is
\eq{
  \int_{Y} \equiv \int\displaylimits_{0}^{1} du_1 \int\displaylimits_{0}^{1-u_1}
  du_2 \int\displaylimits_{0}^{1-u_1-u_2} du_3 \int\displaylimits_{0}^{1}
  dx_2 \int\displaylimits_{0}^{1-x_2} dz_2 \,, \;
  \chi = \left[ 1 - u_1 (1-x_2(1-x_2)) \right]^{1/2}\,,
}
and the denominator $\Delta^{\prime\prime}$ is given by
\eq{
  \Delta^{\prime \prime} = - m_\phi^2 (z_2u_1 + u_2 + u_3 - 1) + z_2^2 m_q^2 u_1
  + \left( \dfrac{x_2 z_2 p_1 u_1 + p_1 u_2 + p_2 u_3}{\chi} \right)^2 \, .
}
The qEDM is given by the coefficient of the dimension$-5$ $CP-$odd
term $\bar{u}(p_2) i \sigma_{\mu \nu} q^\nu \gamma^5 u(p_1)$ in the
limit of vanishing momentum transfer $q^2 \rightarrow 0$. The
$\gamma^5-$version of the Gordon identity
\eq{
  \bar{u}\, (p_2) (p_1 + p_2)^\mu \gamma^5 \, u(p_1)  =
  \bar{u}(p_2) \, i \sigma^{\mu \nu} q_\nu \gamma^5 \, u(p_1)
}
converts the matrix elements \eqref{eq:dn_matel} into a suitable form
to extract the qEDM. To that end the loop functions $X_i^\mu$ and
$Y_i^\mu$ need to be transformed such that they are symmetric in
the external quark momenta $p_1$ and $p_2$:
\eq{
  \mathcal{X}^{\mu} &= \left[ \mathcal{X} \right] (p_1+p_2)^{\mu}
  +  \{ \mathcal{X} \} \, q^\mu ,\\
  \mathcal{Y}^{\mu} &= \left[ \mathcal{Y} \right] (p_1+p_2)^{\mu}
  +  \{ \mathcal{Y} \} \, q^\mu .
}
The parts of the loop functions proportional to $ q^\mu$, denoted by
$\{ \mathcal{X} \}$ and $\{ \mathcal{Y} \}$ respectively, can be
ignored once the external spinors are taken into account by virtue of
the Ward identity. The qEDM generated by the two diagrams is finally
given by
\eq{
  d_q = \dfrac {2 \, e Q_q \, (h_2^q)^3 \, \mu_1 \, m_\phi} {(16 \, \pi^2)^2}
  \lim_{q^2 \rightarrow 0} \Big( \left[ \mathcal{X} \right]
  + \left[ \mathcal{Y} \right] \Big) \, ,
}
where
\eq{ \label{eq:app:two-loop-X}
  \lim\limits_{q^2 \rightarrow 0} \left[ \mathcal{X} \right] = \dfrac {1} {2}
  &\int_{X} -\dfrac {2z_1 m_q^2} {\xi^4} \dfrac {(t_3 + x_1 z_1 t_1)^2 - t_2^2}
  {(\Delta^{\prime}|_{q^2 \rightarrow 0})^2} + \dfrac {4z_1 m_q^2} {\xi^2} \,
  \dfrac {t_3 + x_1 z_1 t_1} {(\Delta^{\prime}|_{q^2 \rightarrow 0})^2}
   \nonumber \\ &-\dfrac {2x_1} {\xi^4} (t_3 + x_1 z_1 t_1)
   \left[ \dfrac {-3} {\Delta^{\prime}|_{q^2 \rightarrow 0} }
     + \dfrac {m_q^2} {\xi^2} \dfrac {(t_3 + t_2+ x_1 z_1 t_1)^2 }
     { (\Delta^{\prime}|_{q^2 \rightarrow 0})^2 }\right] \nonumber \\
   & + \dfrac {2x_1} {\xi^2} \left[ \dfrac {-1}
     {\Delta^{\prime}|_{q^2 \rightarrow 0} } + \dfrac {2m_q^2}{\xi^2}
     \dfrac{ (t_3 + t_2 + x_1 z_1 t_1) (t_3 + x_1 z_1 t_1) }
     {(\Delta^{\prime}|_{q^2 \rightarrow 0})^2} \right] \, ;
 }
\eq{ \label{eq:app:two-loop-Y}
  \lim\limits_{q^2 \rightarrow 0} \left[ \mathcal{Y} \right] = \dfrac {1} {2}
 &\int_{Y} -\dfrac {2z_2 m_q^2} {\chi^4} \dfrac {(u_2 + x_2 z_2 u_1)^2 - u_3^2}
 {(\Delta^{\prime\prime}|_{q^2 \rightarrow 0})^2} + \dfrac {4z_2 m_q^2}
 {\chi^2} \dfrac {u_2 + x_2 z_2 u_1}
 {(\Delta^{\prime\prime}|_{q^2 \rightarrow 0})^2} \nonumber \\
 & - \dfrac {2x_2} {\chi^4} (u_2 + x_2 z_2 u_1) \left[ \dfrac {-3}
   { \Delta^{\prime\prime}|_{q^2 \rightarrow 0} } + \dfrac {m_q^2} {\chi^2}
   \dfrac { (u_3 + u_2 + x_2 z_2 u_1)^2 }
   { (\Delta^{\prime\prime}|_{q^2 \rightarrow 0})^2 }\right] \nonumber \\
 & +\dfrac {2x_2} {\chi^2} \left[ \dfrac {-1}
   {\Delta^{\prime\prime}|_{q^2 \rightarrow 0} } + \dfrac {2m_q^2} {\chi^2}
   \dfrac { (u_3 + u_2 + x_2 z_2 u_1) (u_2 + x_2 z_2 u_1) }
   {(\Delta^{\prime\prime}|_{q^2 \rightarrow 0})^2} \right]\,.
}
The limit $q^2 \rightarrow 0$ implies $p_1 \cdot p_2 \rightarrow m_q^2$,
hence the denominators simplify to
\eq{
  \Delta^{\prime}|_{q^2 \rightarrow 0} \equiv \lim\limits_{q^2 \rightarrow 0}
  \Delta^{\prime} &= - m_\phi^2 \, (z_1 t_1 + t_2 + t_3 - 1) + m_q^2
  \left(z_1^2 t_1 + \dfrac{(t_2 + t_3 + x_1 z_1 t_1)^2}{\xi^2}  \right) \, ;\\
  \Delta^{\prime\prime}|_{q^2 \rightarrow 0} \equiv
  \lim\limits_{q^2 \rightarrow 0} \Delta^{\prime \prime} &= - m_\phi^2 \,
  (z_2u_1 + u_2 + u_3 - 1) + m_q^2 \left(z_2^2 u_1 +
    \dfrac {(u_3 + u_2 + x_2 z_2 u_1)^2} {\chi^2} \right) \, .
}
%

%\paragraph{Note added.} This is also a good position for notes added
%after the paper has been written.

\bibliography{draft-arxiv-final}
\end{document}